\documentclass[12pt]{amsart}
\usepackage{amsfonts,amsmath,amssymb,subeqnarray}
\usepackage{verbatim}
\usepackage{hyperref}
\usepackage[mathscr]{eucal}
\usepackage{graphicx}
\usepackage{graphics}
\usepackage{color}
\usepackage{url}
\usepackage{tikz}
\usepackage{bm}   
\usetikzlibrary{calc}

\newcommand{\detail}[1]{\par\noi{\bf [Proof detail\ }{#1}
\hfill{\bf ]}\par\noi\hspace{-4pt}}
\renewcommand{\detail}[1]{}

\newcommand{\dis}{\displaystyle}

\newcommand{\noi}{\noindent}
\renewcommand{\qed}{{$\square$ \bigskip}}
\newcommand{\med}{\medskip}

\def\proof{\par\medskip\noindent{\sc Proof.\ }}
\def\proofof#1{\bigskip\noindent{\sc Proof of #1.\ }}

\newtheorem{theorem}{Theorem}[section]
\newtheorem{proposition}[theorem]{Proposition}
\newtheorem{corollary}[theorem]{Corollary}
\newtheorem{conjecture}[theorem]{Conjecture}
\newtheorem{lemma}[theorem]{Lemma}

\newtheorem{remark}[theorem]{Remark}
\newtheorem{defi}[theorem]{Definition}
\newcommand{\bt}{\begin{theorem}}
\newcommand{\et}{\end{theorem}}
\newcommand{\bl}{\begin{lemma}}
\newcommand{\el}{\end{lemma}}
\newcommand{\bp}{\begin{proposition}}
\newcommand{\ep}{\end{proposition}}
\newcommand{\bcor}{\begin{corollary}}
\newcommand{\ecor}{\end{corollary}}
\newcommand{\br}{\begin{remark}\rm}
\newcommand{\er}{\end{remark}}
\newcommand{\bcon}{\begin{conjecture}}
\newcommand{\econ}{\end{conjecture}}
\newcommand{\bd}{\begin{defi}}
\newcommand{\ed}{\end{defi}}

\newcommand{\be}{\begin{equation}}
\newcommand{\ee}{\end{equation}}
\newcommand{\ba}{\begin{array}}
\newcommand{\ea}{\end{array}}
\newcommand{\bc}{\be\begin{array}{r@{\,}c@{\,}l}}
\newcommand{\ec}{\end{array}\ee}
\def\reff#1{(\protect\ref{#1})}

\newcommand{\al}{\alpha}
\newcommand{\bet}{\beta}
\newcommand{\ga}{\gamma}
\newcommand{\Ga}{\Gamma}
\newcommand{\de}{\delta}
\newcommand{\De}{\Delta}
\newcommand{\eps}{\epsilon}

\newcommand{\La}{\Lambda}
\newcommand{\sig}{\sigma}

\newcommand{\Ai}{{\mathcal A}}
\newcommand{\Bi}{{\mathcal B}}

\newcommand{\Ji}{{\mathcal J}}

\newcommand{\Si}{{\mathcal S}}

\newcommand{\R}{{\mathbb R}}
\newcommand{\N}{{\mathbb N}}
\newcommand{\Z}{{\mathbb Z}}

\newcommand{\Hb}{{\mathbb H}}
\newcommand{\Sb}{{\mathbb S}}

\newcommand{\E}{{\mathbb E}}

\newcommand{\desd}{\ensuremath{\Leftrightarrow}}
\newcommand{\volgt}{\ensuremath{\Rightarrow}}
\newcommand{\up}{\uparrow}

\newcommand{\beh}{\backslash}
\newcommand{\symdif}{\!\vartriangle\!}

\newcommand{\dgg}{\dagger}
\newcommand{\ov}{\overline}

\newcommand{\subb}[2]{_{\ba{c}\scriptstyle{#1}\\[-.15cm]\scriptstyle{#2}\ea}}

\newcommand{\pa}{\partial}
\newcommand{\ffrac}[2]{{\textstyle\frac{{#1}}{{#2}}}}

\newcommand{\abs}[1]{{\lvert #1\rvert}}

\usepackage{layout}
\newcommand{\ground}{S_{\rm g}}
\newcommand{\Deo}{{\Delta_0}}
\newcommand{\Dei}{{\Delta_1}}

\renewcommand{\emptyset}{\varnothing}

\newcommand{\subq}{\subseteq}
\newcommand{\subs}{\subset}

\definecolor{forestgreen}   {cmyk}{0.91, 0   , 0.88, 0.12}

\definecolor{jans}   {cmyk}{0,0.8,0.88,0}

\numberwithin{equation}{section}

\newcommand{\finite}{}

\begin{document}


\renewcommand{\labelenumi}{{(\roman{enumi})}}

\vspace*{-2cm}
\begin{center}
\Large\bf Accepted for publication in \\
\Large\bf \emph{Communications in Mathematical Physics}
\end{center}

\vspace*{1cm}

\title[Low-temperature
Potts antiferromagnets\hglue0.5cm]{Entropy-driven phase transition
 in  low-temperature  antiferromagnetic \\ Potts models} 
\author[Roman Koteck\'y, Alan D.~Sokal, and Jan~M.~Swart]{}

\thispagestyle{empty} 
\vspace{0.2cm} 

\centerline
{\sc
Roman Koteck\'{y},\footnote{
  Charles University, Prague, Czech Republic,
  and University of Warwick, United Kingdom,
  {\tt R.Kotecky@warwick.ac.uk}}
\/
Alan D.~Sokal,\footnote{
  Department of Physics, New York University, 4 Washington Place, New York,
  NY 10003, USA,
  and   Department of Mathematics,  University College London,
  Gower Street, London WC1E 6BT, United Kingdom,
  {\tt sokal@nyu.edu}}
\/
Jan Swart\footnote{
  Institute of Information Theory and Automation of the ASCR (\' UTIA),
  Pod vod\'arenskou v\v e\v z\' i~4, 18208 Praha 8, Czech Republic,
  {\tt swart@utia.cas.cz}}
}

\maketitle
\vspace*{-3mm}

\begin{abstract}
We prove the existence of long-range order at sufficiently low temperatures,
including zero temperature, for the three-state Potts antiferromagnet on a
class of quasi-transitive plane quadrangulations, including the diced lattice.
More precisely, we show the existence of (at least) three infinite-volume
Gibbs measures, which exhibit spontaneous magnetization in the sense that
vertices in one sublattice have a higher probability to be in one state than
in either of the other two states. For the special case of the diced lattice,
we give a good rigorous lower bound on this probability, based on
computer-assisted calculations that are not available for the other lattices.
\end{abstract}
\bigskip

\noindent
\\[2mm]
\thanks{
%
}


{\small\setlength{\parskip}{0pt}\tableofcontents}

\section{Introduction and main results}

\subsection{Introduction}

\begin{figure}[t]
\begin{center}
\begin{tabular}{c@{\hspace*{1.5cm}}c}
\includegraphics[width=5.5cm,height=5.5cm,clip]{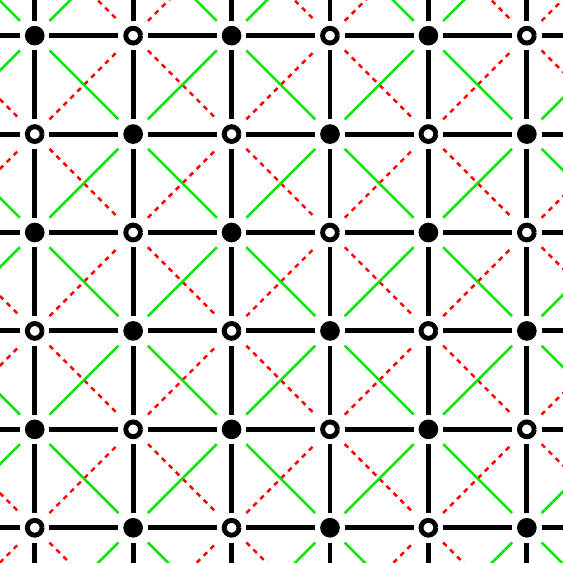}  &
\includegraphics[width=5.5cm,height=5.5cm,clip]{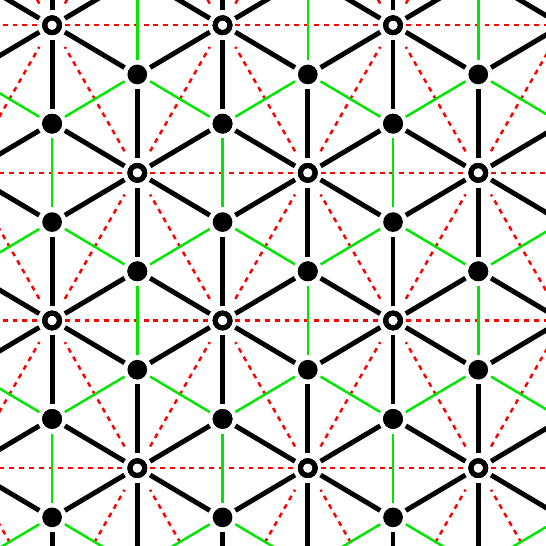}  \\[2mm]
    (a)  &   (b)  \\[1cm]
\includegraphics[width=5.5cm,height=5.5cm,clip]{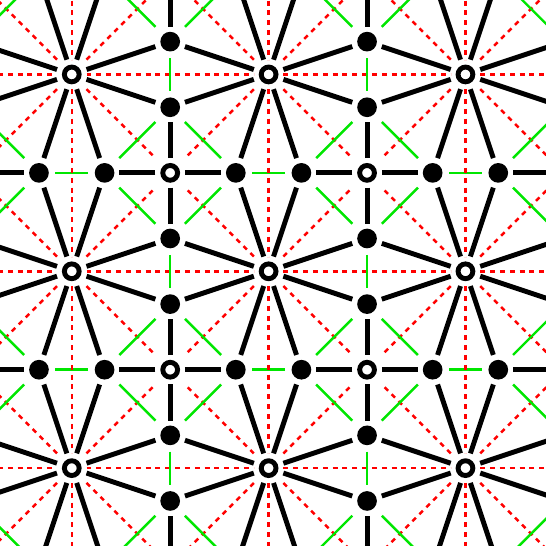} &
\includegraphics[width=5.5cm,height=5.5cm,clip]{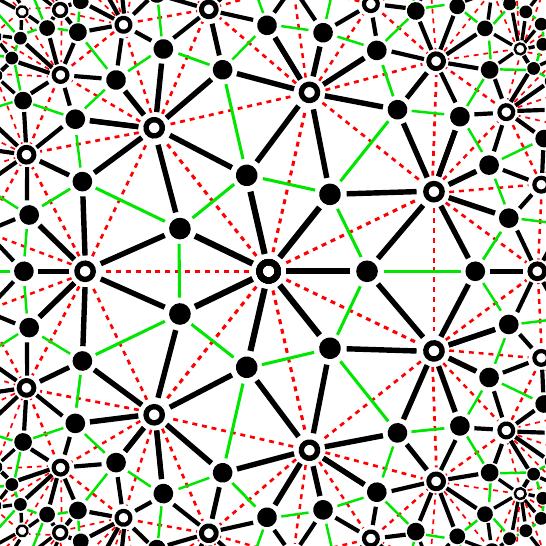} \\[2mm]
    (c)  &   (d)  \\[3mm]
\end{tabular}
\caption{\label{fig:quad}
 Four quasi-transitive quadrangulations with their sublattices
 $G_0$ (open circles joined by dashed red  edges)
 and $G_1$ (filled circles joined by solid green  edges).
 In these examples, the sublattice $G_0$ is
 (a) the square lattice, (b) the triangular lattice,
 (c) the union-jack lattice,
 and (d) the hyperbolic lattice with Schl\"afli symbol $\{3,7\}$.
 Note that in (b)--(d), $G_0$ is a triangulation.
 In (a) and (b), the quadrangulations $G$ are, respectively,
 the square lattice and the diced lattice.
}
\end{center}
\end{figure}

We are interested here in the three-state antiferromagnetic Potts model on a
class of infinite plane quadrangulations. Recall that a graph embedded in the
plane is called a quadrangulation if all its faces are quadrilaterals (i.e.,
have four vertices and four edges).\footnote{
 In this paper we restrict attention to {\em nondegenerate}\/ quadrangulations,
 i.e.\ each face has four {\em distinct}\/ vertices and four {\em distinct}\/
 edges. Some discussion of degenerate plane quadrangulations (in the case of
 finite graphs) can be found in \cite{Jacobsen-Sokal}.
}
Some examples of infinite plane quadrangulations are drawn in
Figure~\ref{fig:quad}: these include the square lattice $\Z^2$ (with
nearest-neighbor edges) and the so-called diced lattice.

On the square lattice, the three-state Potts antiferromagnet at zero temperature
can be mapped into a special case of the six-vertex model that admits an exact
(but nonrigorous) solution \cite[section~8.13]{Bax82}.
This model is therefore believed to be critical at zero temperature
but disordered for any positive temperature.\footnote{
 See e.g.\  the discussion below formula (2.8) in \cite{NNS82}.
 See also \cite{FS99,Salas_98}
 for Monte Carlo data supporting these beliefs.
}

On the diced lattice, by contrast, a proof was outlined in \cite{KSS08}
showing that the three-state Potts antiferromagnet has a phase transition
at nonzero temperature and has long-range order at all sufficiently
low temperatures (including zero temperature).
In the present paper, we present the details of this proof and we extend the
result to a large class of quasi-transitive quadrangulations, including some
hyperbolic lattices.

\begin{figure}[t]
\begin{center}
\begin{tabular}{c@{\hspace*{1.5cm}}c}
\includegraphics[width=5.5cm,height=5.5cm,clip]{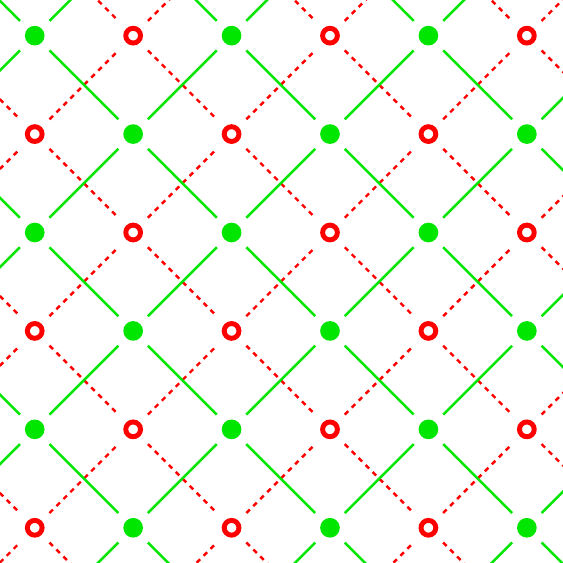}  &
\includegraphics[width=5.5cm,height=5.5cm,clip]{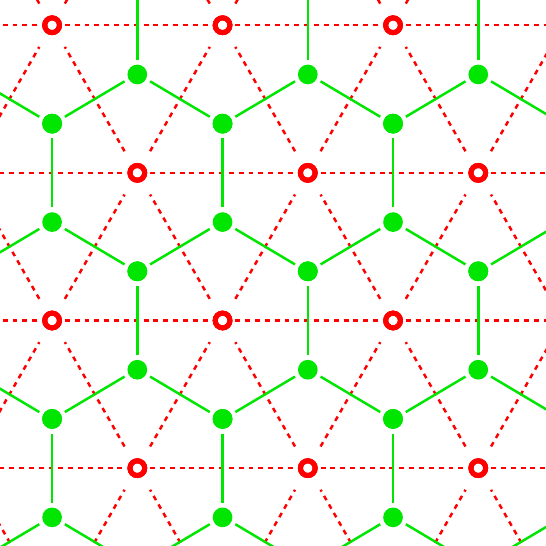}  \\[2mm]
    (a)  &   (b)  \\[1cm]
\includegraphics[width=5.5cm,height=5.5cm,clip]{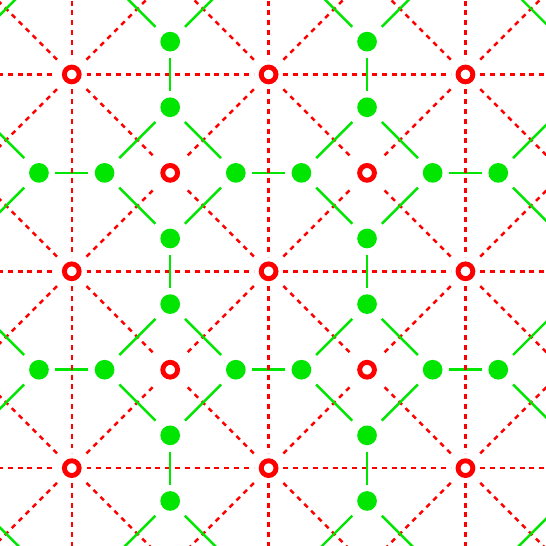} &
\includegraphics[width=5.5cm,height=5.5cm,clip]{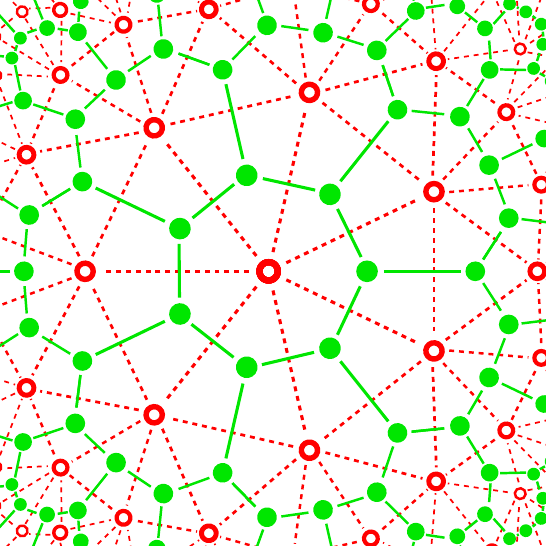} \\[2mm]
    (c)  &   (d)  \\[3mm]
\end{tabular}
\caption{\label{fig:quad2}
 The two sublattices $G_0$ and $G_1$ for each of the four lattices
 from Figure~\ref{fig:quad}.
 The graph $G_0$ (red open circles and dashed edges)
 is dual to the graph $G_1$ (green  filled circles and solid edges).
}
\end{center}
\end{figure}

To explain the class of lattices that we can cover, let us start by observing
that a quadrangulation is a connected bipartite graph $G=(V,E)$, so that the
vertex set has a canonical bipartition $V=V_0\cup V_1$. We may view the two
sublattices $V_0$ and $V_1$ as graphs in their own right by connecting
vertices along the diagonals of the quadrilateral faces of the original
lattice: this yields graphs $G_0=(V_0,E_0)$ and $G_1=(V_1,E_1)$ as shown in
Figure~\ref{fig:quad}. Note that $G_0$ and $G_1$ are duals of each other,
i.e.\ each face of $G_0$ contains a unique vertex of $G_1$, and vice versa;
and each edge of $G_0$ crosses a unique edge of $G_1$, and vice versa (see
Figure~\ref{fig:quad2}). Conversely, given any dual pair of
(finite or infinite) graphs
$G_0 = (V_0,E_0)$ and $G_1 = (V_1,E_1)$ embedded in the plane, we can form a
plane quadrangulation $G=(V,E)$ by setting $V = V_0 \cup V_1$ and placing an
edge between each pair of vertices $v \in V_0$ and $w \in V_1$ where $w$ lies
in a face of $G_0$ that has $v$ on its boundary (or equivalently vice versa).
The main assumption that we will make in this paper
is that one sublattice (say, $G_0$) is a triangulation.
In particular, our proofs cover the lattices shown in (b)--(d)
of Figure~\ref{fig:quad}, but not the square lattice (a).

To explain the nature of the phase transition, note that ground states of the
three-state Potts antiferromagnet are simply proper three-colorings of the
lattice. On any bipartite lattice, we may construct special ground states by
coloring one sublattice (say, $V_0$) in one color and using the other two
colors to color the other sublattice in any possible way. 
Note that in this way, the second sublattice carries all the entropy.
Of course, the special ground states in which the first sublattice uses
{\em only}\/ one color are atypical of the Gibbs measure, even at zero
temperature.  Nevertheless, the underlying idea applies more generally:
there may be a preference for the first sublattice to be colored {\em mostly}\/
in one color because this increases the freedom of choice of colors on the other
sublattice.  Otherwise put, integrating out the colors on the second sublattice
may induce an effective ferromagnetic interaction on the first sublattice.
If this effective interaction is strong enough,
it may result in long-range order on the first sublattice.
We call this an {\em entropy-driven phase transition}\/.\footnote{
 {
 The method for encoding such entropy costs in terms of certain Peierls
 contours was suggested already in \cite{Ko85},
 but in that paper it led to a proof of the transition
 only for some toy models including the three-state Potts antiferromagnet
 on the ``decorated cubic lattice''.
 }
}
In \cite{KSS08} a proof was sketched
along these lines that an entropy-driven transition
indeed occurs on the diced lattice.
Here we will present the details of this proof
and extend it to a large class of plane quadrangulations in which
one sublattice is a triangulation.
The extension uses a variant of the Peierls argument that works
whenever the Peierls sum is finite (even if it is not small),
followed by a random-cluster argument.

In all the cases handled in this paper, there is a strong asymmetry between
the two sublattices, so that it is entropically favorable to ferromagnetically
order the triangulation ($G_0$) and place the entropy on the other sublattice
($G_1$).  By contrast, in the square lattice, where no finite-temperature
phase transition is believed to occur, the two sublattices are isomorphic.
It is therefore natural to ask whether asymmetry is a necessary and/or sufficient
condition for the existence of a finite-temperature phase transition.
This is a subtle question, and we discuss it further in
Section~\ref{subsec.discussion} below.

\subsection{Statement of the results}\label{S:results}

Let us now formulate our results precisely. We first need to define
more precisely the class of graphs we will be considering.
We quickly review here the essential definitions;
a more thorough summary of the needed theory of infinite graphs
can be found in the Appendix.

A graph $G=(V,E)$ is called {\em locally finite}\/ if every vertex has
finitely many neighbors;  in this paper we will consider only locally finite
graphs.  If $G$ has at least $k+1$ vertices, then $G$ is called
{\em $k$-connected}\/ if one needs to remove at least $k$ vertices to
disconnect it.  A graph $G$ is said to have {\em one end}\/ if after the
removal of finitely many edges, there is exactly one
infinite connected component;
note that this implies in particular that $G$ is infinite.

A graph is said to be {\em planar}\/ if it can be drawn in the plane $\R^2$
with vertices represented by distinct points and edges represented by closed
continuous arcs joining their endvertices, mutually disjoint except possibly
at their endpoints. A {\em plane graph}\/ is a planar graph with a given
embedding in the plane.
An embedding of a connected graph in the plane
is called {\em edge-accumulation-point-free}\/
(or {\em EAP-free}\/ for short)
if there are no points in the plane with the property that
each neighborhood of the point
intersects infinitely many edges.%
\footnote{
    The ``one-way-infinite ladder'' plane graph with vertices at points
    $(0,1/n)$ and ($1,1/n)$  [$n=1,2,\dots$]
    and straight-line edges joining the pairs
    $\{(0,1/n), (1,1/n)\}$, $\{(0,1/n), (0, 1/(n+1))\}$ and
    $\{(1,1/n), (1,1/(n+1))\}$
    is an example of a locally finite plane graph that is not EAP-free.
    Indeed, each point $(t,0)$ with $t \in [0,1]$
    is an edge accumulation point.
}
An EAP-free embedding divides
the plane into connected open sets called {\em faces}\/.
The boundary of a face is either a finite cycle
or a two-way-infinite path.%
\footnote{
   By ``two-way-infinite path'' (or ``double ray'')
   we mean simply a graph that is isomorphic to $\Z$
   with nearest-neighbor edges.
   In an EAP-free embedding, a two-way-infinite path
   has no accumulation points in the finite plane $\R^2$,
   but both of its outgoing rays tend to infinity.
   Therefore, in the sphere $\Sb = \R^2 \cup \{\infty\}$,
   the closure of a two-way infinite path is homeomorphic to a circle $\Sb^1$.
   See Appendix~\ref{subsec.planar} for more details
   concerning plane graphs and their faces.
}

Consider an EAP-free embedding of a 3-connected graph $G$.
Then one may define a {\em dual graph}\/ $G^*=(V^*,E^*)$ whose vertex set
$V^*$ is the set of faces of $G$ and with edge set $E^* \simeq E$, where by
definition two faces of $G$  (or equivalently two vertices of $G^*$) are linked by an edge
$e^* \in E^*$ if and only if the corresponding edge $e \in E$
lies in the border of both faces.
Clearly $G^*$ is locally finite if and only if every face of $G$
is bounded by a finite cycle. In this case we can embed $G^*$ in an
EAP-free way in the plane such that each face of $G$ contains exactly
one vertex of $G^*$ and each edge of $G$ is crossed by exactly one edge of
$G^*$, and vice versa; in particular, $G$ is also the dual of $G^*$.
We say that $G$ is a {\em triangulation}\/ (resp.\ {\em quadrangulation}\/)
if every face of $G$ is bounded by a triangle (resp.\ quadrilateral),
or equivalently if each vertex of $G^*$ has degree~3 (resp.\ 4).
It turns out that triangulations and quadrangulations, defined in this way,
are always graphs with at most one end.%
\footnote{
 A more general definition, which allows for multiple ends,
 is discussed in Section~\ref{sec.A.2} of the Appendix.
Briefly, we say that a locally finite 3-connected graph $G$
 is a triangulation (resp.\ quadrangulation) if $G$ has an
 {\em abstract dual}\/ in which each vertex has degree~3 (resp.\ 4);
 see Section~\ref{sec.A.2} for the definition of abstract duals.
 With this more general definition, it can be shown that a triangulation
 (resp.\ quadrangulation) has an EAP-free embedding in the plane
 if and only if it is finite or has one end:
 see Proposition~\ref{P:oneend} in Section~\ref{subsec.planar}.
 }

The final set of definitions we need concerns some form of ``translation
invariance'' of our lattices. An {\em automorphism}\/
of a graph $G=(V,E)$ is a bijection $g\colon\: V\to V$ that preserves the graph
structure. Two vertices $u,v \in V$ are of the {\em same type}\/ if there
exists an automorphism that maps $u$ into $v$. This relation partitions the
vertex set $V$ into equivalence classes called {\em types.}\/ The graph $G$ is
called {\em vertex-transitive}\/ if there is just one equivalence class, and
{\em vertex-quasi-transitive}\/ if there are finitely many equivalence
classes. Edge-transitivity and (for plane graphs) face-transitivity are
defined similarly. The corresponding forms of quasi-transitivity are all
equivalent (see Lemma~\ref{L:quasi} in the Appendix), which is why we simply
talk about quasi-transitivity without specifying whether in the vertex-,
edge- or face-sense.

We will study the three-state antiferromagnetic Potts model on plane
quadrangulations $G$, constructed from mutually dual sublattices $G_0$ and
$G_1$, such that $G_0$ is a locally finite 3-connected quasi-transitive
triangulation with one end.\footnote{ The quasi-transitivity of $G_0$ implies
that also $G_1$ is quasi-transitive: see Theorem~\ref{T:dual}(v) in the
Appendix.  It is not hard to see that now also $G$ must be quasi-transitive.}
Note that quasi-transitivity refers only to the structure of $G_0$ (or $G_1$
or $G$) as an abstract graph (i.e., without reference to any embedding).  It
turns out \cite[Theorem~4.2]{Babai_97} (see also
\cite[Theorem~1]{Servatius_98}) that any locally finite 3-connected
quasi-transitive planar graph with one end can be {\em periodically}\/
embedded in either the Euclidean or hyperbolic plane, i.e., so that the
automorphisms of $G$ correspond to a discrete subgroup of the group of
isometries of the embedding space. But we do not use this fact anywhere in
this paper.

Let us now define the $q$-state Potts antiferromagnet
on an arbitrary infinite graph $G=(V,E)$,
for an arbitrary positive integer $q$.
The state space is the set
\be
S  \;:=\;  [q]^V
\;=\;
\big\{ \sig=(\sig_v)_{v\in V} \colon\:
       \sig_v \in [q] \; \forall v\in V
\big\}
\;,
\ee
where we have used the shorthand notation $[q] = \{1,2,\ldots,q\}$.
We also let
\be
\ground  \;:=\; \big\{\sig\in S \colon\: \sig_u\neq\sig_v\ \forall\{u,v\}\in E\big\}
\ee
denote the set of proper $q$-colorings of $G$.
We sometimes use the terms ``spin configuration'' for $\sig\in S$
and ``ground-state configuration'' for $\sig\in \ground$.
For each finite subset $\La\subs V$ we let
\be
\pa\La  \;:=\; \big\{v\in V\beh\La \colon\: \{v,u\}\in E\mbox{ for some }u\in\La\big\}
\ee
denote the external boundary of $\La$.
For any boundary condition $\tau \colon\: V\to [q]$ and 
any spin configuration $\sig \colon\: \La\to [q]$  on $\La$,
we define the Hamiltonian of $\sig$ under the boundary condition $\tau$ by
\be\label{HLa}
H_\La(\sig\,|\,\tau)  \;:=
\sum\subb{u,v\in\La}{\{u,v\}\in E} \! \delta_{\sig_u,\sig_v}
+\sum\subb{u\in\La,\ v\in\pa\La}{\{u,v\}\in E} \!\!\! \delta_{\sig_u,\tau_v}
\ee
where $\delta_{\sig_u,\sig_v}$ is the Kronecker delta, i.e.
\be
\delta_{ab}  \;=\;
\delta(a,b)   \;=\;    \begin{cases}
                           1  & \text{if $a=b$} \\
                           0  & \text{if $a \neq b$}
                       \end{cases}
\ee
For $\bet \in [0,\infty)$, we define
the Gibbs measure in volume $\Lambda$ with boundary condition $\tau$
at inverse temperature $\bet$:
\be\label{mutau}
\mu^\tau_{\La,\bet}(\sig)  \;:=\;
\frac{1}{Z^\tau_{\La,\bet}}\exp[-\bet H_\La(\sig\,|\,\tau)]
\;.
\ee
For $\bet=\infty$, we define
\be
\mu^\tau_{\La,\infty}(\sig)  \;:=\;
\lim_{\bet\to\infty} \mu^\tau_{\La,\bet}(\sig)
\;.
\ee
That is, $\mu^\tau_{\La,\infty}$ is the uniform distribution on
configurations $\sigma$ that minimize $H_\La(\sig\,|\,\tau)$.
[Note that for some $\tau$ this minimum energy might be strictly positive,
i.e.\ there might not exist proper colorings of $\La \cup \pa\La$
that agree with $\tau$ on $\pa\La$.]
Of course, these definitions actually depend on $\tau$ only via
the restriction $\tau_{\pa\La}:=(\tau_u)_{u\in\pa\La}$ of $\tau$ to $\pa\La$.

We then define infinite-volume Gibbs measures in the usual way through the
Dobrushin--Lanford--Ruelle (DLR) conditions \cite{Georgii_88},
i.e., we say that a probability measure $\mu$ on $S$ is an
infinite-volume Gibbs measure for the $q$-state antiferromagnetic
Potts model at inverse temperature $\bet \in [0,\infty]$
if for each finite $\La\subs V$ its conditional probabilities satisfy
\be
\mu(\sig_{\La} | \,\sig_{V\beh\La} =\tau_{V\beh\La})
\;=\;
\mu^{\tau}_{\La,\bet}(\sig_{\La})
\quad \text{\rm for $\mu$-a.e.\ $\tau$}  \;.
\ee

For the remainder of this paper we specialize to $q=3$.
Here is our main result:

\bt{\bf(Gibbs state multiplicity and positive magnetization)}
\label{T:main}
Let $G=(V,E)$ be a quadrangulation of the plane,
and let $G_0=(V_0,E_0)$ and $G_1=(V_1,E_1)$ be its sublattices,
with edges drawn along the diagonals of quadrilaterals.
Assume that $G_0$ is a locally finite 3-connected quasi-transitive
triangulation with one end.
Then there exist $\bet_0,C<\infty$ and $\eps>0$ such that
for each inverse temperature $\bet\in[\bet_0,\infty]$
and each $k \in \{1,2,3\}$, there exists
an infinite-volume Gibbs measure $\mu_{k,\beta}$ for the
3-state Potts antiferromagnet on $G$ satisfying:
\begin{itemize}
\item[(a)] For all $v_0\in V_0$, we have 
$\mu_{k,\bet}(\sig_{v_0}=k)\geq\frac{1}{3}+\eps$.
\item[(b)] For all $v_1\in V_1$, we have 
$\mu_{k,\bet}(\sig_{v_1}=k)\leq\frac{1}{3}-\eps$.
\item[(c)] For all $\{u,v\}\in E$, we have 
$\mu_{k,\bet}(\sig_u=\sig_v)\leq C e^{-\beta}$.
\end{itemize}
In particular, for each inverse temperature $\beta \in [\beta_0,\infty]$,
the 3-state Potts antiferromagnet on $G$ has at least three distinct
extremal infinite-volume Gibbs measures.
\et

{\bf Remarks.}
1.\ The bound (c) shows in particular that the zero-temperature Gibbs measure
$\mu_{k,\infty}$ is supported on ground states.

2.\ Any subsequential limit as $\beta \to\infty$ of the measures
$\mu_{k,\beta}$ with $\beta < \infty$ also satisfies the bounds (a)--(c).
Therefore, there exist zero-temperature Gibbs measures with these properties
that are limits of finite-temperature Gibbs measures with these
properties.

To see that this is a nontrivial property,
consider on the square lattice $\Z^2$ the configuration $\tau\in S$
of the 3-state Potts antiferromagnet
defined by $\tau_{(i,j)}=1+(i+j \bmod 3)$.
Then $\tau\in \ground$ is a ground-state configuration such
that its restriction to any row and any column, suitably shifted,
is the sequence $(\dots,1,2,3,1,2,3,\dots)$. Since for any
finite $\La\subs\Z^2$ there is precisely one ground-state configuration
that agrees with $\tau$ on $\Z^2\beh\La$, namely $\tau$ itself,
we see that the Dirac measure $\de_\tau$ is a zero-temperature
infinite-volume Gibbs measure.
But this measure is not a limit of positive-temperature Gibbs measures:
the reason can be traced to the fact that at $\beta < \infty$
with boundary condition $\tau$ in a sufficiently large volume $\La$,
there exists a configuration $\bar\tau$
[namely, $\bar\tau_{(i,j)}=1+(i+j \bmod 2)$]
such that we can replace $\tau$ on the internal boundary of $\La$
by $\bar\tau$ at an energetic cost of order $|\partial\Lambda|$,
while gaining a bulk entropic advantage
(with the boundary condition $\bar\tau$, one can color $\La$
 with 1 on one sublattice and arbitrarily 2 or 3 on the other sublattice).

3.\ We construct the infinite-volume Gibbs measures $\mu_{\bet,k}$ as
subsequential limits of finite-volume Gibbs measures. We expect that there is
no need to go to a subsequence and that our approximation procedure yields extremal
infinite-volume Gibbs measures, but we have not proven either of these assertions.

\bigskip

In the special case where $G$ is the diced lattice, we have a good explicit
bound on the probabilities in Theorem~\ref{T:main}(a,b):

\bt{\bf(Quantitative bound for the diced lattice)}
\label{T:dice}
Let $G=(V,E)$ be the diced lattice and let $G_0=(V_0,E_0)$ and $G_1=(V_1,E_1)$
be its triangular and hexagonal sublattices, respectively.
Then there exists $C<\infty$ such that
for each inverse temperature $\bet\in[0,\infty]$
and each $k \in \{1,2,3\}$, there exists 
an infinite-volume Gibbs measure $\mu_{k,\beta}$ for the
3-state Potts antiferromagnet on $G$ satisfying:
\begin{itemize}
\item[(a)] For all $v_0\in V_0$, we have
$\mu_{k,\bet}(\sig_{v_0}=k)\geq 0.90301 - C e^{-\beta}$.
\item[(b)] For all $v_1\in V_1$, we have
$\mu_{k,\bet}(\sig_{v_1}=k)\leq { 0.14549} + C e^{-\beta}$.
\item[(c)] For all $\{u,v\}\in E$, we have
$\mu_{k,\bet}(\sig_u=\sig_v)\leq C e^{-\beta}$.
\end{itemize}
\et

\noindent
The lower bound 0.90301 should be compared with the estimated zero-temperature value
$0.957597 \pm 0.000004$ from Monte Carlo simulations \cite{KSS08}.\footnote{
The value $M_0 = 0.936395 \pm 0.000006$ reported in \cite{KSS08}
is the spontaneous magnetization in the hypertetrahedral representation,
i.e.\ $M_0 = \mu_{1,\infty}(\sigma_v = 1) -
             \frac{1}{2} \mu_{1,\infty}(\sigma_v \neq 1)$.
\label{footnote.M0}
}

\subsection{Discussion}  \label{subsec.discussion}

The phase diagram of non-attractive (i.e., non-ferro\-mag\-ne\-tic)
spin sytems is generally harder to predict
than for attractive (ferromagnetic) spin sytems,
and may sometimes depend subtly on the microscopic details of the model.
In particular, this is true for the two-dimensional 3-state Potts
antiferromagnet, for which we have shown that it has a phase transition
at positive temperature on the diced lattice,
while no such phase transition is believed to occur
on the square lattice --- even though both lattices are bipartite
and are in fact plane quadrangulations.

The existence of a positive-temperature transition in the diced-lattice model
was a surprise when it was first discovered \cite{KSS08}, for the following reason:
Some two-dimensional antiferromagnetic models at zero temperature have the property
that they can be mapped exactly onto a ``height'' model (in general
vector-valued) \cite{Salas_98,Jacobsen_09}. In such cases one can argue
heuristically that the height model must always be in either a ``smooth''
(ordered) or a ``rough'' (massless) phase; correspondingly, the underlying
zero-temperature spin model should either be ordered or critical, never
disordered. Experience teaches us
(or at least seemed to teach us)\ 
that the most common case is criticality.\footnote{
  Some exceptions discussed  in the physics literature prior to \cite{KSS08}
  were the constrained square-lattice 4-state antiferromagnetic
  Potts model \cite{Burton_Henley_97} and the triangular-lattice
  antiferromagnetic spin-$s$ Ising model for large enough
  $s$ \cite{Zeng_Henley_97}, both of which appear to lie in a non-critical
  ordered phase at zero temperature.
}
In particular, when the $q$-state zero-temperature Potts antiferromagnet on a
two-dimensional periodic lattice admits a height representation,
one ordinarily expects that model to have a zero-temperature critical point.
This prediction is confirmed (at least non-rigorously) in most heretofore-studied
cases:  2-state (Ising) triangular \cite{Blote_82,Nienhuis_et_al_84},
3-state square-lattice \cite{NNS82,Kolafa_84,Burton_Henley_97,Salas_98},
3-state kagome \cite{Huse_92,Kondev_96}, 4-state triangular \cite{Moore_00},
and 4-state on the line graph of the square lattice \cite{Kondev_95,Kondev_96}.
Indeed, before the work of \cite{KSS08}, no exceptions were known.

It was furthermore observed in \cite{KSS08}
that the height mapping employed for
the 3-state Potts antiferromagnet on the square lattice \cite{Salas_98}
carries over unchanged to any plane quadrangulation.
One would therefore have expected the 3-state Potts antiferromagnet
to have a zero-temperature critical point
on every periodic plane quadrangulation.
The example of the diced lattice showed that this is not the case;
and the results of the present paper provide further counterexamples.
Clearly, the mere existence of a height representation does {\em not}\/
guarantee that the model will be critical.
Indeed, criticality may well be an exception ---
corresponding to cases with an unusual degree of symmetry ---
rather than the generic case.

The mechanism behind all these transitions is what we have called
an ``entropy-driven phase transition'':
namely, ordering on one sublattice increases the
entropy available to the other sublattice;
or said in a different way, integrating out the spins on the second sublattice
induces an effective ferromagnetic interaction on the first sublattice.
If this effective interaction is strong enough, it may result in
long-range order.
Such a phase transition can therefore occur in principle
in any antiferromagnetic model on any bipartite lattice\footnote{
It can also occur in antiferromagnetic models on
non-bipartite lattices:
for instance, in the 4-state Potts antiferromagnets
on the union-jack and bisected-hexagonal lattices \cite{Deng_11},
which are tripartite,
and for which ordering on one sublattice increases the
entropy available to the other {\em two}\/ sublattices.
However, we are concerned here for simplicity with the bipartite case.
};
whether it actually does occur is a quantitative question
concerning the strength of the induced ferromagnetic interaction.
Thus, such an entropy-driven phase transition is believed not to occur
in the 3-state Potts antiferromagnet on the square lattice $\Z^2$;
but Monte Carlo evidence \cite{WSK90,Gottlob_94a,Gottlob_94b}
suggests that it does occur in this same model on the simple-cubic lattice $\Z^3$
and presumably also on~$\Z^d$ for all $d \ge 3$;
moreover, Peled \cite{Pel10}  and Galvin {\em et al.}\/ \cite{GKRS}
have recently proven this for all sufficiently large $d$
and also for a ``thickened'' version of $\Z^2$  \cite{Pel10}.

{}From the point of view of the Peierls argument, the relevant issue
is the strength of the entropic penalty for domain walls between
differently-ordered regions, compared to the entropy associated to those
domain walls.  In order to successfully carry out the Peierls argument,
one must consider all the relevant ordered phases,
find an appropriate definition of Peierls contours separating spatial regions
resembling those ordered phases, and prove that long Peierls contours $\gamma$
are suppressed like $e^{-c|\gamma|}$ with a sufficiently large constant $c$.

The simplest situation arises when there is an asymmetry between the two sublattices,
so that it is entropically more favorable for one of them (say, $V_0$)
to be ferromagnetically ordered and for the other ($V_1$)
to carry all the entropy.
This situation is expected to occur, for instance,
if $V_1$ has a higher density of points than $V_0$.
The case treated in this paper, in which $G_0$ is a triangulation,
achieves this in the strongest possible way:
namely, for the Euclidean lattices in our class,
it is easy to see using Euler's formula
that the spatial densities%
\footnote{
   We say that a subset $W \subseteq V$ has {\em spatial density $\lambda$}\/
   if, for any sequence of finite sets $V_{(n)}$ increasing to $V$
   such that proportion of vertices in $V_{(n)}$
   that are adjacent to $V \setminus V_{(n)}$ tends to zero
   (i.e., van Hove--F\o{}lner convergence),
   the fraction $|V_{(n)}\cap W| / |V_{(n)}|$ tends to $\lambda$.
}
of the sublattices $V_0$ and $V_1$ are in the proportion 1:2,
which is the most extreme ratio achievable
for two dual periodic Euclidean lattices.

In this asymmetric situation, one knows in advance which sublattice ($V_0$) is
going to be ferromagnetically ordered (if the entropic effect is strong enough
to produce any long-range order at all); therefore, for the 3-state Potts
antiferromagnet on~$G$, one expects at low temperature to have {(at least)}
three distinct ordered phases, corresponding to the three possible choices for
the color that dominates on~$V_0$.  One may therefore define Peierls contours
just as one would for a ferromagnetic Potts model on $G_0$ (see
Section~\ref{S:contour} below for details), and then try to show that long
Peierls contours are sufficiently suppressed, i.e.\ that it is sufficiently
costly to create an interface between regions where one and another color are
used on~$V_0$.  This is a quantitative problem, which is made difficult by the
fact that (unlike in a ferromagnetic model) one does not have any parameter
that can be varied to make the suppression of long contours as large as one
wishes.

The situation is even more delicate for lattices, such as $\Z^d$, where the
two sublattices play a symmetric role (in the sense that there exists an
automorphism of $G$ carrying one sublattice onto the other).  Indeed, for
models with symmetry between the sublattices, for every Gibbs measure where
one sublattice is ordered (in the sense of being colored more often with one
preferred color), there must obviously exist a corresponding Gibbs measure
where the other sublattice is ordered.  Therefore, the system has {\em two}\/
``choices'' to make: first, of the sublattice to be ordered, and then of the
color in which it is ordered --- which leads to a total of {(at least)} six
distinct ordered phases for the 3-state model.  For this sort of long-range
order to occur, it must be sufficiently costly to create an interface between
{\em any}\/ pair of distinct ordered phases; in particular, it must be costly
to create an interface between regions where one and the other sublattice are
ordered (in whatever colors).  To prove such a result will almost certainly
require a different (and more subtle) definition of Peierls contour than is
used in the asymmetric case.

The example of $\Z^d$ for $d$ large  \cite{Pel10,GKRS}
shows that asymmetry is not necessary
for the existence of a finite-temperature phase transition.
But one can nevertheless say heuristically that asymmetry enhances
the effect driving the transition, by increasing the strength of
the effective ferromagnetic interaction on the favored sublattice
(while of course decreasing it on the disfavored sublattice).

Everything said so far holds for an arbitrary bipartite lattice.
But the case in which $G$ is a plane quadrangulation is special,
because $G_0$ and $G_1$ are not merely the two sublattices:
they are a dual pair of plane graphs.\footnote{
 For {\em any}\/ connected bipartite graph $G=(V,E)$
 with vertex bipartition $V = V_0 \cup V_1$,
 one can define graphs $G_0 = (V_0,E_0)$ and $G_1 = (V_1,E_1)$
 by setting $E_0= \big\{ \{u,v\} \colon\: u,v \in V_0 \hbox{ and } d_G(u,v)=2 \big\}$
 and likewise for $E_1$.
 But if $G$ is non-planar, or is planar but not a quadrangulation,
 it is not clear whether these definitions will be useful.
}
In particular, there is a symmetry between $G_0$ and $G_1$
if and only if $G_0$ is {\em self-dual}\/;
and there may be special reasons, connected with the topology of the plane,
that make this self-dual case special (e.g.\ critical at zero temperature).
Now, it is well known that the square lattice is self-dual;
what seems to be less well known\footnote{
 Including to the authors until very recently.
}
is that there exist many other examples
of self-dual periodic plane graphs
\cite{Ashley_91,Okeeffe_92,Okeeffe_96,Servatius_98,Wierman_06,Scullard_06,Ziff_12},
including the ``hextri'' lattice
\cite[Fig.~1]{Okeeffe_92} \cite[Fig.~16]{Servatius_98} \cite[Fig.~1b]{Wierman_06},
the ``house'' lattice \cite[Fig.~2]{Okeeffe_92},
and the martini-B lattice \cite[Fig.~8]{Scullard_06}.
Preliminary results \cite{Deng_private} of Monte Carlo simulations
on a variety of plane quadrangulations suggest that
\begin{quote}
\begin{itemize}
 \item[(a)] If $G_0$ is self-dual, then the 3-state Potts antiferromagnet
    on the associated quadrangulation $G$ has a zero-temperature critical point;
    and
 \item[(b)] If $G_0$ is not self-dual, then the 3-state Potts antiferromagnet
    on $G$ has (always? usually?)\ a finite-temperature phase transition.
\end{itemize}
\end{quote}
In other words, it seems that for plane quadrangulations
--- unlike for general bipartite lattices ---
asymmetry may be both necessary and sufficient for the existence of a
finite-temperature phase transition.
It would be very interesting to find a deeper theoretical explanation,
and ultimately a proof, of this apparent fact.
We conjecture that there is an exact duality mapping that explains why (a) is true.
As for (b), one could argue for it heuristically as follows:
Because the model at zero temperature has a height representation,
it should be either critical or ordered.
If the self-dual cases are critical,
then the non-self-dual cases should be ordered,
since asymmetry enhances the phase transition;
and if the self-dual cases are ordered,
then the non-self-dual cases should be even more strongly ordered.
It goes without saying that this heuristic argument
is extremely vague --- no criterion for comparing lattices is given ---
and hence very far from suggesting a strategy of proof.

Entropy-driven phase transitions are also possible in the
$q$-state Potts antiferromagnet for $q > 3$,
but now one must consider the possibility of Gibbs measures
associated to other partitions $[q] = Q_0 \cup Q_1$,
in which the vertices in $V_0$ (resp.\ $V_1$) take predominantly colors
from $Q_0$ (resp.\ $Q_1$).
Depending on the size and shape of $V_0$ and $V_1$ and the value of $q$,
such measures might be entropically favored.
For instance, such ordering with $|Q_0| = |Q_1| = 2$
has been claimed to occur in the 4-state Potts antiferromagnet
on the simple-cubic lattice $\Z^3$ \cite{Banavar_82,Itakura_99}.
Naive entropic considerations suggest that if the densities of the
sublattices $V_0$ and $V_1$ are in the ratio $\alpha$:$1\!-\!\alpha$,
then the dominant ordering would have $|Q_0| \approx \alpha q$.
In general, one would expect to have
$\binom{q}{|Q_0|}$ ordered phases in the asymmetric case,
and $2 \binom{q}{|Q_0|} = 2 \binom{q}{\lfloor q/2 \rfloor}$ in the symmetric case.
The cases with $|Q_0| > 1$ will require a different (and more subtle)
definition of Peierls contour than the one used here for $|Q_0| = 1$.

The foregoing considerations are purely entropic;
a more complicated phase diagram,
involving tradeoffs between entropy and energy,
can presumably be obtained by adding additional couplings
into the Hamiltonian \reff{HLa}.
Suppose, for instance, that in the 3-state Potts antiferromagnet
on a plane quadrangulation $G$
{where $G_0$ is a triangulation,}
we add an explicit ferromagnetic interaction,
of strength $\lambda$, between adjacent vertices in the sublattice $G_1$.
Then for small $\lambda$ we expect that the favored ordering
at low temperature will be the same as for $\lambda=0$,
namely monocolor on $V_0$ and bicolor on $V_1$;
but for large positive $\lambda$ the favored ordering
will instead be monocolor on $V_1$ and bicolor on $V_0$.
{
It is then natural to guess that for large $\beta$
there is either a switchover between the two orderings
at some particular value $\lambda_t(\beta)$,
or else a pair of phase transitions
$\lambda_{t1}(\beta) < \lambda_{t2}(\beta)$
with a disordered phase in-between
[and possibly $\lambda_{t1}(\infty) = \lambda_{t2}(\infty)$].
It is an interesting open question to determine
the correct qualitative phase diagram
in the $(\lambda,\beta)$-plane
and the order of the phase transition(s).
}

\subsection{Some further open problems}

Here are some further open problems suggested by our work:

\medskip

1) Prove (or disprove) that
\begin{quote}
\begin{quote}
\begin{itemize}
 \item[(a)]  the finite-volume measures $\mu^k_{\La,\bet}$
    used in the proof of Theorem~\ref{T:main} (see Section~\ref{subsec.proofs} below)
    converge as $\La \uparrow V$ (i.e., there is no need to take a subsequence);
 \item[(b)]  the resulting infinite-volume Gibbs measures $\mu_{\bet,k}$
    are {\em extremal}\/ Gibbs measures; and
 \item[(c)]  $\mu_{\bet,k}$ are  invariant with respect to the automorphism of the graph $G$.
\end{itemize}
\end{quote}
\end{quote}

\medskip

2) Prove (or disprove) that for our lattices
there are {\em no more than three}\/ extremal translation-invariant Gibbs measures
at small but strictly positive temperature.
For this, one would need to control more general boundary conditions
than the uniform colorings on $V_0$ that we have used here.

Please note that at {\em zero}\/ temperature,
there are in fact {\em more than three}\/
extremal infinite-volume Gibbs measures on the diced lattice,
since there exist ground-state configurations $\tau$,
similar to the example on $\Z^2$ sketched in
Remark~2 after Theorem~\ref{T:main},
such that for any finite $\La\subs V$ there exists only one ground state
that agrees with $\tau$ on $V\beh\La$ (namely, $\tau$ itself).
The delta measure on such a ground state is therefore
a zero-temperature Gibbs measure;
but by the argument sketched at the end of
that Remark,
this Gibbs measure is not a limit of positive-temperature Gibbs measures.

It is worth pointing out, however, that this latter argument makes essential use of
the fact that the lattice is Euclidean (in particular, its isoperimetric
constant is zero). This raises the question whether on hyperbolic lattices
there might exist delta-measure zero-temperature Gibbs measures that
{\em are}\/ limits of positive-temperature Gibbs measures.

\medskip

{
3) Extend these techniques to the $q$-state Potts antiferromagnet with $q > 3$
on suitable lattices.  For instance, one might hope to prove the existence of
an entropy-driven phase transition in the $q$-state Potts antiferromagnet
on $\Z^d$ for suitable pairs $(q,d)$, i.e., for $q < $ some $q_c(\Z^d)$.
In this case it is not completely clear, even heuristically,
how $q_c(\Z^d)$ should behave as $d \to\infty$.
The example of the infinite $\Delta$-regular tree,
which has multiple Gibbs measures when $q \le \Delta$ \cite{Brightwell_02}
and a unique Gibbs measure when $q \ge \Delta + 1$ \cite{Jonasson_02},
suggests that we might have $q_c(\Z^d) \approx 2d$.
}

\subsection{Plan of this paper}

The remainder of this paper is organized as follows:
In Section~\ref{sec2} we introduce the Peierls-contour representation of our model
and sketch the main ideas underlying our proofs.
In particular, we formulate the key steps in our proof as precise lemmas
(Lemmas~\ref{L:long2}--\ref{L:dice}) that will be proven later,
and we show how they together imply Theorems~\ref{T:main} and \ref{T:dice}.
In Section~\ref{S:zero} we prove Lemmas~\ref{L:long2} and \ref{L:dice}
in the zero-temperature case $\bet=\infty$, using a Peierls argument.
In Section~\ref{S:pos} we extend these proofs to the low-temperature case
$\beta \ge \beta_0$,
and we also prove the technical Lemmas~\ref{L:const2} and \ref{L:ground}.
In Section~\ref{S:magnet} we use a random-cluster argument
to deduce positive magnetization (Lemma~\ref{L:magnet}).
In the Appendix we review the needed theory of infinite graphs.

\section{Main structure of the proofs}  \label{sec2}

\subsection{Contour model}\label{S:contour}

Our proofs of Theorems~\ref{T:main} and \ref{T:dice} are based on suitable
bounds for finite-volume Gibbs measures, uniform in the system size and in the
inverse temperature above a certain value. We will concentrate on
finite-volume Gibbs measures with uniform $1$ boundary conditions on the
sublattice $V_0$; by symmetry, all statements immediately imply analogous
results for boundary conditions $2$ or $3$. We will always employ finite sets
$\Lambda \subs V$ whose external boundary lies entirely in the sublattice
$V_0$, i.e.\ $\partial\Lambda \subs V_0$. We will also assume that $\La\subs
V$ is {\em simply connected,}\/ by which we mean that both $\La$ and
$V\beh\La$ are connected in $G$. Thus, let us fix any
configuration $\tau$ that equals $1$ on $V_0$, and let $\mu^1_{\La,\bet}$
denote the finite-volume Gibbs measure in $\La$ with boundary condition $\tau$
and inverse temperature $\beta \in [0,\infty]$. (Since
$\partial\Lambda \subs V_0$, this measure is the same for all configurations
$\tau$ that equal $1$ on $V_0$.)

Our proofs are based on a version of Peierls argument
relying on a contour reformulation of the measure $\mu^1_{\La,\bet}$.
Our goal is to prove that the sublattice $V_0$ exhibits
ferromagnetic order of a suitable kind.
Therefore we will define Peierls contours just as one would
for studying the ferromagnetic Potts model on $G_0$.
Thus, for any configuration $\sigma$,
we look only at the restriction of $\sigma$ to $V_0$,
and we define $E_0(\sig)$ to be the set of ``unsatisfied edges'', i.e.
\be
E_0(\sig)
\;:=\;
\big\{\{u,v\}\in E_0 \colon\: u,v\in V_0\cap (\La \cup \partial\La)
       \hbox{ and } \sig_u\neq\sig_v \big\}
\;.
\label{def.E0}
\ee
Letting
\be
E_1(\sig)
\;:=\;
\{e\in E_1 \colon\: e\text{ crosses some }f\in E_0(\sig)\}
\label{def.E1}
\ee
denote the edges in the dual graph $G_1$ that cross an edge in $E_0(\sig)$,
we see that edges in $E_1(\sig)$ correspond to boundaries separating areas
where the vertices of $G_0$ are uniformly colored in one of the colors
$1,2,3$. Note that since $\pa\La\subs V_0$ and $\pa\La$ is
uniformly colored, every edge of
$E_1(\sig)$ has {\em both}\/ of its endpoints in $\Lambda$.

Since $G_0$ is a triangulation, each vertex of $G_1$ is of degree~3. If the
three vertices of $G_0$ surrounding a vertex $v\in V_1$ are colored with three
different colors, then one of these vertices must have the same color as
$v$. This is clearly not possible for a ground state $\sig$ (i.e., a proper
coloring), so at $\beta=\infty$ {\em at most two}\/ different colors can surround
any vertex $v \in V_1$. It follows that at zero temperature, either
zero or two edges of $E_1(\sig)$ emanate from the vertex $v$. Hence
$E_1(\sig)$ consists of a collection $\Ga(\sig)$ of disjoint simple circuits
that we call contours.

At positive temperature, we define contours to be connected components of 
$E_1(\sigma)$, which can be much more complicated than a circuit. Nevertheless,
we will show that at low temperatures, contours that are not simple
circuits are rare.

\subsection{The basic lemmas}

Let us now sketch in broad lines the main ideas of our proofs,
and formulate a number of precise lemmas, to be proven later,
that together will imply our main results.
We have seen that uniformly colored areas in the sublattice $G_0$
are separated by contours in the sublattice $G_1$, which at zero temperature
are simple circuits. The number of different simple circuits of a given length $L$
surrounding a given point is roughly of order $\al^L$, where $\al$ is the
connective constant of the lattice $G_1$. Since each vertex in $G_1$ has
degree~3, a contour entering a vertex has two possible directions in which to
continue. In view of this, it is easy to see that $\al\leq 2$. With a bit more
work using quasi-transitivity, this can be improved to $\al<2$.
On the other hand, each vertex $v$ in
$G_1$ that lies on a contour is surrounded by vertices in $G_0$ of two
different colors. At zero temperature, this means that there is only one color
available for $v$, compared to two for a vertex in $G_1$ that does not lie on
a contour. As a result, for each contour of length $L$ we have to pay an
entropic price $2^{-L}$. In view of this, we will prove in
Section~\ref{S:zero} below that the expected number of contours surrounding a
given site is of order $\sum_L\al^L2^{-L}$, which is finite.

Note that this reasoning tells us that the Peierls sum is {\em finite}\/,
but not necessarily that it is {\em small}\/.
In a traditional Peierls argument
(such as, for example, the proof of \cite[Theorem~IV.3.14]{Lig85}),
one argues that if the Peierls contour sum is smaller than a certain
model-dependent threshold (typically a number somewhat less than 1),
then the model has spontaneous magnetization.
This is indeed how we will prove Theorem~\ref{T:dice} for the diced lattice.
But for the general class of lattices in Theorem~\ref{T:main},
all one can hope to prove is that the Peierls sum is finite; it need not be small.
To handle this situation, we use a trick that we learned from
\cite[section~6a]{Dur88}, where it is used for percolation.
We observe that if $\Deo\subs V_0$ is connected in $G_0$,
then $\Deo$ is uniformly colored in one color if and only if no contours
cut through $\Deo$.  On the other hand, if $\Deo$ is sufficiently large,
then by the finiteness of the Peierls sum, $\Deo$ is unlikely to be
surrounded by a contour.  It follows that,
{\em conditional on $\Deo$ being uniformly colored}\/
(which is of course a rare event),
it is much more likely for $\Deo$ to be uniformly colored in the color~1
than in either of the other two colors.

More precisely, for each $k\in\{1,2,3\}$ and each finite set $\Deo\subs V_0$,
let $\Ji_{k,\Deo}$ denote the event that all sites in $\Deo$ have the color $k$,
and let $\Ji_\Deo =\bigcup_{k=1}^3 \Ji_{k,\Deo}$ denote the event that all sites
in $\Deo$ are colored with the same color.
By the arguments sketched above for the case of zero temperature,
together with a careful estimate of non-simple contours
at small positive temperatures, we are able to prove the following lemma:

\bl{\bf(Long-range dependence)}
\label{L:long2}
There exists $\bet_0<\infty$ such that for each $\eps>0$,
there exists $M_\eps<\infty$ such that for every finite set $\Deo\subs V_0$
that is connected in $G_0$ and satisfies $|\Deo|\geq M_\eps$, one has
\be\label{aim2}
\mu^1_{\La,\bet}(\Ji_{1,\Deo}\mid\Ji_{\Deo}) \;\geq\; 1-\eps
\ee  
uniformly for all $\beta \in [\beta_0,\infty]$
and all simply connected finite sets $\La\supseteq\Deo$
such that $\pa\La\subs V_0$.
\el

In order for Lemma~\ref{L:long2} to be of any use, we need to show that the
event on which we are conditioning in (\ref{aim2}) has positive probability,
uniformly in the system size:

\bl{\bf(Uniformly colored sets)}\label{L:const2}
Let $\Deo\subs V_0$ be finite and connected in $G_0$. Then there
exists a constant $\de>0$ such that
\be
\mu^1_{\La,\bet}(\Ji_\Deo) \;\geq\; \de
\ee
uniformly for all $0\leq\bet\leq\infty$ and all finite and simply connected
$\La\supseteq\Deo$ such that $\pa\La\subs V_0$.
\el

\noindent
Of course $\delta$ gets very small as $\Deo$ gets large, but we do not care,
as we will take $\Deo$ to be large but {\em fixed}\/.

Let us note that Lemmas~\ref{L:long2} and \ref{L:const2} are sufficient,
by themselves, to prove the existence of at least three distinct
infinite-volume Gibbs measures at all $\bet\in[\bet_0,\infty]$.\footnote{
 To see this, just follow the proof of Theorem~\ref{T:main}
 given in Section~\ref{subsec.proofs} below
 and disregard all references to Lemma~\ref{L:magnet}.
 The bound \reff{eq.thm1.1.bound} and its analogues for $k=2,3$
 survive to the (subsequential) infinite-volume limit
 and hence show that the Gibbs measures $\mu_{k,\beta}$ for $k=1,2,3$
 are distinct.
}
These infinite-volume Gibbs measures may or may not have spontaneous magnetization,
but they do at least have long-range order of a special kind:
namely, they assign unequal probabilities to the (rare) events $\Ji_{k,\Deo}$
($k=1,2,3$) for some large but finite set $\Deo$.
This part of the argument is quite general and  applies
to  other models as well, as long as it can be shown that
the Peierls sum is finite, even if it is not necessarily small.\footnote{
For example, consider the Ising model on $\Z^2$,
  with the contours constructed as in \cite[Lemma~6.14]{Georgii_88}
  by taking a circuit on the dual lattice that is the external boundary
  of the connected (by nearest-neighbor edges in $\Z^2$)
  region of constant spin containing origin.
  Then this argument gives the existence of at least two distinct
  infinite-volume Gibbs measures whenever $e^{-2\bet} \alpha_{\rm SQ}<1$
  (i.e. $\bet>\frac12\log \alpha_{\rm SQ}$),
  where $\alpha_{\rm SQ}$ is the connective constant for
  self-avoiding polygons
  (or equivalently, self-avoiding walks \cite[Corollary~3.2.5]{Madras-Slade})
  on the square lattice,
  and $\beta$ is the inverse temperature in the standard Ising normalization.
  (It is known that $\alpha_{\rm SQ} < 2.679192495$ \cite{PT00};
   the current best numerical estimate is
   $\alpha_{\rm SQ} \approx 2.63815853031(3)$ \cite{Jensen_03}.
}

But for our particular model, we can actually do better and prove
that there is spontaneous magnetization,
thanks to the following lemma, which says that if a
sufficiently ``thick'' block is more likely to be uniformly colored
in one color than in the other two colors,
then the same must be true for single sites within that block.
Let us say that a set $\Delta \subset V$ is {\em thick}\/\footnote{
 Notice that the sublattices $V_0$ and $V_1$ enter the definition of
 thickness in an asymmetric way.
}
if there exists a nonempty finite subset $\Delta_1 \subset V_1$
that is connected in $G_1$ and such that
$\Delta = \{ v \in V \colon\: d_G(v,\Delta_1) \le 1 \}$.
Then $\Delta$ is connected in $G$, and we have $\Delta_1 = \Delta \cap V_1$;
we write $\Delta_0 := \Delta \cap V_0$. 

\bl{\bf(Positive magnetization)}\label{L:magnet}
Fix $\bet_0 > 0$ and let $\Delta\subset V$ be thick.
Then there exists $\eps>0$ such that for each $v_0\in\Deo$,
\be
\mu^1_{\La,\bet}(\sig_{v_0}=1)-\mu^1_{\La,\bet}(\sig_{v_0}=2)
\;\geq\;
\eps\big[\mu^1_{\La,\bet}(\Ji_{1,\Deo})-\mu^1_{\La,\bet}(\Ji_{2,\Deo})\big]
 \;,
\ee
and similarly, for each $v_1\in\Dei$,
\be
\mu^1_{\La,\bet}(\sig_{v_1}=2)-\mu^1_{\La,\bet}(\sig_{v_1}=1)
\;\geq\;
\eps\big[\mu^1_{\La,\bet}(\Ji_{1,\Deo})-\mu^1_{\La,\bet}(\Ji_{2,\Deo})\big]
 \;,
\ee
uniformly for all $\bet\in[\bet_0,\infty]$ and all simply connected finite
sets $\La\supseteq\Delta$ such that $\pa\La\subs V_0$.
\el

The proof of Lemma~\ref{L:magnet} is not very complicated but is very much
dependent on the specific properties of our Potts model. Inspired by the
cluster algorithm introduced in \cite{WSK89,WSK90}, we condition on the
position of the 3's and use the random-cluster representation for the Ising
model of 1's and 2's on the remaining diluted lattice.
We show that the difference of probabilities that
$\Deo$ is uniformly colored in the color~1 or in the color~2
equals the probability that
$\Deo$ is uniformly colored and
there is a 1-2 {random}-cluster connection between
$\Deo$ and the boundary of $\Lambda$.
Using this latter quantity,
it is then easy to produce (by a finite-energy {argument})
a lower bound on the probability that there is a
1-2 {random}-cluster connection
between a fixed lattice site $v_0$ and the boundary of $\Lambda$;
{and this, in turn,} equals the magnetization.

The last main missing ingredient of Theorem~\ref{T:main} is the following lemma,
which shows that improperly colored edges are rare when $\beta$ is large;
in particular it shows that any limit as $\bet\to\infty$ of the finite-temperature
infinite-volume Gibbs measures that we will construct is concentrated on the
set $\ground$ of ground states.

\bl{\bf(Rarity of improperly colored edges)}
\label{L:ground}
There exists $C<\infty$ such that
\be
\mu^1_{\La,\bet}\big(\sig_u=\sig_v\big)
\;\le\;  Ce^{-\beta}
\ee
for all $\beta\in[0,\infty]$, all $\{u,v\} \in E$,
and all finite and simply connected $\La\ni u,v$ such that $\pa\La\subs V_0$.
\el

Finally, to prove Theorem~\ref{T:dice} for the diced lattice,
we need the following quantitative bound:

\bl{\bf(Explicit Peierls bound for the diced lattice)}
\label{L:dice}
If $G$ is the diced lattice, then there exists $C < \infty$ such that
\be
 \mu^1_{\La,\bet}(\sig_{v_0}=1) \;\geq\;  0.90301 - C e^{-\bet}
\label{dicest2}
\ee
uniformly for all $\beta\in[0,\infty]$, all $v_0 \in V_0$,
and all simply connected finite sets $\La\ni v_0$ such that $\pa\La\subs V_0$.
\el

\subsection{Proof of the main theorems, given the basic lemmas}  \label{subsec.proofs}

Let us now show how to prove Theorems~\ref{T:main} and \ref{T:dice}, given
Lemmas~\ref{L:long2}--\ref{L:dice}.

\proofof{Theorem~\ref{T:main}}
Fix $\eps>0$, let $\bet_0,M_\eps$ and $\Deo$ be as in Lemma~\ref{L:long2},
and let $\de$ be as in Lemma~\ref{L:const2}.
Since the colors $2$ and $3$ play a symmetric role
under the measure $\mu^1_{\La,\bet}$, we have
\be
\mu^1_{\La,\bet}(\Ji_{2,\Deo}\mid\Ji_\Deo)
\;=\;
\ffrac{1}{2}\big[1-\mu^1_{\La,\bet}(\Ji_{1,\Deo}\mid\Ji_\Deo)\big]
\ee
and hence
\begin{eqnarray}
 \dis\mu^1_{\La,\bet}(\Ji_{1,\Deo}) \,-\, \mu^1_{\La,\bet}(\Ji_{2,\Deo})
 & \:=\: &
 \ffrac{1}{2}\big[3\mu^1_{\La,\bet}(\Ji_{1,\Deo}\mid\Ji_\Deo)-1\big] \,
   \mu^1_{\La,\bet}(\Ji_\Deo)
        \nonumber \\[2mm]
 & \:\ge\: &
 \ffrac{1}{2} [3(1-\eps)-1]\de \;=\; \ffrac{1}{2}(2-3\eps)\de
\;,
\label{eq.thm1.1.bound}
\end{eqnarray}
which is positive for $\eps<2/3$ (which we henceforth assume).
Then, for any $v_0\in V_0$, we may choose a thick set $\De \subset V$
such that $|\Deo|\geq M_\eps$ and $v_0\in\Deo$
(with $\Delta_0:=\Delta\cap V_0$ as defined earlier).
By \reff{eq.thm1.1.bound} together with Lemma~\ref{L:magnet},
there exists $\ov\eps(v_0) >0$ such that
\be
 \mu^1_{\La,\bet}(\sig_{v_0}=1) \,-\, \mu^1_{\La,\bet}(\sig_{v_0}=2)
 \;\ge\;
 \ov\eps(v_0)
\ee  
uniformly for all $\bet \in [\beta_0,\infty]$ and all finite and
simply connected $\La\supset\Deo$ such that $\pa\La\subs V_0$.
Since the measure $\mu^1_{\La,\bet}$ treats the colors $2$ and $3$
symmetrically, it follows that
\be
 \mu^1_{\La,\bet}(\sig_{v_0}=1)  \,-\,
 \ffrac{1}{2}\big[1-\mu^1_{\La,\bet}(\sig_{v_0}=1)\big]
 \;\ge\;
 \ov\eps(v_0)
\ee
and hence
\be
\mu^1_{\La,\bet}(\sig_{v_0}=1)
\;\geq\;
\ffrac{1}{3}+\ffrac{2}{3}\ov\eps(v_0)  \;.
\label{aim3}
\ee
Similarly, for any $v_1\in V_1$, we may choose a thick set $\De$
such that $|\Deo|\geq M_\eps$ and $v_1\in\Dei$.
An analogous argument then shows that
\be
\mu^1_{\La,\bet}(\sig_{v_1}=1)
\;{\leq}\;
\ffrac{1}{3}-\ffrac{2}{3}\ov\eps(v_1),
\label{aim3b}
\ee
again uniformly in $\Lambda$.

In this argument $\ov\eps$ depends on $v_0$ or $v_1$.
But since all the quantities under study are invariant under graph automorphisms,
$\ov\eps$ actually depends only on the type of $v_0$ or $v_1$.
And since by quasi-transitivity there are only finitely many types,
we may choose $\ov\eps$ such that \reff{aim3} and \reff{aim3b}
hold uniformly for all $v_0 \in V_0$ and $v_1 \in V_1$.

To construct the desired infinite-volume Gibbs measures,
we use a compactness argument.
For any $\bet \in [0,\infty]$ and finite $\La\subs V$, let $\ov\mu_{\La,\bet}
:=\mu^1_{\La,\bet}\otimes\de_{\tau_{V\beh\La}}$, i.e., if
$\sig\in\{1,2,3\}^V$ is distributed according to $\ov\mu_{\La,\bet}$, then
$(\sig_v)_{v\in\La}$ is distributed according to
$\mu^1_{\La,\bet}$ and $\sig_v=\tau_v$ for all $v\in V\beh\La$.
Choose finite and simply connected $\La_n\up V$ such that
$\pa\La_n\subs V_0$. Since $S=\{1,2,3\}^V$ is a compact space, the
set of measures $\{\ov\mu_{\La_n,\bet}\}$ is automatically tight.
It follows from \cite[Theorem~4.17]{Georgii_88}
that each weak subsequential limit $\mu_\bet$ as $\La_n\up\La$
is an infinite-volume Gibbs measure at inverse temperature $\bet$.
Taking the limit $\La_n\up V$ in \reff{aim3}/\reff{aim3b},
we see that
\begin{eqnarray}
\mu_\bet(\sig_{v_0}=1)  & \ge &  \ffrac{1}{3}+\ov\eps  \qquad\hbox{for } v_0\in V_0
   \label{1eps} \\
\mu_\bet(\sig_{v_1}=1)  & \le &  \ffrac{1}{3}-\ov\eps  \qquad\hbox{for } v_1\in V_1
   \label{2eps}
\end{eqnarray}
Taking the limit $\La_n\up V$ in Lemma~\ref{L:ground}, we obtain
\be
\mu_\bet\big(\sig_u=\sig_v\big)  \;\leq\; C e^{-\beta}  \qquad\hbox{for }
    \{u,v\}\in E  \;.
\label{3eps}
\ee
\qed

\proofof{Theorem~\ref{T:dice}}
(a) and (c) follow from the same arguments as in the proof of Theorem~\ref{T:main},
but with the inequality (\ref{aim3}) replaced by (\ref{dicest2}).

{
To prove (b), consider any $v_1\in V_1\cap\La$
and let $w_1,w_2,w_3\in V_0 \cap (\La \cup \pa\La)$
be its neighbors in $G$.
Then the DLR equations for the volume $\{v_1\}$ imply that
\begin{subeqnarray}
\mu^{1}_{\La,\bet}(\sigma_{v_1} = 1 |
                   \sigma_{w_1} = \sigma_{w_2} = \sigma_{w_3} = 1)
& \;=\; &
\frac{e^{-3\bet}}{2+e^{-3\bet}}
         \\[3mm]
\mu^{1}_{\La,\bet}(\sigma_{v_1} = 1 |
                   \sigma_{w_1} = \sigma_{w_2} = 1,\, \sigma_{w_3} \neq 1)
& \;=\; &
\frac{e^{-2\bet}}{1+e^{-\bet}+e^{-2\bet}}
\label{eq.DLR.1}
\end{subeqnarray}
(we call these the ``good'' cases).
In the ``bad'' cases (i.e., those with two or three spins $\sigma_{w_i} \neq 1$)
we will use only that the conditional probability is $\le 1$.
On the other hand, using (a) we can bound the probability of the ``bad'' cases by
\begin{eqnarray}
 \mu^{1}_{\La,\bet}(\hbox{two or three spins $\sigma_{w_i} \neq 1$})
 & \;\le\; &
 \ffrac{1}{2} \, \E_{\mu^{1}_{\La,\bet}}(\hbox{\# spins $\sigma_{w_i} \neq 1$})
        \nonumber \\[1mm]
 & \;\le\; &
 \ffrac{1}{2} \,\times\, 3 \,\times\, (1 - 0.90301 + C e^{-\beta}) \qquad
        \nonumber \\[1mm]
 & \;\le\; &
 0.14549 \,+\, C' e^{-\beta}  \;,
\label{eq.DLR.2.uncond}
\end{eqnarray}
and we bound the probability of the ``good'' cases trivially by 1.
Putting together (\ref{eq.DLR.1}a,b) and \reff{eq.DLR.2.uncond},
we conclude that
$\mu^{1}_{\La,\bet}(\sigma_{v_1} = 1) \le 0.14549 \,+\, C' e^{-\beta} + e^{-2\beta}$.
}
\hfill\hfill\qed

\section{The zero-temperature case}\label{S:zero}

In the present section, we will prove Lemmas~\ref{L:long2} and \ref{L:dice}
in the zero-temperature case $\bet=\infty$.
Then, in Section~\ref{S:pos}, we will show how our arguments can be
adapted to cover the (more complicated) case of low positive temperatures;
there we will also prove the technical Lemmas~\ref{L:const2} and \ref{L:ground}.
The proof of Lemma~\ref{L:magnet} is postponed to
Section~\ref{S:magnet}.

\subsection{Contour model for zero temperature}\label{S:contour2}

Let $\La\subs V$ be finite and simply connected in $G$ and such that $\pa\La\subs
V_0$. Recall that $\mu^1_{\La,\infty}$ is the uniform distribution on the set
$(\ground)^1_\La$ of all proper 3-colorings of $\La\cup\pa\La$ that take the color
1 on $\pa\La$, i.e.\ all configurations $\sig\in\{1,2,3\}^{\La\cup\pa\La}$
such that $\sig_u\neq\sig_v$ for all $u,v\in\La\cup\pa\La$ with $\{u,v\}\in E$
and $\sig_v=1$ for all $v\in\pa\La$. Since
$\pa\La\subs V_0$, the set $(\ground)^1_\La$ is nonempty: for
instance, it includes all configurations in which all sites of $V_0$ are
colored 1 and all sites of $V_1$ are colored 2 or 3. 

As explained in Section~\ref{S:contour}, for any $\sig\in(\ground)^1_\La$, we let
$E_1(\sig)$ be the collection of edges in $G_1$ that separate areas where the
vertices of $G_0$ are uniformly colored in one of the colors $1,2,3$.
And since at zero temperature at most two different colors on
$V_0$ can meet at any vertex in $V_1$, the set $E_1(\sig)$ consists of a
collection $\Ga(\sig)$ of disjoint simple circuits that we call contours.
[This is what makes the zero-temperature case so easy to handle.
At positive temperature, a connected component of $E_1(\sigma)$
can be much more complicated than a circuit:
see Section~\ref{S:pos} below.]

In the zero-temperature case, therefore, we use the term {\em contour}\/
to denote any simple circuit in $G_1$.
We write $\abs{\ga}$ to denote the length of a contour $\ga$,
defined as the number of its edges (or equivalently the number of its vertices).
For a collection $\Gamma$ of disjoint contours,
we write $\abs{\Ga}:=\sum_{\ga\in\Ga}\abs{\ga}$
for the total length of the contours in $\Ga$,
and $\#\Ga$ for the number of contours in $\Ga$.
Each contour $\ga$ divides $V_0$ into two connected (in
the sense of $G_0$) components, of which one is infinite and the
other is finite and simply connected.
We call these the {\em exterior}\/ ${\rm Ext}(\ga)$
and {\em interior}\/ ${\rm Int}(\ga)$ of $\ga$, respectively.
We will say that a contour $\ga$ {\em surrounds}\/ $\Deo$
if $\Deo\subq{\rm Int}(\ga)$.
We say that a contour lies in $\La$ if all its vertices are in $V_1\cap\La$.
Note that if $\gamma$ lies in $\La$, then by our assumption that
$\La$ is simply connected, we have ${\rm Int}(\ga) \subq \La$.

To each configuration $\sig\in (\ground)^1_\La$, there thus corresponds
a unique collection $\Ga(\sig)$ of disjoint contours in $\La$.
Conversely, to each collection $\Ga$ of disjoint contours,
there are $2^{\#\Ga} 2^{\abs{V_1\cap\La}-\abs{\Ga}}$ distinct
configurations $\sig\in(\ground)^1_\La$ that yield the collection
$\Ga(\sig)=\Ga$. Here the first and second factor are the number
of restrictions $\sig_{V_0\cap\La}$ and $\sig_{V_1\cap \La}$,
respectively, that are consistent with the specified collection of contours
and the fixed boundary condition $\sigma_v = 1$ for $v \in \pa\La$.
To understand the first factor, observe that in passing through any contour
(from outside to inside)
we have 2 ($=q-1$) independent alternatives for the choice of the color
on $V_0$ just inside the contour. The second factor comes from the fact that
there are either one ($=q-2$) or two ($=q-1$) colors
available for a vertex in $G_1$,
depending on whether this vertex lies on a contour or not. Notice
that, given $\Ga$, this latter number is independent of the configuration
$\sig_{V_0\cap\La}$.

Let us therefore introduce the probability measure 
\be\label{nudef}
\nu_\La(\Ga) \;=\; \frac{1}{Z_{\La}} \, 2^{\#\Ga} \, 2^{-|\Ga|}
\ee
on the set of all collections $\Ga$ of disjoint contours in $\La$,
where $Z_{\La}=\sum_\Ga2^{\#\Ga} \, 2^{-|\Ga|}$ is the normalizing constant.
We have just shown that, under the probability measure
$\mu^1_{\La,\infty}$, the contour configuration $\Ga(\sig)$
is distributed according to $\nu_\La$.

{
The first step in any Peierls argument is to obtain
an upper bound on the probability that $\Ga$ contains a given contour $\ga$:

\begin{lemma}{\bf(Bound on probability of a contour)}
 \label{lemma.gamma.T=0}
Let $\Lambda \subs V$ be a simply connected finite set
such that $\partial\Lambda \subs V_0$.
Then, for any contour $\gamma$ in $\Lambda$,
\be
 \nu_{\La}(\{\Ga \colon\: \ga\in\Ga\})
 \;\leq\;
 {2^{1-|\ga|} \over 1+2^{1-|\ga|}}  \;.
\label{eq.lemma.gamma.T=0}
\ee
\end{lemma}

\proof
We have
\begin{eqnarray}
 & &
\dis\nu_{\La}(\{\Ga \colon\: \ga\in\Ga\})
\;=\; \sum_{\Ga\ni\ga}\nu_{\La}(\Ga)
\;=\; 2^{1-|\ga|}\sum_{\Ga\ni\ga}\nu_{\La}(\Ga\beh\{\ga\})
    \nonumber \\[3mm]
 & & \qquad\qquad
 \leq\; 2^{1-|\ga|}\sum_{\Ga\not\ni\ga}\nu_{\La}(\Ga)
\;=\; 2^{1-|\ga|}\big[1-\nu_{\La}(\{\Ga \colon\: \ga\in\Ga\})\big]
 \;,
\label{eq.proof.lemma.gamma.T=0}
\end{eqnarray}
which proves \reff{eq.lemma.gamma.T=0}.
\qed

Now
}
let $\Deo\subq \La\cap V_0$ be connected in $G_0$. Let us say that a contour
$\ga$ {\em cuts}\/ $\Deo$ if $\ga$ contains an edge that separates
some pair of vertices $v,w\in\Deo$ that are adjacent in $G_0$.
Then, obviously,
the event that $\Deo$ is uniformly colored corresponds to the event
that no contour $\ga\in\Ga$ cuts $\Deo$. Let $\nu_{\La|\Deo}$ denote the measure
$\nu_\La$ from (\ref{nudef}) conditioned on this event.
Let $S_\Deo(\Ga)$ denote the number of contours in a contour configuration $\Ga$
that surround $\Deo$.
We obtain a lower bound for the conditional probability in (\ref{aim2})
by writing
\be\ba{l}\label{munu}
\dis\mu^1_{\La,\infty}\big(\Ji_{1,\Deo}\,\big|\,\Ji_\Deo)
\;\geq\; \nu_{\La|\Deo}(\{\Ga \colon\: S_\Deo(\Ga)=0\})\\[8pt]
\dis\quad\geq 1-\sum_\Ga\nu_{\La|\Deo}(\Ga)S_\Deo(\Ga)
\;=\; 1-\!\!\sum_{\ga\colon\, {\rm Int}(\ga) \supseteq \Deo}
\nu_{\La|\Deo}(\{\Ga \colon\: \ga\in\Ga\})
\;.
\ec
Then the probability that $\Ga$ contains a given contour $\ga$
is bounded under $\nu_{\La|\Deo}$ in exactly the same way
as it was bounded under $\nu_\La$ [cf.\ \reff{eq.proof.lemma.gamma.T=0}],
yielding
\be\label{prec}
\nu_{\La|\Deo}(\{\Ga \colon\: \ga\in\Ga\}) \;\leq\;  {2^{1-|\ga|} \over 1+2^{1-|\ga|}}
\ee
for every $\ga$ that surrounds $\Deo$.
Inserting this into (\ref{munu}) yields:

\begin{lemma}{\bf(Peierls bound for zero temperature)}
\label{lemma.peierls}
Let $\Lambda \subs V$ be a simply connected finite set
such that $\partial\Lambda \subs V_0$,
and let $\Deo \subq \Lambda \cap V_0$ be connected in $G_0$.
Then
\be
 1 - \mu^1_{\La,\infty}\big(\Ji_{1,\Deo}\,\big|\,\Ji_\Deo)
 \;\leq\;
 \sum_{\ga\colon\, {\rm Int}(\ga) \supseteq \Deo} {2^{1-|\ga|} \over 1+2^{1-|\ga|}}
 \;=\;
 \sum_{L=3}^\infty N_\Deo(L) \, {2^{1-L} \over 1+2^{1-L}}
 \;,
\label{basic2}
\ee
where $N_\Deo(L)$ denotes the number of contours of length $L$ surrounding
$\Deo$.
\end{lemma}

Our proofs of Lemmas~\ref{L:long2} and \ref{L:dice} in the zero-temperature
case will be based on the estimate (\ref{basic2}) and suitable bounds on the
numbers $N_\Deo(L)$.\med

{\bf Remarks.}
1.  In the special case that $\Deo$ is a singleton, the event
$\Ji_\Deo$ is trivially fulfilled and the conditional probability in
\reff{basic2} reduces to an unconditional probability.

2.  The simpler but slightly weaker bound
\be
1-\mu^1_{\La,\infty}\big(\Ji_{1,\Deo}\,\big|\,\Ji_\Deo)
\;\le\; \sum_{L=3}^\infty N_\Deo(L) \, 2^{1-L}
\label{basic}
\ee
is sufficient for nearly all purposes.
Indeed, even for quantitative bounds the difference between
\reff{basic2} and \reff{basic} is very small:
for instance, when $G$ is the diced lattice and $G_1$ is the hexagonal lattice,
we have $L \ge 6$, so one sees immediately that the difference between
\reff{basic2} and \reff{basic} cannot be more than about 3\%.
See also the proof of Lemma~\ref{L:dice} for $\bet=\infty$
in Section~\ref{S:diced} below.

\subsection{Bounds on contours for zero temperature}  \label{S:contourbounds}

The main ingredient in the proof of Lemma \ref{L:long2}
will be a bound on the number of simple circuits in $G_1$
of a given length surrounding a given vertex in $G_0$.
We start by bounding the number of self-avoiding paths in $G_1$,
or more generally in quasi-transitive graphs of bounded degree.
We then use this bound to obtain a bound on self-avoiding polygons,
i.e.\ simple circuits.

Let $H=(V,E)$ be any graph.
It will be convenient to view $H$ as a directed graph,
by introducing a pair of directed edges (one in each direction)
corresponding to each edge of the undirected graph $H$.
So let $A$ be the set of directed edges of $H$,
i.e., $A$ is the set of all ordered pairs $(v,w)$ of vertices
such that $\{v,w\}\in E$. By definition, a {\em self-avoiding path}\/
in $G$ of length $n$ is a finite sequence of vertices $v_0,\ldots,v_n\in V$,
all different from each other, such that $(v_{k-1},v_k)\in A$ for all $k=1,\ldots,n$.
We call $(v_0,v_1)$ the {\em starting edge}\/
and $(v_{n-1},v_n)$ the {\em final edge}\/ of the path.
For $n \ge 1$ and $a,b \in A$, we denote by $C_n(a,b)$ the number of
self-avoiding paths in $G$ of length $n$ with starting edge $a$ and final edge $b$.
We then set
\begin{equation}
 C_n(a)   \;:=\;   \sum_{b \in A}  C_n(a,b)
 \quad \text{and} \quad
 C_n^*    \;:=\;  \sup_{a \in A} \, C_n(a)  \;.
\end{equation}

\bl{\bf(Exponential bound on self-avoiding paths)}
\label{L:cn}
Let $H=(V,E)$ be an infinite connected graph in which each vertex has
degree at most $k$.  Then the limit
\be
\al(H)  \;:=\; \lim_{n\to\infty} \, (C_n^*)^{1/n}
\label{def.alpha}
\ee
exists and equals $\inf\limits_{n \ge 1} \, (C_{n+1}^*)^{1/n}$;
it satisfies $1\leq\al(H)\leq k-1$.
Furthermore, if $H$ is quasi-transitive and is anything other than
a tree in which every vertex has degree~$k$, then $\al(H) <k-1$.
\el

\proof
For $m,n \ge 1$ and $a,c \in A$ we have
\be
 C_{m+n-1}(a,c)  \;\le\;  \sum_{b \in A}  C_m(a,b) \, C_n(b,c)
\ee
because any self-avoiding path of length $m+n-1$ can be decomposed uniquely
into its  first $m$ steps and
 its  last $n$ steps, each of which is a self-avoiding path,
which overlap in a single directed edge (here called $b$).
This implies the submultiplicativity
\be
 C_{m+n-1}^*  \;\le\;  C_m^* \, C_n^*
 \;.
\label{submultiplicativity}
\ee
We see that $n\mapsto\log C_{n+1}^*$ ($n\ge 0$) is subadditive,
which implies (see, e.g., \cite[Theorem~B.22]{Lig99}) that the limit
\be
\label{Cnsub}
\al(H) \;:=\; \lim_{n\to\infty} (C_{n+1}^*)^{1/n}
  \;=\; \inf_{n\geq 1} \, (C_{n+1}^*)^{1/n}
\ee
exists, with $0\leq\al(H)<\infty$.

By Lemma~\ref{L:geodesic}(a), there exists an infinite self-avoiding path
$(v_0,v_1,v_2,\ldots)$;  so taking $a = (v_0,v_1)$ we see that $C_n(a) \ge 1$
for all $n \ge 1$.  Hence $\alpha(H) \ge 1$.

Since each $v\in V$ is of degree at most $k$,
self-avoidance trivially implies that
\be
 C_{n+1}^*  \;\le\; (k-1)^n
 \;,
\label{eq.k-1_bound}
\ee
so that $\al(H) \leq k-1$.

If $H$ is anything other than a $k$-regular tree, then since $H$ is connected,
for each $a\in A$ there exists an integer $m$
(depending only on the equivalence class of $a$ under the automorphism group of $H$)
such that $C_{m+1}(a)<(k-1)^m$:
it suffices to walk to a vertex of degree $<k$ and then one step more,
or else walk into and around a circuit.
Using the submultiplicativity \reff{submultiplicativity}
together with \reff{eq.k-1_bound},
it follows that $C_{n+1}(a)<(k-1)^n$ for all $n\geq m$.
If now $H$ is (vertex-)quasi-transitive,
then it is  not hard to see
that it is also directed-edge-quasi-transitive
(see Lemma~\ref{L:quasi} for a  proof),
i.e.\ there are finitely many equivalence classes of directed edges,
so we can choose an $m$ that works for all $a \in A$.
It follows that $C_{n+1}^* <(k-1)^n$ for some $n$
(in fact for all sufficiently large $n$),
which shows that the infimum in \reff{Cnsub} is strictly less than $k-1$.
\qed

{\bf Remark.}
Most of this proof can alternatively be carried out in terms of
the more familiar vertex-to-vertex counts $c_n(u,v)$ for $n \ge 0$
and the corresponding quantities
$c_n^* = \sup_{u \in V} \sum_{v \in V} c_n(u,v)$.
(Since $C_n^* \le c_n^* \le k C_n^*$,
the two counts have identical asymptotic growth.)
Indeed for $m,n \ge 0$ we have
\be
  c_{m+n}(u,w)  \;\le\;  \sum_{v \in V}  c_m(u,v) \, c_n(v,w)
\ee
and hence $c_{m+n}^* \le c_m^* \, c_n^*$,
from which it follows that
\be
  \al(H)
  \;=\; \lim_{n\to\infty} \, (c_n^*)^{1/n}
  \;=\; \inf_{n\geq 1} \, (c_n^*)^{1/n}
\ee
exists.
But it is more difficult in this framework to prove that $\alpha(H) < k-1$,
since the bound $c_n^* \le k(k-1)^{n-1}$ has an extra factor $k/(k-1)$
that we must somehow overcome.
It is for this reason that we found it convenient to work with directed edges
instead of vertices.

\bigskip

It follows from \reff{def.alpha} that for each $\epsilon > 0$
there exists $K_\epsilon < \infty$ such that
\be
 C_n^*  \;\le\;  K_\epsilon \, [\alpha(H) + \epsilon]^n
 \qquad\hbox{for all } n\geq 0  \;.
\label{algrow}
\ee

Now let $G=(V,E)$ be as in Theorem~\ref{T:main} and let $G_0=(V_0,E_0)$ and
$G_1=(V_1,E_1)$ be its sublattices. Recall from Section~\ref{S:contour2} that
$N_\Deo(L)$ denotes the number of simple circuits of length $L$ in $G_1$
surrounding a set $\Deo\subs V_0$.
Lemma~\ref{L:cn} applied to $H=G_1$ implies the following bound on $N_\Deo(L)$:

\bl{\bf(Exponential bound on circuits surrounding a point)}\label{L:circbd}
We have $\al(G_1)<2$. Moreover, for every $\eps > 0$, there exists a constant
$C_\eps<\infty$ such that
\be
N_{\{v\}}(L)  \;\leq\;  C_\eps \, [\al(G_1)+\eps]^L
\ee
for all $v\in V_0$ and all $L\geq 1$.
\el

\proof
Since every vertex in $G_1$ has degree~3 and $G_1$ is not a tree (indeed,
each vertex in $G_0$ is surrounded by a circuit in $G_1$), it follows
from Lemma~\ref{L:cn} that $\al(G_1)<2$.

Since $G$ is infinite, connected and locally finite,
it is not hard to show [see Lemma~\ref{L:geodesic}(a) in the Appendix]
that for each $v \in V_0$ we can find an infinite self-avoiding path
$\pi=(v_0,v_1,\ldots)$ in $G_0$ starting at $v_0=v$ such that the graph
distance (in $G_0$) of $v_n$ to $v$ is $n$. It is not hard to see that any
simple circuit surrounding $v$ must cross some edge of $\pi$. With a bit more
work, we can get a quantitative bound on how far this edge can be from the starting
point of $\pi$. Indeed, it follows from Proposition~\ref{P:surround}
that there exists a constant $K<\infty$, depending only on the graph $G_0$,
such that any simple circuit of length $L$ surrounding $v$ must cross one of
the first $N$ edges of $\pi$, where
\be\label{NKL}
N \;:=\; 1+K+\ffrac{1}{2}(\ffrac{3}{2}-1)L \;=\; 1+K+L/4  \;.
\ee

So let $\ga$ be a simple circuit of length $L$ surrounding $v$.  Let
$(v_{k-1},v_k)$ be the first edge of $\pi$ that is crossed by $\ga$, and let
$a$ be the corresponding (dual) edge in $\ga$. We can view $a$ as a directed
edge by agreeing that we turn $(v_{k-1},v_k)$ anticlockwise to get $a$.  Then we
can specify $\ga$ completely by specifying the first edge of $\pi$ to be
crossed by $\ga$ and by specifying the self-avoiding path formed by the first
$L-1$ edges of $\ga$, starting with $a$. By (\ref{NKL}), this yields the bound
\be
N_{\{v\}}(L) \;\leq\; (1+K+L/4) \, C_{L-1}^*  \;.
\ee
By (\ref{algrow}), the claim follows:
it suffices to absorb the factor $(1+K+L/4)$
into a change of the base of the exponential term.
\qed

\subsection{Long-range dependence for zero temperature}  \label{S:long}

We are now ready to prove Lemma~\ref{L:long2} for zero temperature.

\proofof{Lemma~\ref{L:long2} for $\bet=\infty$}
It follows from Proposition~\ref{P:surround} that for each $L_0<\infty$, there
exists $M<\infty$ such that each finite, $G_0$-connected set $\Deo\subs V_0$
with $|\Deo|\geq M$ has the property that any simple circuit in $G_1$
surrounding $\Deo$ must be of length at least $L_0$.

Then the weak Peierls bound (\ref{basic}) and Lemma~\ref{L:circbd}
imply that for any $\eps>0$ there exists $C_\eps<\infty$
such that for every finite and simply connected $\La\supset\Deo$
with $\pa\La\subs V_0$, we have
\be\label{mulong}
\mu^1_{\La,\infty}\big(\Ji_{1,\Deo}\,\big|\Ji_\Deo\big)
\;\geq\; 1-C_\eps\sum_{L=L_0}^\infty2^{1-L}[\al(G_1)+\eps]^L.
\ee
Since $\al(G_1)<2$, by choosing first $\eps$ small enough and then $L_0$ large
enough (and $M$ appropriately), we can make the conditional probability in
(\ref{mulong}) as close to 1 as we wish, uniformly in $\La$.
\qed

\subsection{Quantitative bound for the diced lattice}  \label{S:diced}

Let $p_L$ denote the number of simple circuits (i.e., self-avoiding polygons)
of length $L$ in the hexagonal lattice, modulo translation.
And let $q_L$ denote the number of simple circuits of length $L$
in the hexagonal lattice that surround a given vertex
of the triangular lattice.  We have the following bounds:

\begin{lemma}\hspace{8pt}
{\bf(Supermultiplicativity of hexagonal-lattice polygons)}
\hspace{7pt}\label{lemma:supermultiplicativity}
The number $p_L$ of hexagonal-lattice self-avoiding polygons
of length $L$, modulo translation, satisfies
\be
p_{L+M-2} \;\ge\; p_L \, p_M  \;.
\ee
\end{lemma}

\begin{corollary}{\bf(Bound on hexagonal-lattice circuits surrounding a point)}
\label{cor:qL}
The number of simple circuits in the hexagonal lattice $G_1$
surrounding a given vertex in $G_0$ is bounded as
\be
\label{E:qL<}
q_L \;\le\; (L^2/36) \, (2+\sqrt2)^{(L-2)/2}.
\ee
\end{corollary}

\proofof{Lemma~\ref{lemma:supermultiplicativity}}
We use concatenation:
Consider two polygons $\gamma_1$ and $\gamma_2$ contributing to
$p_L$ and $p_M$, respectively.
Let $(x,x+e_2)$ be the highest vertical edge of $\gamma_1$
in its rightmost column,
and let $(y,y+e_2)$ be the lowest vertical edge of $\gamma_2$
in its leftmost column, where $e_1:=(1,0)$ and $e_2:=(0,1)$ denote the
natural basisvectors of $\R^2$.
Uniting the polygon $\gamma_2$ with $\gamma_1$ shifted by $y-x$
and erasing the edges $(y,y+e_2)$,
we get a contour $\gamma=T_{y-x}(\gamma_1)\cup \gamma_2\setminus (y,y+e_2)$
contributing to $p_{L+M-2}$.
To complete the argument, we must show that different choices of $\ga_1$
and/or $\ga_2$ lead to a different $\ga$ (modulo translation), i.e., we can
reconstruct $\ga_1$ and $\ga_2$ (modulo translation) from $\ga$.
To this aim, we observe that $(y,y+e_2)$ is the only vertical edge
in its column that cuts the interior of $\gamma$. Also, if another column cuts
the interior of $\ga$ in a single edge, then the contours $\ga'_1$
and $\ga'_2$ obtained by cutting $\ga$ at this edge into a left and right
piece will have lengths
different from $L$ and $M$. Thus, for fixed $L$ and $M$, each different
(modulo translations) ordered pair $(\ga_1,\ga_2)$ of polygons of lengths $L$
and $M$ yields a different (modulo translations) polygon of length $L+M-2$.
\qed

\proofof{Corollary~\ref{cor:qL}}
The proof combines three ingredients.
The first is the fact,
conjectured in \cite{Nienhuis_82} and proven in \cite{DC-C10},
that the connective constant of the hexagonal lattice is exactly
$\alpha = \sqrt{2+\sqrt2} \approx 1.847759$.
The second ingredient is the isoperimetric inequality for the hexagonal
lattice: the number of faces surrounded by a circuit of length $L$
is at most $L^2/36$. The third ingredient is a bound on the number $p_L$
of $L$-step hexagonal-lattice self-avoiding polygons modulo translation
in terms of the connective constant $\alpha$ for self-avoiding walks
on the hexagonal lattice,
namely \cite{KSS08}
\be
p_L  \;\le\; \alpha^{L-2}  \;.
\label{eq.pL.bound}
\ee
Indeed, the supermultiplicativity $p_{L+M-2} \ge p_L p_M$
implies, by standard arguments,
that $\alpha_{\rm SAP} = \lim_{L \to\infty} (p_L)^{1/L}$ exists
and that $p_L \le (\alpha_{\rm SAP})^{L-2}$.
On the other hand, since $p_L \le c_{L-1}/(2L)$
where $c_n$ is the number of $n$-step self-avoiding paths
starting at a given vertex,
we manifestly have $\alpha_{\rm SAP} \le \alpha$.
\qed

{\bf Remarks.}
1.  The supermultiplicativity $p_{L+M-2} \ge p_L p_M$
for the hexagonal lattice is stronger than the inequality
$p_{L+M} \ge p_L p_M$ that holds for the square lattice
\cite[Theorem~3.2.3]{Madras-Slade}.
As a consequence, we are able to prove $p_L \le \alpha^{L-2}$
rather than just $p_L \le \alpha^L$.

2.  For self-avoiding paths and polygons on $\Z^d$
it is known \cite[Corollary~3.2.5]{Madras-Slade}
that $\alpha_{\rm SAP} = \alpha$.
The same presumably holds also for the hexagonal lattice
and for other lattices periodically embedded in Euclidean space,
but we are not aware of any proof.
Since we need only an upper bound on $\alpha_{\rm SAP}$,
we have refrained from addressing this question.
Note also that $\alpha_{\rm SAP} < \alpha$ on hyperbolic lattices 
(with the possible exception of eight such lattices) \cite{Madras_05},
so the equality $\alpha_{\rm SAP} = \alpha$ is a somewhat delicate matter.

\bigskip

\proofof{Lemma~\ref{L:dice} for $\bet=\infty$}
We use the explicit values of $q_L$ for $L=6,8,\ldots,140$
obtained by Jensen's computer-assisted enumerations \cite{J06}\footnote{
The relevant series is called there the ``first area-weighted moment''
for honeycomb-lattice polygons
and is contained in the file {\tt hcsapmom1.ser}.
}
together with the bound \reff{E:qL<} for even $L \ge 142$.
{}From \cite{J06} we get
\be
\sum_{L=6}^{140} q_L \, 2^{-L}
\;=\;
\frac{22074233899340881133583692519761872405249}
     {2^{139}}
\;<\;
0.03168
\;.
\label{eq.sum_jensen}
\ee
On the other hand, we have
\be
\sum_{{\rm even\,} L \ge 142}  (L^2/36) \, (2+\sqrt2)^{(L-2)/2} \, 2^{-L}
\;=\;
\frac{(2+\sqrt2)^{70} \, (2907 + 1531 \sqrt{2})}
    {9 \cdot 2^{139}}
\;<\;
0.01731
\;.
\label{eq.sum_tail}
\ee
Putting these together, we have
\be
\sum_{L=6}^{\infty} q_L \, 2^{-L}
\;<\;
0.04899
\;.
\label{eq.sum_total}
\ee
Inserting this into the weak Peierls bound \reff{basic} specialized
to $\Deo = \{v\}$, we obtain
\be
\mu^1_{\La,\infty}(\sigma_v = 1)
\;>\;
1 \,-\, 2 (0.04899)
\;=\;
0.90202
\;.
\label{eq.M0_lower_bound}
\ee

A slight improvement of \eqref{eq.M0_lower_bound}
can be obtained by using \reff{basic2}
in place of \reff{basic}:  we have
\be
\sum_{L=6}^{140} q_L \, \frac{2^{-L}}{1 + 2^{1-L}}
\;<\;
0.03119
\;.
\label{eq.sum_jensen.bis}
\ee
(The improvement in the tail sum $L \ge 142$ is of course utterly negligible.)
The final result \reff{eq.M0_lower_bound}
is then improved from 0.90202 to 0.90301.
\qed

{\bf Remarks.}
1.  Jensen \cite{J06} conjectured, based on his enumerations for $L \le 140$,
that the large-$L$ asymptotic asymptotic behavior of $q_L$ is
\be
q_L  \;=\;   \frac{1}{4\pi} \, (2+\sqrt{2})^{L/2} \, L^{-1}
             \big[ 1 + o(1) \big]
\;.
\ee
(At $L=140$ the exact value for $q_L$ is already within 0.4\%
of this asymptotic form.)
Using this formula in place of the bound \reff{E:qL<},
we find for the tail
\be
\sum_{{\rm even\,} L \ge 142}  q_L \, 2^{-L}
\;\approx\;
4.7 \times 10^{-8}
\;\;(\ll 0.01731)
\;.
\label{eq.sum_tail.bis}
\ee
It follows that {\em if}\/ we could know $q_L$ exactly for all $L$,
then our Peierls argument using \reff{basic2}
would be capable of proving a lower bound 0.93762 in \reff{eq.M0_lower_bound}.
This should be compared with the actual zero-temperature value
$0.957597 \pm 0.000004$ obtained by Monte Carlo simulations
\cite{KSS08}.\footnote{
See footnote~\ref{footnote.M0} above.
}

2.  When \cite{KSS08} was written, the exact result
$\alpha = \sqrt{2+\sqrt2} \approx 1.847759$ was not yet a rigorous theorem,
so we used instead the bound $\alpha < 1.868832$
due to Alm and Parviainen \cite{Alm_2004}.
Then, to get a sufficient final estimate, 
the additional factor $\alpha^{-2}$ from the improved bound
\eqref{eq.pL.bound} implied by stronger supermultiplicativity
(see Remark~1 after the proof of Corollary~\ref{cor:qL})
was crucial.

\section{The positive-temperature case}\label{S:pos}

In this section we extend the Peierls argument to positive temperature,
allowing us to complete the proof of Lemmas~\ref{L:long2} and \ref{L:dice}.
We also prove the technical Lemmas~\ref{L:const2} and \ref{L:ground}.

\subsection{Contour model for positive temperature}

As before, we consider a graph $G=(V,E)$ satisfying the conditions of
Theorem~\ref{T:main} and take a finite and simply connected set
$\La\subs V$ such that $\pa\La\subs V_0$. Our aim is to derive bounds on the
probabilities of certain events under the finite-volume Gibbs measures
$\mu^1_{\La,\bet}$ which correspond to uniform color-1 boundary conditions on
$\pa\La$.

We recall from Section~\ref{S:contour} that every color configuration $\sig$
on $\La$  \reff{def.E0}/\reff{def.E1} a collection $E_1(\sig)$
of edges in the sublattice $G_1$ that separate differently colored vertices
in $G_0$ (or equivalently faces in $G_1$).
Since $\pa\La \subset V_0$ and $\pa\La$ is uniformly colored (in color~1),
each edge of $E_1(\sig)$ has both its endvertices in $V_1\cap\La$.
In general, we define contours to be connected components of $E_1(\sig)$.
[More precisely, we define contours to be the connected components of the graph
$(V_1 \cap \La,\, E_1(\sig))$ other than isolated vertices.]
If $\sig$ is a ground state, then at each
$v\in V_1\cap\La$, either zero or two edges of $E_1(\sig)$ are incident, hence
the connected components of $E_1(\sig)$ are simple circuits in $G_1$.
But for general color configurations $\sig$,
the connected components of $E_1(\sig)$ may be more complicated. In
particular, it is possible that three edges of $E_1(\sig)$ are incident to a
vertex $v\in V_1\cap\La$. Recall that a connected graph is called {\em bridgeless}\/
(or {\em 2-edge-connected}\/) if it contains no {\em bridges}\/ (i.e., single
edges the removal of which disconnects the graph). We observe that for any
color configuration $\sig$, the connected components of $E_1(\sig)$ must be
bridgeless, since otherwise there would be a uniformly colored region of $G_0$
that bounds such a bridge on both sides, contradicting the definition of
$E_1(\sig)$. 

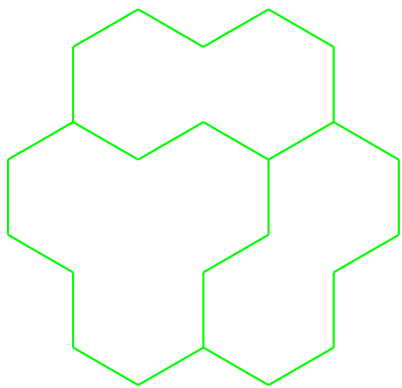
\begin{figure}[t]
\centering
\begin{tikzpicture}
\draw[green,thick] (0,0) -- ++(30:1cm);
\draw[green,thick] (30:1cm) -- ++(90:1cm);
\draw[green,thick] (30:1cm) ++ (90:1cm) -- ++(30:1cm);
\draw[green,thick] (30:2cm |- 30:4cm) -- (30:2cm |- 30:6cm);
\draw[green,thick] (30:2cm |- 30:6cm) -- (30:1cm |- 30:7cm);
\draw[green,thick] (30:1cm) ++ (90:3cm) -- ++(-150:1cm);
\draw[green,thick] (90:3cm) -- ++(150:1cm);
\draw[green,thick] (150:1cm)  ++ (90:3cm) -- ++(-150:1cm);
\draw[green,thick] (150:2cm)  ++ (90:2cm) -- ++(150:1cm);
\draw[green,thick] (30:-1cm) ++ (90:5cm) -- ++(150:1cm);
\draw[green,thick] (30:-2cm) ++ (90:6cm) -- ++(-150:1cm);
\draw[green,thick] (30:-3cm) ++ (90:6cm) -- ++(-90:1cm);
\draw[green,thick] (30:-3cm) ++ (90:5cm) -- ++(-150:1cm);
\draw[green,thick] (30:-4cm) ++ (90:5cm) -- ++(-90:1cm);
\draw[green,thick] (30:-4cm) ++ (90:4cm) -- ++(-30:1cm);
\draw[green,thick] (30:-3cm) ++ (90:3cm) -- ++(-90:1cm);
\draw[green,thick] (30:-3cm) ++ (90:2cm) -- ++(-30:1cm);
\draw[green,thick] (30:-2cm) ++ (90:1cm) -- ++(30:1cm);
\draw[green,thick] (30:-1cm) ++ (90:1cm) -- ++(-30:1cm);
\draw[green,thick] (30:0cm) ++ (90:3cm) -- ++(-90:1cm);
\draw[green,thick] (30:-1cm) ++ (90:2cm) -- ++(30:1cm);
\draw[green,thick] (30:-1cm) ++ (90:2cm) -- ++(-90:1cm);
\draw[green,thick] (30:1cm |- 30:7cm) -- ++(90:1cm);
\draw[green,thick] (30:1cm |- 30:9cm) -- ++(150:1cm);
\draw[green,thick] (30:0cm |- 30:10cm) -- ++(-150:1cm);
\end{tikzpicture}
\vglue-1cm
\caption{\label{fig:MB}
   A contour $\ga$ with $\chi(\ga)=0$.
}
\end{figure}

In view of this, in the positive-temperature model let us define a
{\em contour}\/ to be a finite connected bridgeless subgraph $\ga$ of $G_1$
containing at least one edge.
Note that each vertex of such a contour has degree~2 or 3.  It is easy to see
that the number of vertices of degree~3 must be even
(just notice that twice the number of vertices of degree two
plus three times the number of vertices of degree three
equals twice the number of edges).
We let $|\ga|$ denote the number of edges of $\ga$,
to which we will refer as the {\em length}\/ of $\ga$.
We let $t(\ga)$ be the number such that $\ga$ has
$2t(\ga)$ vertices of degree~3.  Then $\ga$ has $|\ga| - 3t(\ga)$ vertices
of degree~2.  Moreover, $\ga$ divides $V_0$ into $2+t(\ga)$ connected components,
of which one is infinite and the others are finite and simply connected.
We say that a contour $\ga$ {\em surrounds}\/ a set $\Deo\subs V_0$,
denoted as $\ga\circlearrowright\Deo$, if $\Deo$ is contained in one of
the finite components. We call the infinite component the {\em exterior}\/
${\rm Ext}(\ga)$ of $\ga$, and we refer to the union of all the finite
components as the {\em interior}\/ ${\rm Int}(\ga)$ of $\ga$.
[Please note that saying that $\ga$ surrounds a set $\Deo$ is stronger than
saying that $\Deo\subq{\rm Int}(\ga)$, since ``surrounding'' is defined as
$\Deo$ being entirely contained in {\em one}\/ of the finite components.]
Given that the exterior of $\ga$ is colored in one particular color, we let
$\chi(\ga)$ denote the number of possible three-colorings of the connected
components of ${\rm Int}(\ga)$ in such a way that along each edge of $\ga$,
two different colors meet.\footnote{
 Otherwise put, $\chi(\ga)$ is $\ffrac{1}{3}$ times the number of proper
 3-colorings of the dual graph $\ga^*$.
}
Please note that it is possible to have $\chi(\ga)=0$:  see Figure~\ref{fig:MB}.
Obviously, such contours are ``not allowed'',
and we shall soon see that their probability is zero.\footnote{
We could, if we wanted, redefine the term ``contour'' to include
only those having $\chi(\ga) > 0$.
But there is little to be gained from complicating the definition
in this way, since our counting of contours (Lemma~\ref{L:contourbd} below)
is too crude to distinguish between those having $\chi(\ga) > 0$
or $\chi(\ga)=0$.}
Finally, let us observe that $\chi(\ga) \le 2^{t(\ga)+1}$.

We now claim that if $\sig$ is distributed according to $\mu^{1}_{\La,\bet}$,
and $\Ga(\sig)$ is the collection of connected components of
$(V_1 \cap \La,\, E_1(\sig))$ other than isolated vertices,
then $\Ga(\sig)$ is distributed according to the law
\be
\label{posnudef}
\nu_{\La,\bet}(\Ga) \;=\; \frac{1}{Z_{\La,\bet}}
\prod_{\ga\in\Ga}\chi(\ga)\,p_\bet^{|\ga|}\,q_\bet^{t(\ga)}  \;,
\ee
where $Z_{\La,\bet}$ is a normalizing constant and
\begin{subeqnarray}
\dis p_\bet  &:=&  \dis  {1+e^{-\bet}+e^{-2\bet}  \over 2+e^{-3\bet}}  \\[2mm]
\dis q_\bet  &:=&  \dis {
                       {9e^{-2\bet} (2+e^{-3\bet}) \over (1+e^{-\bet}+e^{-2\bet})^3 }
                      }
\label{eq.pbeta.qbeta}
\end{subeqnarray}
To see this, note that there are $\prod_{\ga\in\Ga}\chi(\ga)$ ways of
coloring the sites in $V_0\cap\La$ in a way that is consistent with $\Ga$.
Given a coloring of $V_0\cap\La$, summing the probabilities of all
possible colorings of $V_1\cap\La$ yields for each site in $V_1\cap\La$ a factor
\be
1+1+e^{-3\bet},\quad 1+e^{-\bet}+e^{-2\bet} \quad
\mbox{or}\quad e^{-\bet} +e^{-\bet}+e^{-\bet}
\ee
depending on whether the site has neighbors with
one, two or three different colors, respectively.
These cases correspond, respectively, to sites not on a contour,
sites of degree~2 on a contour, and sites of degree~3 on a contour.
Absorbing the factor $1+1+e^{-3\bet}$ into the normalization contant $Z_{\La,\bet}$,
we get a factor $(1+e^{-\bet}+e^{-2\bet})/(2+e^{-3\bet})$
for each of the $|\ga| - 3t(\ga)$ sites of degree~2,
and a factor $3e^{-\bet}/(2+e^{-3\bet})$
for each of the $2t(\ga)$ sites of degree~3.
Putting this all together, we arrive at (\ref{posnudef})/(\ref{eq.pbeta.qbeta}).

In the limit $\bet\to\infty$, we have $p_\bet\to \ffrac{1}{2}$ and $q_\bet\to 0$;
in particular, the only countours that get nonzero weight in this limit are
simple circuits, for which $\chi(\ga) = 2$.  Then the contour law
(\ref{posnudef}) reduces to (\ref{nudef}), as expected.

More generally, it can be easily verified that $p_\bet$ decreases monotonically
from 1 to $\ffrac{1}{2}$ as $\bet$ runs from 0 to $\infty$,
and behaves for large $\beta$ as $\ffrac{1}{2} + O(e^{-\beta})$;
and that $q_\bet$ decreases monotonically
from 1 to 0 as $\bet$ runs from 0 to $\infty$,
and behaves for large $\beta$ as $O(e^{-2\beta})$.

By the same arguments as in (\ref{munu})--(\ref{prec}), and using
$\chi(\ga) \le 2^{t(\ga)+1}$, we find:

\begin{lemma}{\bf(Peierls bound for positive temperature)}
\label{lemma.peierls.bis}
Let $\Lambda \subs V$ be a simply connected finite set
such that $\partial\Lambda \subs V_0$,
and let $\Deo \subq \Lambda \cap V_0$ be connected in $G_0$.
Then
\be
\label{posbasic}
 1-\mu^{1}_{\La,\bet}\big(\Ji_{1,\Deo}\,\big|\,\Ji_\Deo)
 \;\le\;
 \sum_{T=0}^\infty\sum_{L=3}^\infty N_\Deo(L,T)\,
    \frac{2^{T+1}\,p_\bet^L\,q_\bet^T}{1+2^{T+1}\,p_\bet^L\,q_\bet^T}  \;,
 \;
\ee
where $N_\Deo(L,T)$ denotes the number of contours $\ga$ surrounding $\Deo$
satisfying $|\ga|=L$ and $t(\ga)=T$.
\end{lemma}

\subsection{Bounds on contours for positive temperature}

In this section, we prove Lemmas~\ref{L:long2} and \ref{L:dice}. We first need
to generalize Lemma~\ref{L:circbd} to contours that are not simple
circuits. Recall that $N_\Deo(L,T)$ denotes the number of contours $\ga$
surrounding $\Deo$ satisfying $|\ga|=L$ and $t(\ga)=T$.  Recall also from
Lemma~\ref{L:cn} that $\alpha(H)$ denotes the connective constant of
a graph $H$ as defined in \reff{def.alpha},
and from Lemma~\ref{L:circbd} that $\alpha(G_1) < 2$.

\bl{\bf(Bound on number of contours)}\label{L:contourbd}
For every $\eps > 0$ there exists a constant $C'_\eps<\infty$ such that
\be
N_{\{v\}}(L,T) \;\leq\; (L^{T}/T!)^2 \, (C'_\eps)^{T+1} \, [\al(G_1)+\eps]^L
\ee
for all $v \in V_0$ and all $L,T\geq 0$.
\el

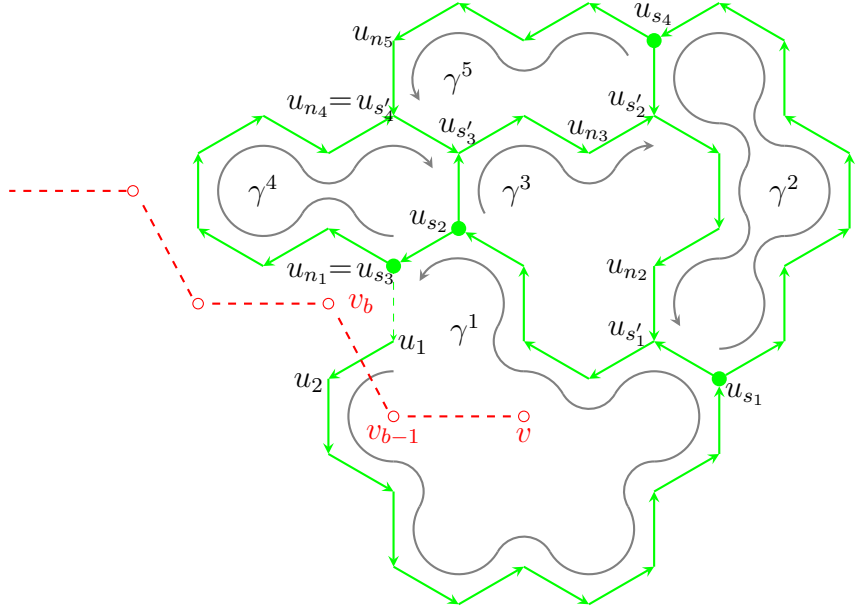
\begin{figure}[t]
\centering
\begin{tikzpicture}[>=stealth]
\draw[green,thick,->] (0,0) -- ++(30:1cm);
\draw[green,thick,->] (30:1cm) -- ++(90:1cm);
\draw[green,thick,->] (30:1cm) ++ (90:1cm) -- ++(30:1cm);
\draw[green,thick,->] (30:2cm |- 30:4cm) -- (30:2cm |- 30:5.8cm);
\draw[black] (30:2.4cm |- 30:5.6cm) node {$u_{s_1}$};
\draw[green,thick,->] (30:2cm |- 30:6cm) -- (30:1cm |- 30:7cm);
\draw[green,thick,->] (30:1cm) ++ (90:3cm) -- ++(-150:1cm);
\draw[green,thick,->] (90:3cm) -- ++(150:1cm);
\draw[green,thick,->] (30:-1cm) ++ (90:4cm) -- ++(90:1cm);
\draw[black] (30:1cm |- 30:15.7cm) node {$u_{s_4}$};
\draw[black] (30:-2.45cm |- 30:10.2cm) node {$u_{s_2}$};
\draw[green,thick,->] (30:-1cm) ++ (90:5cm) -- ++(150:0.9cm);
\draw[green,thick,->] (30:-2cm) ++ (90:6cm) -- ++(-150:0.9cm);
\draw[green,dashed,very thin,->] (30:-3cm) ++ (90:6cm) -- ++(-90:1cm);
\draw[black] (30:-2.7cm |- 30:6.9cm) node {$u_1$};
\draw[black] (30:-4.3cm |- 30:5.9cm) node {$u_2$};
\draw[green,thick,->] (30:-3cm) ++ (90:5cm) -- ++(-150:1cm);
\draw[green,thick,->] (30:-4cm) ++ (90:5cm) -- ++(-90:1cm);
\draw[green,thick,->] (30:-4cm) ++ (90:4cm) -- ++(-30:1cm);
\draw[green,thick,->] (30:-3cm) ++ (90:3cm) -- ++(-90:1cm);
\draw[green,thick,->] (30:-3cm) ++ (90:2cm) -- ++(-30:1cm);
\draw[green,thick,->] (30:-2cm) ++ (90:1cm) -- ++(30:1cm);
\draw[green,thick,->] (30:-1cm) ++ (90:1cm) -- ++(-30:1cm);
\draw[gray,thick] (30:-3cm |- 30:5cm) ++ (90:0.6cm) arc (90:270:0.6cm);
\draw[gray,thick] (30:-3cm |- 30:3cm) ++ (90:0.4cm) arc (90:-30:0.4cm);
\draw[gray,thick] (30:-2cm |- 30:2cm) ++ (150:0.6cm) arc (150:330:0.6cm);
\draw[gray,thick] (30:-1cm |- 30:1cm) ++ (150:0.4cm) arc (150:30:0.4cm);
\draw[gray,thick] (30:0cm |- 30:2cm) ++ (-150:0.6cm) arc (-150:30:0.6cm);
\draw[gray,thick] (30:1cm |- 30:3cm) ++ (210:0.4cm) arc (210:90:0.4cm);
\draw[gray,thick] (30:1cm |- 30:5cm) ++ (-90:0.6cm) arc (-90:150:0.6cm);
\draw[gray,thick] (30:0cm |- 30:6cm) ++ (330:0.4cm) arc (330:210:0.4cm);
\draw[gray,thick] (30:-1cm |- 30:5cm) ++ (30:0.6cm) arc (30:90:0.6cm);
\draw[gray,thick] (30:-1cm |- 30:7cm) ++ (270:0.4cm) arc (270:150:0.4cm);
\draw[gray,thick,->] (30:-2cm |- 30:8cm) ++ (-30:0.6cm) arc (-30:150:0.6cm);
\draw[black] (30:-1.9cm |- 30:7.3cm) node {$\gamma^1$};
\fill[green] (30:2cm |- 30:6cm) circle (1mm);

\draw[red] (30:-1cm |- 30:5cm) circle (.7mm);
\draw[red] (30:-3cm |- 30:5cm) circle (.7mm);
\draw[red] (30:-4cm |- 30:8cm) circle (.7mm);
\draw[red] (30:-6cm |- 30:8cm) circle (.7mm);
\draw[red] (30:-7cm |- 30:11cm) circle (.7mm);
\draw[red] (30:-1cm |- 30:4.5cm) node {$v$};
\draw[red] (30:-3cm |- 30:4.5cm) node {$v_{b-1}$};
\draw[red] (30:-3.5cm |- 30:8cm) node {$v_b$};
\draw[red,dashed,thick] (30:-1.2cm |- 30:5cm) -- (30:-2.8cm |- 30:5cm);
\draw[red,dashed,thick] (30:-3.1cm |- 30:5.2cm) -- (30:-3.9cm |- 30:7.8cm);
\draw[red,dashed,thick] (30:-4.2cm |- 30:8cm) -- (30:-5.8cm |- 30:8cm);
\draw[red,dashed,thick] (30:-6.1cm |- 30:8.2cm) -- (30:-6.9cm |- 30:10.8cm);
\draw[red,dashed,thick] (30:-7.1cm |- 30:11cm) -- (30:-8.9cm |- 30:11cm);

\draw[green,thick,->] (30:2.1cm |- 30:6.1cm) -- ++(30:0.9cm);
\draw[green,thick,->] (30:3cm |- 30:7cm) -- ++(90:1cm);
\draw[green,thick,->] (30:3cm |- 30:9cm) -- ++(30:1cm);
\draw[green,thick,->] (30:4cm |- 30:10cm) -- ++(90:1cm);
\draw[green,thick,->] (30:4cm |- 30:12cm) -- ++(150:1cm);
\draw[green,thick,->] (30:3cm |- 30:13cm) -- ++(90:1cm);
\draw[green,thick,->] (30:3cm |- 30:15cm) -- ++(150:1cm);
\draw[green,thick,->] (30:2cm |- 30:16cm) -- ++(-150:0.9cm);
\draw[green,thick,->] (30:1cm |- 30:15cm) -- ++(-90:1cm);
\draw[green,thick,<-] (30:2cm |- 30:12cm) -- ++(150:1cm);
\draw[green,thick,<-] (30:2cm |- 30:10cm) -- ++(90:1cm);
\draw[green,thick,<-] (30:1cm |- 30:9cm) -- ++(30:1cm);
\draw[green,thick,<-] (30:1cm |- 30:7cm) -- ++(90:1cm);

\draw[black] (30:3cm |- 30:11cm) node {$\gamma^2$};
\draw[gray,thick] (30:2cm |- 30:8cm) ++ (-90:0.6cm) arc (-90:30:0.6cm);
\draw[gray,thick,->] (30:2cm |- 30:8cm) ++ (90:0.6cm) arc (90:210:0.6cm);
\draw[gray,thick] (30:3cm |- 30:11cm) ++ (210:0.6cm) arc (210:150:0.6cm);
\draw[gray,thick] (30:3cm |- 30:11cm) ++ (-90:0.6cm) arc (-90:90:0.6cm);
\draw[gray,thick] (30:2cm |- 30:14cm) ++ (270:0.6cm) arc (270:-30:0.6cm);
\draw[gray,thick] (30:2cm |- 30:12cm) ++ (-30:0.4cm) arc (-30:90:0.4cm);
\draw[gray,thick] (30:3cm |- 30:13cm) ++ (270:0.4cm) arc (270:150:0.4cm);
\draw[gray,thick] (30:2cm |- 30:10cm) ++ (30:0.4cm) arc (30:-90:0.4cm);
\draw[gray,thick] (30:3cm |- 30:9cm) ++ (90:0.4cm) arc (90:210:0.4cm);

\draw[green,thick,<-] (30:1cm |- 30:13cm) -- ++(-150:1cm); 
\draw[green,thick,<-] (30:-1cm |- 30:13cm) -- ++(-150:1cm);
\draw[green,thick,<-] (30:0cm |- 30:12cm) -- ++(150:1cm);
\draw[green,thick,<-] (30:-2cm |- 30:12cm) -- ++(-90:1cm);
\draw[black] (30:-1.1cm |- 30:11cm) node {$\gamma^3$};
\draw[gray,thick] (30:-1cm |- 30:11cm) ++ (210:0.6cm) arc (210:30:0.6cm);
\draw[gray,thick] (30:0cm |- 30:12cm) ++ (-150:0.4cm) arc (-150:-30:0.4cm);
\draw[gray,thick,->] (30:1cm |- 30:11cm) ++ (150:0.6cm) arc (150:90:0.6cm);

\fill[green] (30:-2cm |- 30:10cm) circle (1mm); 

\draw[green,thick,->] (30:1cm |- 30:15cm) -- ++(150:1cm);
\draw[green,thick,->] (30:0cm |- 30:16cm) -- ++(-150:1cm);
\draw[green,thick,->] (30:-1cm |- 30:15cm) -- ++(150:1cm);
\draw[green,thick,->] (30:-2cm |- 30:16cm) -- ++(-150:1cm);
\draw[green,thick,->] (30:-3cm |- 30:15cm) -- (30:-3cm |- 30:13cm);

\draw[black] (30:-2cm |- 30:14cm) node {$\gamma^5$};
\draw[gray,thick,<-] (30:-2cm |- 30:14cm) ++ (210:0.6cm) arc (210:30:0.6cm);
\draw[gray,thick] (30:-1cm |- 30:15cm) ++ (-150:0.4cm) arc (-150:-30:0.4cm);
\draw[gray,thick] (30:0cm |- 30:14cm) ++ (150:0.6cm) arc (150:30:0.6cm);

\fill[green] (30:-3cm |- 30:9cm) circle (1mm);

\draw[green,thick,->] (30:-3cm |- 30:13cm) -- ++(-30:1cm);
\draw[green,thick,<-] (30:-3cm |- 30:13cm) -- ++(-150:1cm);
\draw[green,thick,<-] (30:-4cm |- 30:12cm) -- ++(150:1cm);
\draw[green,thick,<-] (30:-5cm |- 30:13cm) -- ++(-150:1cm);
\draw[green,thick,<-] (30:-6cm |- 30:12cm) -- ++(-90:1cm);
\draw[green,thick,<-] (30:-6cm |- 30:10cm) -- ++(-30:1cm);
\draw[green,thick,<-] (30:-5cm |- 30:9cm) -- ++(30:1cm);
\draw[green,thick,<-] (30:-4cm |- 30:10cm) -- ++(-30:1cm);

\draw[black] (30:-3.8cm |- 30:8.8cm) node {$u_{n_1} \!\!=\! u_{s_3}$};
\draw[black] (30:0.6cm |- 30:8.9cm) node {$u_{n_2}$};
\draw[black] (30:0.6cm |- 30:7.2cm) node {$u_{s'_1}$};
\draw[black] (30:0cm |- 30:12.6cm) node {$u_{n_3}$};
\draw[black] (30:0.6cm |- 30:13.3cm) node {$u_{s'_2}$};
\draw[black] (30:-3.8cm |- 30:13.2cm) node {$u_{n_4} \!\!=\! u_{s'_4}$};
\draw[black] (30:-3.3cm |- 30:15.1cm) node {$u_{n_5}$};
\draw[black] (30:-2cm |- 30:12.6cm) node {$u_{s'_3}$};

\draw[black] (30:-5cm |- 30:11cm) node {$\gamma^4$};
\draw[gray,thick,->] (30:-3cm |- 30:11cm) ++ (150:0.6cm) arc (150:30:0.6cm);
\draw[gray,thick] (30:-4cm |- 30:12cm) ++ (-150:0.4cm) arc (-150:-30:0.4cm);
\draw[gray,thick] (30:-5cm |- 30:11cm) ++ (30:0.6cm) arc (30:330:0.6cm);
\draw[gray,thick] (30:-4cm |- 30:10cm) ++ (150:0.4cm) arc (150:30:0.4cm);
\draw[gray,thick] (30:-3cm |- 30:11cm) ++ (-150:0.6cm) arc (-150:-90:0.6cm);

\fill[green] (30:1cm |- 30:15cm) circle (1mm);

\end{tikzpicture}
\caption{\label{fig:poscont}
 A contour $\ga$ in the case when $G$ is the diced lattice and
 $G_1$ is the hexagonal lattice. This contour $\ga$ contains $8$
 vertices of degree~3, hence $t(\ga)=4$. It is not hard to check that
 $\chi(\ga)=2$.
}
\end{figure}

\proof
Let $\ga$ be a contour surrounding $\{v\}$ such that $|\ga|=L$ and $t(\ga)=T$.
We need a suitable way to encode $\ga$.  We begin, as in the proof of
Lemma~\ref{L:circbd}, by letting $\pi=(v_0,v_1,\ldots)$ be an infinite
self-avoiding path in $G_0$ starting at $v_0=v$ such that the graph distance
(in $G_0$) of $v_n$ to $v$ is $n$.
According to Proposition~\ref{P:surround} and formula (\ref{NKL}),
the contour $\gamma$ intersects an edge $(v_{b-1},v_b)$ of $\pi$
with $b\le N:=1+K+L/4$, where $K$ is a
constant depending only on the graph $G_0$. Thus, we can find some directed
simple circuit $\ga^* =(u_1,\ldots,u_{n_1},u_1)$ contained in $\ga$,
such that $(u_1,u_2)$ crosses the edge
$(v_{b-1},v_b)$ in the anticlockwise direction
(see Figure~\ref{fig:poscont}).

Let us write $\gamma^1 = (u_1,\ldots,u_{n_1})$, which is a self-avoiding path.
If $T=0$, then $\ga=\ga^*$ and our encoding is complete.
Otherwise, let
$s_1:=\min\{i\geq 1 \colon\: u_i\mbox{ is of degree 3}$ $\mbox{in }\ga\}$.
Then we can find a self-avoiding path
\be
\ga^2 \;=\; (u_{s_1},u_{n_1+1},\ldots,u_{n_2},u_{s'_1})
\ee
in $\ga$ such that only the starting and ending points
$u_{s_1}$ and $u_{s'_1}$ are in $\ga^1$.
If $T=1$, then $\ga=\ga^*\cup\ga^2$ and we are done.
Otherwise, let
$s_2:=\min\{i>s_1 \colon\: i\neq s'_1
         \mbox{ and $u_i$ is of degree 3 in }\ga\}$.
Then we can find another self-avoiding path
\be
\ga^3 \;=\; (u_{s_2},u_{n_2+1},\ldots,u_{n_3},u_{s'_2})
\ee
in $\ga$ such that only the starting and ending points
$u_{s_2}$ and $u_{s'_2}$ are in $\ga^1\cup\ga^2$.
Continuing in this way, we see that we can code all the information
needed to construct $\ga$ by specifying numbers
\be
b\leq N,\quad
2=n_0<n_1<\cdots<n_{T+1}=L-T \quad\mbox{and}\quad
0<s_1<\cdots<s_{T}<L-T
\ee
and self-avoiding paths $\ga^1,\ldots,\ga^{T+1}$ of lengths
$n_1-n_0+1, n_2-n_1+1,\ldots,n_{T+1}-n_T+1$
whose starting edges are uniquely determined
by the information previously coded.
By Lemma~\ref{L:cn} and its consequence \eqref{algrow},
for each $\eps>0$ there exists $K_{\eps}<\infty$ such that
the number of self-avoiding paths of length $n$ with a
specified starting edge is bounded from above by
$K_{\eps}[\al(G_1)+\eps]^n$.

Therefore, there are at most
\be
\prod_{i=1}^{T+1} K_{\eps}[\al(G_1)+\eps]^{n_i-n_{i-1}+1}
\;=\; (K_{\eps})^{T+1} \, [\al(G_1)+\eps]^{L-1}
\ee
different contours $\ga$ associated with given data
$b,n_1,\ldots,n_{T},s_1,\ldots,s_T$.
Since there are
$\displaystyle \binom{L-T-3}{T}$ and $\displaystyle \binom{L-T-1}{T}$
ways of choosing numbers $2<n_1<\cdots<n_{T}<L-T$ and $0<s_1<\cdots<s_T<L-T$,
respectively, and since $b\leq N=1+K+L/4$, summing over all ways to choose the
numbers $b,n_1,\ldots,n_{T},s_1,\ldots,s_T$ shows that the total number of
contours $\ga$ surrounding $v$ with given $|\ga|=L$ and $t(\ga)=T$ is bounded by
\be\ba{l}
\dis\big(1+K+L/4\big)\binom{L-T-3}{T}\binom{L-T-1}{T}
(K_{\eps})^{T+1} \, \big[\al(G_1)+\eps\big]^{L-1}\\[15pt]
\dis\quad\leq\; (C'_\eps)^{T+1}{\binom{L-T}{T}}^2
\big[\al(G_1)+2\eps\big]^{L-1}
\;\leq\; (C'_\eps)^{T+1}\big(L^{T}/T!\big)^2
\big[\al(G_1)+2\eps\big]^L  \;,
\ec
where the factor $1+K+L/4$ was absorbed into a change of base
of the exponential term followed by the change of constant into $C'_\eps$.
\hfill\hfill\qed

\subsection{Long-range dependence for positive temperature}  \label{S:long.pos}

\quad\par

\proofof{Lemmas~\ref{L:long2} and \ref{L:dice} in the positive-temperature case}
In the zero-temperature case, both lemmas have already been proven 
in Sections~\ref{S:long} and \ref{S:diced}, respectively,
by showing that for some sufficiently large $\Deo$
(respectively for $\Deo=\{v\}$)
the right-hand side of (\ref{basic2})
can be made sufficiently small.
To generalize the two lemmas to small positive temperatures,
it therefore suffices to show that the right-hand side of (\ref{posbasic})
converges to the right-hand side of (\ref{basic2})
as $\bet\to\infty$
[for Lemma~\ref{L:dice} we should also show that the error is $O(e^{-\beta})$].
In view of this, Lemmas~\ref{L:long2} and \ref{L:dice}
are consequences of the following lemma.
\qed

\bl{\bf(Large-$\bm{\bet}$ behavior of the Peierls bound)}
\label{L:bdconv}
There exist $\beta_0, C < \infty$ such that
\be
 0
 \;\le\;
 \sum_{T=0}^\infty \sum_{L=3}^\infty N_\Deo(L,T)\,
    \frac{2^{T+1}\, p_\bet^L\, q_\bet^T}{1+2^{T+1}\, p_\bet^L\, q_\bet^T}
 \:-\:
 \sum_{L=3}^\infty N_\Deo(L) \, \frac{2^{1-L}}{1+2^{1-L}}
 \;\le\;
 C e^{-\beta}
\label{bdconv}
\ee
and
\be
 \sum_{T=1}^\infty \sum_{L=3}^\infty N_\Deo(L,T)\,
    \frac{2^{T+1}\, p_\bet^L\, q_\bet^T}{1+2^{T+1}\, p_\bet^L\, q_\bet^T}
 \;\le\;
 {
 \sum_{T=1}^\infty \sum_{L=3}^\infty N_\Deo(L,T)\,
    2^{T+1}\, p_\bet^L\, q_\bet^T
 }
 \;\le\;
 C e^{-2\beta}
\label{bd.Tge1}
\ee
uniformly for $\beta \in [\beta_0,\infty]$
and for nonempty finite $G_0$-connected sets $\Deo\subs V_0$.
\el

\proof
The lower bound in \reff{bdconv} is a trivial consequence of
$p_\beta \ge \ffrac{1}{2}$ and $q_\beta \ge 0$.
To prove the upper bounds,
we split the double sum in \reff{bdconv} into its contributions $T=0$ and $T\geq 1$
and bound them separately,
using $p_\beta = \ffrac{1}{2} + O(e^{-\beta})$ and $q_\beta = O(e^{-2\beta})$.

$\bm{T=0.}$ \quad
By Lemma~\ref{L:circbd}, there exist $C < \infty$ and $\alpha < 2$
such that $N_\Deo(L) \le C \alpha^L$.
The term $T=0$ can therefore be bounded
{ (using $p_\beta \ge \ffrac{1}{2}$)}
as 
\be
 \sum_{L=3}^\infty N_\Deo(L) \, \frac{2p_\bet^L}{1+2p_\bet^L}
 \;\le\;
 \sum_{L=3}^\infty N_\Deo(L) \, \frac{2^{1-L}}{1+ 2^{1-L}}
 \:+\:
 2C \sum_{L=3}^\infty \alpha^L \,
        \frac{p_\bet^L - (\ffrac{1}{2})^L}{1+ 2^{1-L}}
 \;.
\label{T0term}
\ee
Choosing $\beta_0$ large enough so that $\alpha \, p_{\beta_0} < 1$,
it is easy to see, using $p_\beta = \ffrac{1}{2} + O(e^{-\beta})$,
that the last term in \reff{T0term} is $O(e^{-\beta})$.

$\bm{T \ge 1.}$ \quad
By Lemmas~\ref{L:circbd} and \ref{L:contourbd},
there exist $C<\infty$ and $\al<2$
such that $N_\Deo(L,T) \le (L^{T}/T!)^2 C^{T+1} \alpha^L$.
Therefore the terms $T\geq 1$ in \reff{bdconv} can be bounded as
\begin{eqnarray}
 \sum_{T=1}^\infty \sum_{L=3}^\infty
    N_\Deo(L,T) \,
    \frac{2^{T+1} \, p_\bet^L \, q_\bet^T}{1+ 2^{T+1} \, p_\bet^L \, q_\bet^T}
 & \;\le\; &
 \sum_{T=1}^\infty \sum_{L=3}^\infty
    N_\Deo(L,T) \, 2^{T+1} \, p_\bet^L \, q_\bet^T
        \nonumber \\
 & \;\le\; &
 2C \sum_{L=3}^\infty (\al p_\beta)^L
    \sum_{T=1}^\infty {(2C q_\beta)^T  \over (T!)^2} \, L^{2T}
        \nonumber \\
 & \;\le\; &
 2C \sum_{L=3}^\infty (\al p_\beta)^L
    \sum_{T=1}^\infty {(8C q_\beta)^T  \over (2T)!} \, L^{2T}
        \nonumber \\
 & \;=\; &
 16C^2 q_\beta \sum_{L=3}^\infty L^{2} \, (\al p_\beta)^L
    \sum_{T=0}^\infty {(8C q_\beta)^T  \over (2T+2)!} \, L^{2T}
        \nonumber \\
 & \;\le\; &
 16C^2 q_\beta \sum_{L=3}^\infty L^{2} \,
                \big(\al p_\beta \, e^{\sqrt{8C q_\beta}} \big)^L
\label{T1term}
\end{eqnarray}
where we used $\displaystyle{ \frac{(2T)!}{(T!)^2} = \binom{2T}{T} \le 2^{2T} }$.
Choosing $\beta_0$ large enough so that one has
$\alpha \, p_{\beta_0} \, e^{\sqrt{8C q_{\beta_0}}} < 1$,
we see that \reff{T1term} is $O(q_\beta) = O(e^{-2\beta})$,
which proves \reff{bd.Tge1} and completes the proof of \reff{bdconv}.
\qed

The bound \reff{bd.Tge1} from Lemma~\ref{L:bdconv} has a useful corollary.
Let us say that a contour $\ga$ is {\em simple}\/ if it is a simple circuit,
i.e.\ $t(\ga)=0$.
For any contour configuration $\Ga$ and any $v\in V_0$,
let $S^{\rm t}_v(\Ga)$ denote the number of non-simple contours in $\Ga$
that surround $\{v\}$.
We then have the following bound showing that non-simple contours
are rare at low temperature:

\bcor{\bf(Rarity of non-simple contours)}\label{C:simple}
Let $\nu_{\La,\bet}$ be the contour measure from (\ref{posnudef}).
Then there exist $\beta_0,C < \infty$ such that
\be
\sum_\Ga \nu_{\La,\bet}(\Ga) \, S^{\rm t}_v(\Ga)
\;\le\;
C {e^{-2\beta}}
\ee
uniformly for $\beta \in [\beta_0,\infty]$,
for finite simply connected $\La \subset V$ such that $\pa\La\subs V_0$,
and for $v \in \La \cap V_0$.
\ecor

\proof
This is an immediate consequence of \reff{bd.Tge1}
together with the positive-temperature analogue
of Lemma~\ref{lemma.gamma.T=0}.
\qed

\subsection{Proof of the technical lemmas}\label{S:tech}

In this section we prove Lemmas~\ref{L:const2} and \ref{L:ground}.

\proofof{Lemma~\ref{L:const2}}
It is easy to show that for each $\bet_0<\infty$ there exists an $\eps>0$ such
that $\mu^{1}_{\La,\bet}(\Ji_\Deo)\geq\eps$, uniformly for all
$0\leq\bet\leq\bet_0$ and all finite and simply connected $\La\supset\Deo$ such
that $\pa\La\subs V_0$. Indeed, this follows from a ``finite energy''
argument: given any {configuration}  $\sig\in\{1,2,3\}^\La$, we can recolor the sites in
$\Deo$ in any color of our choice at an energetic cost of at worst
$e^{-\beta |\pa\Deo|}$ and an entropic cost of at worst $3^{-|\Deo|}$.
Note that here $\Deo$ is fixed and finite, so the precise dependence
of the costs on $\Deo$ is irrelevant.
The only difficulty is that the bound one obtains in this way is not
uniform in $\bet$ as $\bet\to\infty$. Therefore, to complete the proof, it
suffices to show that there exists some $\bet_0<\infty$ such that
$\mu^{1}_{\La,\bet}(\Ji_\Deo)$ can be estimated from below uniformly
in $\bet_0\leq\bet\leq\infty$ and $\La$.

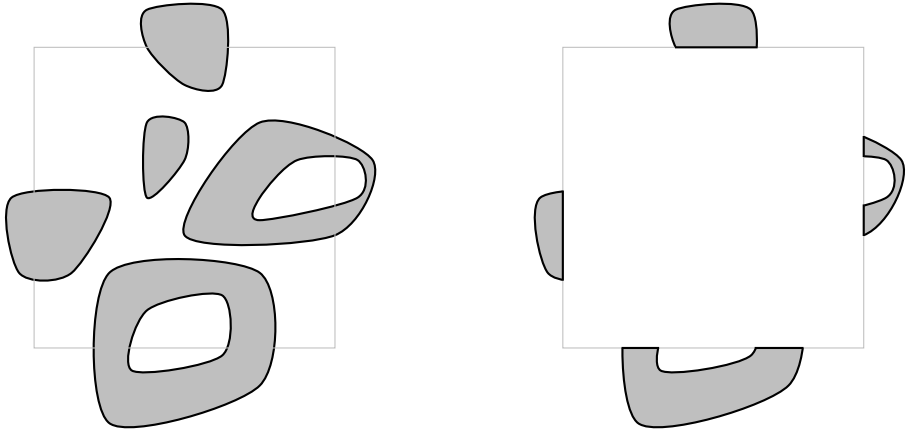
\begin{figure}[ht]
\centering
\begin{tikzpicture}
\fill[lightgray] plot[smooth cycle] coordinates{(0,-0.5) (1,1) (2.5,0.5) (2,-0.5) };
\draw[thick] plot[smooth cycle] coordinates{(0,-0.5) (1,1) (2.5,0.5) (2,-0.5) };
\fill[white] plot[smooth cycle] coordinates{(1,0) (1.5,0.5) (2.3,0.5) (2.3,0) (1,-0.3) };
\draw[thick] plot[smooth cycle] coordinates{(1,0) (1.5,0.5) (2.3,0.5) (2.3,0) (1,-0.3) };
\fill[lightgray] plot[smooth cycle] coordinates{(-2.2,-1) (-2.3, 0) (-1,0) (-1.5,-1) };
\draw[thick] plot[smooth cycle] coordinates{(-2.2,-1) (-2.3, 0) (-1,0) (-1.5,-1) };
\fill[lightgray] plot[smooth cycle] coordinates{(-0.5,2) (-0.5,2.5) (0.5,2.5) (0.5,1.5) (0,1.5) };
\draw[thick]  plot[smooth cycle] coordinates{(-0.5,2) (-0.5,2.5) (0.5,2.5) (0.5,1.5) (0,1.5) };
\fill[lightgray] plot[smooth cycle] coordinates{(-0.5,0) (-0.5,1) (0,1) (0,0.5)};
\draw[thick] plot[smooth cycle] coordinates{(-0.5,0) (-0.5,1) (0,1) (0,0.5)};
\fill[lightgray] plot[smooth cycle] coordinates{(-1,-1) (1,-1) (1,-2.5) (-1,-3)};
\draw[thick] plot[smooth cycle] coordinates{(-1,-1) (1,-1) (1,-2.5) (-1,-3)};
\fill[white] plot[smooth cycle] coordinates{(-0.5,-1.5) (0.5,-1.3) (0.5,-2.1) (-0.7,-2.3)};
\draw[thick] plot[smooth cycle] coordinates{(-0.5,-1.5) (0.5,-1.3) (0.5,-2.1) (-0.7,-2.3)};
\draw[lightgray]  (-2,-2) rectangle (2,2);

\pgftransformshift{\pgfpoint{200}{0}}
\fill[lightgray] plot[smooth cycle] coordinates{(0,-0.5) (1,1) (2.5,0.5) (2,-0.5) };
\draw[thick] plot[smooth cycle] coordinates{(0,-0.5) (1,1) (2.5,0.5) (2,-0.5) };
\fill[white] plot[smooth cycle] coordinates{(1,0) (1.5,0.5) (2.3,0.5) (2.3,0) (1,-0.3) };
\draw[thick] plot[smooth cycle] coordinates{(1,0) (1.5,0.5) (2.3,0.5) (2.3,0) (1,-0.3) };
\fill[lightgray] plot[smooth cycle] coordinates{(-2.2,-1) (-2.3, 0) (-1,0) (-1.5,-1) };
\draw[thick] plot[smooth cycle] coordinates{(-2.2,-1) (-2.3, 0) (-1,0) (-1.5,-1) };
\fill[lightgray] plot[smooth cycle] coordinates{(-0.5,2) (-0.5,2.5) (0.5,2.5) (0.5,1.5) (0,1.5) };
\draw[thick]  plot[smooth cycle] coordinates{(-0.5,2) (-0.5,2.5) (0.5,2.5) (0.5,1.5) (0,1.5) };
\fill[lightgray] plot[smooth cycle] coordinates{(-0.5,0) (-0.5,1) (0,1) (0,0.5)};
\draw[thick] plot[smooth cycle] coordinates{(-0.5,0) (-0.5,1) (0,1) (0,0.5)};
\fill[lightgray] plot[smooth cycle] coordinates{(-1,-1) (1,-1) (1,-2.5) (-1,-3)};
\draw[thick] plot[smooth cycle] coordinates{(-1,-1) (1,-1) (1,-2.5) (-1,-3)};
\fill[white] plot[smooth cycle] coordinates{(-0.5,-1.5) (0.5,-1.3) (0.5,-2.1) (-0.7,-2.3)};
\draw[thick] plot[smooth cycle] coordinates{(-0.5,-1.5) (0.5,-1.3) (0.5,-2.1) (-0.7,-2.3)};
\fill[white] (-2,-2) rectangle (2,2);
\draw[lightgray] (-2,-2) rectangle (2,2);
\draw[thick] (-2,0.095) -- (-2,-1.105);
\draw[thick] (-1.22,-2) -- (-0.72,-2);
\draw[thick] (0.55,-2) -- (1.2,-2);
\draw[thick] (2,-0.51) -- (2,-0.09);
\draw[thick] (2,0.54) -- (2,0.82);
\draw[thick] (-0.51,2) -- (0.59,2);
\end{tikzpicture}
\caption{\label{fig:white}
   Simple contours intersecting the square $\ov\Deo$ colored with two colors:
   1 (white) and 2 (gray). 
   After flipping all sites in $\Deo$ to the color 1,
   no contour is intersecting $\Deo$ and no triple point was created.
   The same would be true when flipping all sites in $\Deo$ to the color 2.
}
\end{figure}

In order to prove this, let $\Deo\subs V_0$ be finite and $G_0$-connected, and let
$\ov\Deo$ denote the union of $\Deo$ with its boundary in $G_0$, i.e.,
$\ov\Deo:=\Deo\cup\{v\in V_0:\exists u\in\Deo\mbox{ s.t.\ }\{u,v\}\in E_0\}$.
By Corollary~\ref{C:simple}, the probability that a non-simple contour
intersects
$\ov\Deo$ tends to zero as $\bet\to\infty$, uniformly in $\La$. Thus, we may
choose $\bet_0<\infty$ such that
\be\label{simpcut}
\nu_{\La,\bet}\big(\{\Ga \colon\: \nexists
\ga\in\Ga\mbox{ s.t.\ }t(\ga)\geq
1,\ \ga\mbox{ intersects }\ov\Deo\}\big)\geq 1/2,
\ee
uniformly in $\La$ {and $\bet_0\leq\bet\leq\infty$.} If all
contours intersecting $\ov\Deo$ are simple, then we claim
that we can change our contour configuration at a finite energetic cost {\em
uniformly} in $\bet_0\leq\bet\leq\infty$, so that no contour intersects $\Deo$.
{To describe the algorithm of changing a contour configuration
$\Ga$ into a configuration $\Ga'$ with no contour intersecting $\Deo$, we first
observe that relying on } the fact that all contours intersecting $\ov\Deo$ are
simple, we can color the vertices in $\La\cap V_0$ in three colors in such a
way that boundaries between different colors correspond to contours and only
two different colors occur in $\ov\Deo$. (Note that this part of the argument
uses that the contours intersecting $\ov\Deo$ are simple everywhere and not
just that there are no triple points inside $\ov\Deo$.) Now we change our
coloring by painting $\Deo$ uniformly in one of these two colors, {
defining thus the new contour configuration $\Ga'$} (see
Figure~\ref{fig:white}). Since { in the construction of $\Ga'$ a
two-color configuration in $\ov\Deo$ was changed using the same two colors}
{\em no new triple points}\/ are introduced. This, together with
(\ref{posnudef}) and a standard finite-energy argument proves our claim.

For completeness, we write down this finite-energy argument in detail.  Let
$\Ga$ and $\Ga'{=\psi(\Ga)}$ denote the old and new contour configuration obtained by the
procedure described above.  We need to estimate the relative probability of
$\Ga'$ with respect to $\Ga$ and the number of different configurations $\Ga$
that can be mapped onto the same $\Ga'$, {$\abs{\Psi^{-1}(\Ga')}$}. Let $|\Deo|$ be the number of sites in
$\Deo$, let $M_\Deo$ be the number of edges in $E_1$ that separate sites in
$\Deo$ from each other and let $M_{\pa\Deo}$ be the number of edges in $E_1$
that separate sites in $\Deo$ from sites in $\ov\Deo\beh\Deo$. {Further}, let
\be
\chi(\Ga):=\prod_{\ga\in\Ga}\chi(\ga),\quad
|\Ga|:=\sum_{\ga\in\Ga}|\ga|
\quad\mbox{and}\quad t(\Ga):=\sum_{\ga\in\Ga}t(\ga).
\ee
{Since all contours we remove or alter are simple contours with $\chi(\gamma)=2$ and we remove or alter} no more than $M_\Deo$ contours from our configuration and add
no more than $M_{\pa\Deo}$ edges, we have
\be
\chi(\Ga')\geq 2^{-M_\Deo}\chi(\Ga)
\quad\mbox{and}\quad|\Ga'|\leq|\Ga|+M_{\pa\Deo},
\ee
while $t(\Ga')=t(\Ga)$, which by (\ref{posnudef}) implies that
\be
\nu_{\La,\bet}(\Ga')\geq 2^{-M_\Deo}p_\bet^{M_{\pa\Deo}}\nu_{\La,\bet}(\Ga).
\ee
Moreover, since there are $2^{|\Deo|}$ ways of coloring the vertices in $\Deo$
using only two colors, we see that there are at most $2^{|\Deo|}$ different
contour configurations $\Ga$ {in $\Psi^{-1}(\Ga')$}. Recall that
$\Ji_\Deo=\{\Ga \colon\: \mbox{ no contour in $\Ga$ intersects }\Deo\}$ corresponds to the
event that $\Deo$ is uniformly colored in one color. Let $\Si_{\ov\Deo}$ be the
event that all contours intersecting $\ov\Deo$ are simple. Then
\be\ba{l}
\dis\nu_{\La,\bet}(\Ji_\Deo)
=\sum_{\Ga'\in\Ji_\Deo}\nu_{\La,\bet}(\Ga')
\geq 2^{-|\Deo|}\sum_{\Ga\in\Si_{\ov\Deo}}\nu_{\La,\bet}({\Psi(\Ga)})\\[5pt]
\dis\quad\geq 2^{-|\Deo|-M_\Deo}p_\bet^{M_{\pa\Deo}}
\sum_{\Ga\in\Si_{\ov\Deo}}\nu_{\La,\bet}(\Ga)
\geq 2^{-1-|\Deo|-M_\Deo}p_\bet^{M_{\pa\Deo}},
\ec
where we have used (\ref{simpcut}) in the last step.
\qed

\proofof{Lemma~\ref{L:ground}}
Consider any $v\in V_1\cap\La$ and let $w_1,w_2,w_3\in V_0$
be its neighbors in $G$.
Then the DLR equations for the volume $\{v\}$ imply that
\be\ba{l}
\dis \mu^{1}_{\La,\bet}(\exists i \hbox{ with }\sigma_v = \sigma_{w_i} |
  \sigma_{w_1}, \sigma_{w_2}, \sigma_{w_3}) \\[5mm]
\;=\;\: 
\begin{cases}
    \displaystyle \frac{e^{-3\bet}}{2+e^{-3\bet}}   &
           \text{if $\sigma_{w_1}=\sigma_{w_2}=\sigma_{w_3}$} \\[3mm]
    \displaystyle \frac{e^{-\bet}+e^{-2\bet}}{1+e^{-\bet}+e^{-2\bet}}  &
         \text{if $|\{\sigma_{w_1},\sigma_{w_2},\sigma_{w_3}\}|=2$} \\[3mm]
    1  & \text{if $|\{\sigma_{w_1},\sigma_{w_2},\sigma_{w_3}\}|=3$}
\end{cases}
\qquad
\label{eq.v1.DLR}
\ec
Let $\Bi:=\{\sig \colon\: |\{\sig_{w_1},\sig_{w_2},\sig_{w_3}\}| = 3\}$
be the (``bad'') event that $w_1,w_2,w_3$ are colored in three different colors.
It follows from Corollary~\ref{C:simple} that
\be
\mu^{1}_{\La,\bet}(\Bi)  \;\le\;  C e^{-2\beta}
\label{eq.v1.bad}
\ee
uniformly for $\beta \in [\beta_0,\infty]$ and
for finite and simply connected $\La\ni v$ such that $\pa\La\subs V_0$;
and by increasing $C$ we can make this hold uniformly for $\beta \in [0,\infty]$.
It then follows from \reff{eq.v1.DLR} and \reff{eq.v1.bad} that
\be
\mu^{1}_{\La,\bet}(\exists i \hbox{ with }\sigma_v = \sigma_{w_i} )
\;\le\;
 C' e^{-{\beta}}
\ee
uniformly for $\beta \in [0,\infty]$
and for finite and simply connected $\La\ni v$ such that $\pa\La\subs V_0$.
\qed

\section{Positive magnetization}\label{S:magnet}

In this section we prove Lemma~\ref{L:magnet}:
by using this lemma we can
improve the statement that sufficiently large blocks are more likely
to be uniformly colored in the color~1 than in any other color,
to the ``positive magnetization'' statements in Theorem~\ref{T:main}(a,b),
which say that single vertices in the sublattices $V_0$ and $V_1$
are colored with the color~1 with a probability that is
strictly larger (resp.\ strictly smaller) than $1/3$.

We fix an arbitrary $\bet_0 > 0$ throughout this section;
our estimates will be uniform in $\bet\in[\bet_0,\infty]$.
We will later also fix a finite $G_0$-connected set $\Deo \subset V_0$.
As in all our proofs, we work with the finite-volume Gibbs measures
$\mu^1_{\La,\bet}$, where $\La\subs V$ is finite and simply connected in $G$
and satisfies $\pa\La\subs V_0$. We aim to derive bounds that are uniform in such
$\La$ with $\La\supseteq\Deo$.

Unlike what was done in the preceding subsections,
we will not make use of the contour description of $\mu^1_{\La,\bet}$,
nor will we integrate out one sublattice.
Rather, we will work directly with the Potts antiferromagnet on our original
quadrangulation $G=(V,E)$.

Note first that the measures $\mu^1_{\La,\bet}$ are invariant under global interchange
of the colors~2 and 3.  In particular, we have
$\mu^1_{\La,\bet}(\sig_v=2)=\mu^1_{\La,\bet}(\sig_v=3)$ for all $v\in\La$.
Thus, to show that $\mu^1_{\La,\bet}(\sig_v=1)>1/3$
(resp.\ $<1/3$), we may equivalently show that
$\mu^1_{\La,\bet}(\sig_v=1)-\mu^1_{\La,\bet}(\sig_v=2)>0$ (resp.\ $<0$).
Because of the antiferromagnetic nature of our model, it is in fact
already nontrivial to show that these quantities are nonnegative
(resp.\ nonpositive) for $v\in V_0$ (resp.\ $v\in V_1$).
This problem has been solved, however, in \cite[Appendix~A]{FS99},
where a { first} Griffiths inequality
for antiferromagnetic Potts models on bipartite graphs
is proven using ideas based on the cluster
algorithm from \cite{WSK89,WSK90}.

We will elaborate on these ideas.
The main step will be to give a random-cluster representation
for the law of the 1's and 2's conditional on the 3's.
In this representation, we will see that for $v_0\in V_0$,
the difference between the probabilities that $v_0$ is colored~1 or colored~2
equals the probability that $v_0$ percolates,
i.e., that $v_0$ is in the same random cluster of 1's and 2's
as the boundary $\pa\La$.
{Moreover,} a similar statement holds for the probability that
$\Deo$ is uniformly colored in the color~1 minus the probability
that it is uniformly colored in the color~2.
Thus, by showing that both {of these}
quantities are related to percolation of the 1's and 2's,
we can prove that if one {of them}
is strictly positive, then so must be the other.
Note that conditioning on the positions of the 3's
would not in general be a very useful thing
to do when trying to prove statements about our model, since we have no
{\em a priori}\/ knowledge of the distribution of the 3's.
Nevertheless, as we see here, it can be used to show that a certain statement
that has already been proved is equivalent to another statement
for which we have no direct control.

So let $G=(V,E)$ be our original quadrangulation,
and let $E_\La$ be the set of edges in~$E$
that have at least one endvertex in $\La$.
We define
the measure $\rho^1_{\La,\bet}$ on $\{1,2,3\}^\La\times\{0,1\}^{E_\La}$
so that the marginal distribution of $\rho^1_{\La,\bet}(\sigma,\eta)$
on $\sigma$ is the Gibbs measure $\mu^1_{\La,\bet}$ and so that,
conditional on $\sig$, independently for each $e\in E_\La$,
one has $\eta_e=1$ with probability $p := 1-e^{-\bet}$
if $\sig_u,\sig_v\in\{1,2\}$ and $\sig_u\neq\sig_v$,
and $\eta_e=0$ {in all other cases.}
That is,
\bc
\dis\rho^1_{\La,\bet}(\sig,\eta)&:=&\dis\frac{1}{Z^1_{\La,\bet}}
\exp\big[-\bet H_\La(\sig\,|\,\tau)\big]  \\[3mm]
&&\dis\times\prod_{\{u,v\}\in E_\La} \!\!
\big(1_{\Ai_{u,v}}\big[p1_{\{\eta_{\{u,v\}}=1\}}+(1-p)1_{\{\eta_{u,v}=0\}}\big]
+1_{\Ai^{\rm c}_{u,v}}1_{\{\eta_{u,v}=0\}}\big)
\ec
where $\Ai_{u,v}$ is the event
\be
\Ai_{u,v} \;:=\; \big\{\sig_u,\sig_v\in\{1,2\}\mbox{ and }\sig_u\neq\sig_v\big\}  \;,
\ee
$\Ai^{\rm c}_{u,v}$ is its complement, $\tau$ is any spin configuration that
assumes the value 1 on $\pa\La$, $H_\La(\sig|\tau)$ is defined in (\ref{HLa}),
and $Z^1_{\La,\beta}$ is the same normalizing constant as in (\ref{mutau}).

Now let
\begin{subeqnarray}
\Lambda^{12}  &:=&  \big\{v\in\La\cup\pa\La \colon\: \sig_v\in\{1,2\}\big\}  \\
\Lambda^3     &:=&  \big\{v\in\La \colon\: \sig_v=3\big\}
\end{subeqnarray}
be the sets of vertices in $\La \cup \pa\La$
where $\sig$ assumes the values 1 or 2 (resp.\ 3), and set
\be
E^{12} \;:=\; \{e\in E_\La \colon\: \eta_e=1\} \;.
\ee
Conditionally on $\Lambda^3$,
the spins $(\sig_v)_{v\in \Lambda^{12}}$ are distributed
as an antiferromagnetic Ising model, with 1 boundary conditions, on the diluted
lattice $\Lambda^{12}$. Since $\Lambda^{12}$ is bipartite and the boundary conditions
lie entirely on the sublattice $V_0$, we may flip the spins on the other sublattice
(i.e., on $\Lambda^{12} \cap V_1$) to obtain a ferromagnetic Ising model
$(\sig'_v)_{v\in \Lambda^{12}}$
on $\Lambda^{12}$.  After this flipping, the conditional joint law of
$(\sig'_v)_{v\in \Lambda^{12}}$ and $\eta$ given $\Lambda^3$
is just the standard
coupling of this ferromagnetic Ising model and its corresponding
random-cluster model on $\Lambda^{12}$
(see \cite{ES88} and \cite[Section~1.4]{Gri06}).
(Notice that for all edges $\{u,v\}$ such that $\{u,v\}\cap \Lambda^3 \neq \emptyset$,
we have $\eta_{\{u,v\}}=0$.)
Returning to the original (unflipped) spins $(\sig_v)_{v\in \Lambda^{12}}$,
we see from \cite[Theorem~1.13]{Gri06} that, conditional on $\Lambda^3$ and $\eta$,
the connected components of the graph $G^{12}=(\Lambda^{12},E^{12})$ are independently
given proper 2-colorings (with colors 1 and 2) as follows:
for any component not connected to the boundary $\pa\La$,
each of the two proper 2-colorings arises with probability $1/2$;
and any component connected to the boundary is given the unique
proper 2-coloring that is compatible with the boundary conditions
(namely, color~1 on $V_0$ and color~2 on $V_1$).
In particular, for points $v_0\in V_0\cap\La$ one has
\begin{eqnarray}
 \rho^1_{\La,\bet}\big(\sig_{v_0}=1\,\big|\,\Lambda^3,\eta\big)
 & \:=\: &
 \begin{cases}
      1             & \text{if $v_0\leftrightarrow_\eta\pa\La$}   \\[1mm]
      \ffrac{1}{2}  & \text{if $v_0 \in \Lambda^{12}$ and 
                         $v_0\not\hspace{-1.5pt}\leftrightarrow_\eta\pa\La$} \\[1mm]
      0             & \text{if $v_0 \in \Lambda^{3}$}
 \end{cases}
      \\[3mm]
 \rho^1_{\La,\bet}\big(\sig_{v_0}=2\,\big|\,\Lambda^3,\eta\big)
 & \:=\: &
 \begin{cases}
      0             & \text{if $v_0\leftrightarrow_\eta\pa\La$}   \\[1mm]
      \ffrac{1}{2}  & \text{if $v_0 \in \Lambda^{12}$ and 
                         $v_0\not\hspace{-1.5pt}\leftrightarrow_\eta\pa\La$} \\[1mm]
      0             & \text{if $v_0 \in \Lambda^{3}$}
 \end{cases}
\end{eqnarray}
where $v\leftrightarrow_\eta\pa\La$ denotes the event that $v$ is connected to
$\pa\La$ through a path of edges with $\eta_e=1$
[note that $v_0 \in \Lambda^{3}$ implies
$v_0\not\hspace{-1.5pt}\leftrightarrow_\eta\pa\La$].
For the unconditional law, it follows that
\be
\label{E:ro1-ro2}
\rho^1_{\La,\bet}(\sig_{v_0}=1)-\rho^1_{\La,\bet}(\sig_{v_0}=2)
\;=\;
\rho^1_{\La,\bet}\big(v_0\leftrightarrow_\eta\pa\La\big)  \;.
\ee
For $v_1\in V_1\cap\La$, one has similar equations with the roles of
colors~1 and 2 interchanged, so that
\be
\label{E:ro2-ro1}
\rho^1_{\La,\bet}(\sig_{v_1}=2)-\rho^1_{\La,\bet}(\sig_{v_1}=1)
\;=\;
\rho^1_{\La,\bet}\big(v_1\leftrightarrow_\eta\pa\La\big)  \;.
\ee

Now consider a finite set $\Deo \subset V_0$,
and recall that $\Ji_{k,\Deo}$ denotes the event that $\Deo$ is uniformly
colored in the color~$k$,
and that $\Ji_\Deo =\bigcup_{k=1}^3 \Ji_{k,\Deo}$ denotes the event that all sites
in $\Deo$ are uniformly colored in some color.
Let $\Deo\leftrightarrow_\eta\pa\La$ denote the
event that there is at least one site in $\Deo$ that is connected to $\pa\La$
through a path of edges with $\eta_e=1$.
Since
\be
\rho^1_{\La,\bet}\big(\Ji_{1,\Deo}\,\big|\,\Lambda^3,\eta\big)
\;=\; \rho^1_{\La,\bet}\big(\Ji_{2,\Deo}\,\big|\,\Lambda^3,\eta\big)
\quad\mbox{a.s.\ on}\quad
\Deo\not\hspace{-1.5pt}\leftrightarrow_\eta\pa\La
\ee
and
\be
\rho^1_{\La,\bet}\big(\Ji_{2,\Deo}\,\big|\,\Lambda^3,\eta\big) \;=\;  0
\quad\mbox{a.s.\ on}\quad
\Deo\leftrightarrow_\eta\pa\La,
\ee
we see that
\be
\label{E:Ji1-Ji2}
\rho^1_{\La,\bet}(\Ji_{1,\Deo})-\rho^1_{\La,\bet}(\Ji_{2,\Deo})
\;=\; \rho^1_{\La,\bet}\big(\Ji_{\Deo}
\cap\{\Deo\leftrightarrow_\eta\pa\La\}\big) \;.
\ee

Now, recall that a finite set $\Delta \subset V$ is termed {\em thick}\/
if there exists a nonempty finite subset $\Delta_1 \subset V_1$
that is connected in $G_1$ and such that
$\Delta = \{ v \in V \colon\: d_G(v,\Delta_1) \le 1 \}$.
We therefore fix some thick set $\Delta\subs V$
and define $\Deo:=\Delta\cap V_0$
(since $G$ is bipartite we have $\Dei=\Delta\cap V_1$).
Comparing \reff{E:ro1-ro2}/\reff{E:ro2-ro1}/\reff{E:Ji1-Ji2}
and noting that $\rho^1_{\La,\bet}$ can be replaced by
$\mu^1_{\La,\bet}$ on the left-hand sides,
we see that Lemma~\ref{L:magnet} is implied by the following claim:

\bl{\bf({Comparison lemma})}
\label{lemma_perc12}
Fix $\bet_0 > 0$ and let $\Delta\subs V$ be thick.
Then there exists an $\eps>0$ such that
\be
\rho^1_{\La,\bet}\big(v\leftrightarrow_\eta\pa\La \hbox{ for all } v \in\Delta \big)
\;\geq\;
\eps \: \rho^1_{\La,\bet}\big(\Ji_{\Deo} \cap\{\Deo\leftrightarrow_\eta\pa\La\}\big)
\label{eq.lemma_perc12}
\ee
uniformly for all $\bet\in[\bet_0,\infty]$ and all simply connected finite
sets $\La\supseteq\Delta$ such that $\pa\La\subs V_0$.
In fact, we can choose $\epsilon = 3^{-|\Dei|} \, (1-e^{-\beta_0})^{|E_\Delta|}$
where $E_\Delta = \big\{ \{u,v\} \in E \colon\: u,v \in \Delta \big\}$.
\el

\proof
{
The proof is by a finite-energy argument:
that is, to each $(\sig,\eta) \in \Ji_{\Deo} \cap\{\Deo\leftrightarrow_\eta\pa\La\}$
we associate a
$(\sig'',\eta'') \in \{v\leftrightarrow_\eta\pa\La \hbox{ for all } v \in\Delta\}$;
we then compute a lower bound on the ratio of $\rho^1_{\La,\bet}(\sig'',\eta'')$
to the total $\rho^1_{\La,\bet}$-weight of the configurations $(\sig,\eta)$
that map onto it.
The construction is in two steps
$(\sig,\eta) \mapsto (\sig',\eta') \mapsto (\sig'',\eta'')$.
In the first step we recolor all spins $(\sig_v)_{v \in \Dei}$ to $\sig'_v = 2$
(leaving all other variables as is).
In the second step we set all bond variables $(\eta_e)_{e \in E_\Delta}$
to $\eta''_e = 1$ (again leaving all other variables as is).
Let us now compute a lower bound on the ratio of weights, as follows:

Since $(\sig,\eta) \in \Ji_{\Deo} \cap\{\Deo\leftrightarrow_\eta\pa\La\}$
and $\pa\La$ is colored~1, it follows that $\sig_v = 1$ for all $v \in \Deo$.
Since $\Delta$ is thick, every vertex in $\Dei$ has all its neighbors in $\Deo$.
Therefore we can recolor all sites in $\Dei$ with the color~2
without increase in energy,
i.e.\ $\rho^1_{\La,\bet}(\sig',\eta') \ge \rho^1_{\La,\bet}(\sig,\eta)$.\footnote{
 In fact, \reff{eq.lemma_perc12} would still hold (with a worse $\eps$)
 even if there were an energy cost associated to this operation,
 provided that this energy cost is uniformly bounded.
}
We lose a factor $3^{|\Dei|}$ because $3^{|\Dei|}$ configurations
$(\sig,\eta)$ map onto the same configuration $(\sig',\eta')$.

We now have $\sig'_u \neq \sig'_v$ for all $\{u,v\} \in E_\Delta$.
Therefore $\rho^1_{\La,\bet}(\sig'',\eta'')$ is precisely $p^{E_\Delta}$
times the total $\rho^1_{\La,\bet}$-weight of the configurations $(\sig',\eta')$
that map onto it, where $p = 1 - e^{-\beta} \ge 1 - e^{-\beta_0}$.
}
\qed

\medskip

{\bf Remark.}
The ideas in this section
---  in particular, formulas \eqref{E:ro1-ro2}, \eqref{E:ro2-ro1}
and \eqref{E:Ji1-Ji2} ---
have an obvious generalization to the $q$-state Potts antiferromagnet
for any $q \ge 2$ on any bipartite graph (not necessarily a
plane quadrangulation),
where we condition on $q-2$ colors and use the random-cluster representation
for the remaining 2 colors.
Indeed, as in \cite[Appendix~A]{FS99}, we may consider an even more general
situation:
Suppose that the vertex set $V$ is partitioned as $V = V_0 \cup V_1$;
then we can consider a Potts model with antiferromagnetic interactions
on edges connecting $V_0$ to $V_1$
and ferromagnetic interactions on edges $V_0$--$V_0$ and $V_1$--$V_1$.

\appendix

\section{Some facts about infinite planar graphs}

The purpose of this appendix is to collect some facts about infinite graphs,
and in particular about infinite graphs embedded in the plane, that will be
needed in the main part of the paper. An excellent general introduction to
the theory of infinite graphs can be found in \cite[Chapter~8]{Diestel_10};
but we shall require here some further facts that are scattered throughout the
recent research literature, plus a few that appear to be new.

\subsection{Basic facts and definitions}

Recall that a graph is a pair $G=(V,E)$ consisting of a
(not necessarily finite) vertex set $V$ and edge set $E$.  Unless mentioned
otherwise, when we say ``graph'' we will always mean
{a {\em simple}\/ graph, i.e.,}
an undirected graph that has no loops or multiple edges.
Thus, the elements of $E$ (the edges) are
unordered sets $\{v,w\}$ containing two distinct elements of $V$.
Two vertices $v,w\in V$ are called {\em adjacent}\/ if $\{v,w\}\in E$. An
edge $e$ containing a vertex $v$ is said to be {\em incident}\/ to $v$.  The
{\em degree}\/ of a vertex $v \in V$ is the number of edges incident to it.
We say that $G$ is {\em finite}\/ (resp.\ {\em countable}\/)
if both $V$ and $E$ are finite (resp.\ countable).
We say that $G$ is {\em locally finite}\/ (resp.\ {\em locally countable}\/)
if every vertex has finite (resp.\ countable) degree.

A graph $G'=(V',E')$ such that $V'\subq V$ and $E'\subq E$
is called a {\em subgraph}\/ of $G=(V,E)$;
we also say that $G$ {\em contains}\/ $G'$.
If $E'$ contains all edges $\{v,w\}\in E$ with $v,w\in V'$,
then $G'$ is called the subgraph of $G$ {\em induced}\/ by $V'$.
Likewise, if $V'=\{v\in V \colon\: \exists w\in V\mbox{ s.t.\ }\{v,w\}\in E'\}$,
then we call $G'$ the subgraph {\em induced}\/ by $E'$.

We will say that a graph $G$ is {\em connected}\/ if for every {\em proper}\/
subset $W\subs V$ (the word ``proper'' means that $W\neq\emptyset,V$) there
is an edge $\{v,w\}\in E$ with $v\in V\beh W$, $w\in W$.
Let us note that every locally countable, connected graph
is countable.\footnote{
 Indeed, if $G=(V,E)$ is connected
 and $W$ is the set of all vertices at finite graph distance
 from a given vertex, then connectedness implies $W=V$.
}
A connected graph in which each vertex has degree $\le 2$
will be called a {\em generalized path}\/.
The {\em length}\/ of a generalized path is the number of its edges.
Vertices of degree~2 are called {\em internal vertices}\/
of the generalized path, while
vertices of degree one or zero are called {\em endvertices}\/.\footnote{
 Note that a vertex of degree zero can occur in a connected graph
 (and more particularly in a path)
 only if the graph has precisely one vertex.
}
An infinite generalized path with one endvertex is called a one-way-infinite
path or {\em ray}\/;
an infinite generalized path without endvertices is called
a two-way-infinite path or {\em double ray}\/;
a finite generalized path without endvertices is called a {\em cycle}\/;
and a finite generalized path with one or two endvertices is called
a {\em path}\/.
In particular, a graph consisting of a single vertex and no edges
is a path of length zero.

Two vertices $v,w$ in a graph $G$ are {\em linked}\/ by a path
if $G$ contains a path that has $v$ and $w$ as its endpoints.
Then it is easy to see that a graph is connected
(according to our definition above) if and only if every
pair of vertices in $G$ is linked by a path.
The {\em graph distance}\/ $d(v,w)$ between two vertices $v,w \in V$
is the length of a shortest path linking $v$ and $w$ if such a path exists,
and $\infty$ otherwise.
An edge $\{v,w\}$ in a path $P$ is said to be a {\em final edge}\/ of $P$
if either $v$ or $w$ (or both) has degree one.
The {\em graph distance}\/ $d(e,f)$ between two edges $e,f\in E$
is defined as the minimal length minus one of a path
that has $e$ and $f$ as final edges.
Note, in particular, that $d(e,f)=0$ iff $e=f$,
and that $d(e,f)=1$ iff $e\neq f$ but $e$ and $f$ share a vertex.
It is easy to check that, whenever $G$ is connected,
the graph distance between vertices (resp.\ edges)
defines a metric on $V$ (resp.\ $E$).

Let $G=(V,E)$ be a connected graph.
A set $B\subq E$ is called {\em separating}\/
if $G\beh B$ is not connected.
A set $C\subq E$ is called a {\em cutset}\/ if there exists a partition
$\{V_1,V_2\}$ of $V$ into two nonempty sets such that
$C = E(V_1,V_2) := \{\{v_1,v_2\}\in E \colon\: v_1\in V_1,\ v_2\in V_2\}$.
(Since $G$ is assumed connected, $\emptyset$ is not a cutset.)
A separating set (resp.\ cutset) is called {\em minimal}\/ if it
contains no proper subset that is a separating set (resp.\ a cutset).
In fact, the two concepts are the same: each minimal separating set
is also a minimal cutset and vice versa. Moreover, a set $C\subq E$ is a
minimal cutset if and only if $(V,E\beh C)$ has exactly two connected
components.\footnote{
   In the graph-theory literature,
   minimal cutsets are sometimes called {\em bonds}\/.
   Alas, in the statistical-mechanics literature,
   ``bond'' is often used as a synonym for ``edge''.
   To forestall confusion, we prefer to avoid the term ``bond'' altogether.
}

Two rays in an infinite graph $G$ are said to be {\em end-equivalent}\/
(or {\em equivalent}\/ for short)
if one (hence all) of the following equivalent conditions holds:
\begin{quote}

\noindent
1) there exists a third ray whose intersection with both of them is infinite;

\noindent
2) there are infinitely many disjoint paths in $G$ joining the two rays;

\noindent
3) for every finite set $S\subset V$, the two rays are
  eventually contained in the same connected component of $G \beh S$.
\end{quote}
It is easy to see that end-equivalence is an equivalence relation.
The corresponding equivalence classes are termed the
{\em ends}\/ of the graph $G$.\footnote{
 This definition of the ends of a graph is
 due to Halin \cite{Halin_64} in 1964
 (see also Freudenthal \cite{Freudenthal_45} for the locally finite case).
 The notion of ``end'' of a topological space was introduced
 much earlier by Freudenthal \cite{Freudenthal_31}, in 1931.
 If one identifies $G$ with the ``topological realization''
 defined below (at the beginning of Section~\ref{subsec.planar}),
 then the two notions coincide for {\em locally finite}\/ graphs
 but are different in general \cite{Diestel-Kuhn_03}.
 See \cite{Kron_05} for an elementary introduction
 to the theory of ends of locally finite graphs;
 and see \cite{Kron-Teufl} for a survey treating both graphs
 and topological spaces.
}
It is not hard to see that
a connected, locally finite graph $G=(V,E)$ has one end
if and only if for every finite $E' {\subq} E$,
the   subgraph  $(V,E\beh E')$ has exactly one infinite component;
or equivalently, if every finite minimal {cutset} divides $V$ into two connected
components, of which exactly one is of infinite size.

For $k \ge 1$, a graph $G=(V,E)$ is called {\em $k$-connected}\/ if $|V|\geq
k+1$ and the subgraph induced by $V\beh W$ is connected for all $W\subs V$
satisfying $|W|< k$.  (Otherwise put, to disconnect $G$ one must remove at
least $k$ vertices.)
Two vertices $v,w$ in $G$ are said to be {\em $k$-edge-connected}\/
if one needs to remove at least $k$ edges to unlink them;
a graph is called {\em $k$-edge-connected}\/ if every pair of vertices in it
is $k$-edge-connected.
It is easy to see that $k$-edge-connectedness of vertices
is an equivalence relation;
the corresponding equivalence classes of vertices (and their induced subgraphs)
are called the {\em $k$-edge-connected components}\/ of the graph.
Two paths are called {\em vertex-disjoint}\/ (resp.\ {\em edge-disjoint}\/)
if their sets of internal vertices (resp.\ edges) are disjoint.
By Menger's theorem, two vertices are $k$-edge-connected if and only if they
are linked by $k$ edge-disjoint paths, and a graph is $k$-connected if and
only if every two vertices are linked by $k$ vertex-disjoint paths.

An {\em automorphism}\/ of a graph $G=(V,E)$ is a bijection
$g\colon\: V\to V$ such that $\{g(v),g(v')\}\in E$
if and only if $\{v,v'\}\in E$.
We say that two vertices $v,w\in V$ are of the same {\em type}\/,
denoted $v\sim w$, if there exists an automorphism $g$ of $G$ such that
$g(v)=w$. Then $\sim$ is an equivalence relation that divides the vertex set
$V$ into equivalence classes called types.
A graph is called {\em vertex-transitive}\/
if there is only one type of vertex,
and {\em vertex-quasi-transitive}\/
if there are only finitely many types of vertices.
Similarly, we say that two edges $\{v,v'\},\, \{w,w'\} \in E$
are of the same type if there exists an automorphism $g$ of $G$
such that $\{g(v),g(v')\}=\{w,w'\}$;
and we say that two directed edges $(v,v'),\, (w,w') \in V \times V$
with $\{v,v'\},\, \{w,w'\} \in E$
are of the same type if there exists an automorphism $g$ of $G$
such that $g(v) = w$ and $g(v') = w'$.
Edge- and directed-edge- transitivity or quasi-transitivity
are then defined in the obvious way.
We shall need the following fairly easy result:

\begin{lemma}
\label{L:quasi}
For a locally finite graph $G$,
the following are equivalent:
\begin{itemize}
\item[(a)]  $G$ is vertex-quasi-transitive.
\item[(b)]  $G$ is edge-quasi-transitive.
\item[(c)]  $G$ is directed-edge-quasi-transitive.
\end{itemize}
\end{lemma}

\proof
(b) $\desd$ (c): Obviously, if two directed edges $(v,v')$ and
$(w,w')$ are of the same type, then the corresponding undirected edges
$\{v,v'\}$ and $\{w,w'\}$ are also of the same type. This shows that there are
as most as many types of edges as there are types of directed
edges. Conversely, since there are only two ways to order a set with two
elements, there are at most twice as many types of directed edges as there are
types of edges.

(c) $\volgt$ (a): If two directed edges $(v,v')$ and $(w,w')$ are of the same
type, then obviously $v$ and $w$ are of the same type. Since all isolated
vertices (i.e., vertices of degree zero) are of the same type, this shows that
the number of types of vertices is at most the number of types of directed
edges plus one.

(a) $\volgt$ (c):
Assume that there are $m$ types of vertices and that these have
degrees $d_1,\ldots,d_m$. Pick representatives $v_1,\ldots,v_m$ of these
equivalence classes. For any directed edge $(v,v')$, there exists a
$k\in\{1,\ldots,m\}$ and a graph automorphism that maps $v$ to $v_k$.
Since a graph automorphism preserves the graph structure, $w'$ must be mapped
into one of the $d_k$ vertices adjacent to $v_k$. Thus, we have found
$d_1+\cdots+d_m$ directed edges such that each directed edge can be mapped
into one of these by a graph automorphism. In particular, the number of types
of directed edges is at most $d_1+\cdots+d_m$.
\qed

In view of Lemma~\ref{L:quasi},
we usually talk about quasi-transitive graphs without specifying
whether we mean in the vertex, edge or directed-edge sense.\footnote{
  The term ``almost transitive'' is also used
  as a synonym for ``quasi-transitive''.
}

A {\em geodesic}\/ in a graph $G$ is a generalized path $P$
such that for each pair of vertices $v,w$ in $P$,
the graph distance from $v$ to $w$ in $P$ coincides with the
graph distance from $v$ to $w$ in $G$.
It is not hard to see that any path of minimal length
linking two vertices $v',w'$ is a geodesic.
For completeness, we prove the following simple fact.
Part~(a) of this lemma can be found, for example, in \cite[Prop.~1]{Nor91};
{it is a simple corollary of K\"onig's Infinity Lemma
\cite[Lemma~8.1.2]{Diestel_10}.}
We did not find a reference for part~(b),
but it is presumably well known.

\bl{\bf(Infinite geodesics)}\label{L:geodesic}
Let $G=(V,E)$ be a locally finite connected graph with infinite
vertex set $V$.  Then:
\begin{itemize}
\item[{\rm(a)}] For each $v\in V$, there exists a geodesic ray whose
endpoint is $v$.
\item[{\rm(b)}] If $G$ is moreover quasi-transitive, then $G$ contains a
geodesic double ray.
\end{itemize}
\el

\proof
Since each vertex is of finite degree, the set of vertices at distance $k$ from
$v$ is finite for each $k\geq 0$. Therefore, since $V$ is infinite and $G$ is
connected, for each $n\geq 1$ we can find a vertex $v_n\in V$ at distance
$d(v,v_n)=n$.  Consider a path $(v_0^{(n)}=v, v_1^{(n)},\dots, v_{n-1}^{(n)},v_n^{(n)}=v_n )$
and define the function $f_n \colon\: \N\to V$ by taking $f_n(k)=v_{k}^{(n)}$ for $k\le n$
and $f_n(k)=v_n$ for all $k\geq n$. 
Since the set of points at distance $k$ from $v$ is finite for each
$k\geq 0$, we may select a subsequence $f_{n_m}$
that converges {pointwise} in the discrete topology.
It is easy to see that the limit of such a subsequence is a geodesic
ray starting at $v$, proving part~(a) of the lemma.

To prove also part~(b), we use that by quasi-transitivity, the geodesic ray
constructed in part~(a) must pass through at least one type of vertex
infinitely often. It follows that for a vertex $v$ of this type, for each
$n\geq 0$ we can find a function $f_n \colon\, \Z\to V$ such that $f_n(0)=v$,
$f_n(k)=f_n(-n)$ for all $k\leq-n$, the vertices $\{f_n(k) \colon\: k\geq-n\}$
are all different, and the graph induced by this set is a geodesic ray. 
(It suffices to employ the corresponding automorphism
to shift the original geodesic ray by identifying a 
vertex of given type sufficiently far  on it with the vertex $v$.)
Now the
statement follows from the same sort of compactness argument as used in the
proof of part~(a).
\qed

\subsection{Duality}  \label{sec.A.2}

In the main part of this paper, we make extensive use of the fact that the two
sublattices $G_0$ and $G_1$ are each other's dual in the sense of planar graph
duality. Such duals may be defined abstractly, using only basic concepts of
graph theory, without any reference to embeddings of graphs in the plane.
In fact, this abstract  notion of duality is
sufficient for all our proofs,
as we shall show in the present subsection.
Nevertheless, in the next subsection we will complement
this abstract theory
by showing that sufficiently ``nice'' embeddings of a graph in the plane
give rise to abstract duals,
and conversely that every locally finite abstract dual arises in this way.

The basic theory of duals of infinite graphs was
developed by Thomassen \cite{Tho80,Tho82},
which we largely follow here;
see also Bruhn and Diestel \cite{BD06,Diestel_survey,Bruhn-Stein}
for a partially alternative approach.\footnote{
  The approach of Thomassen \cite{Tho80,Tho82}
  is based on the study of {\em finite}\/ cycles and minimal cutsets,
  as explained below.
  The alternative approach of Bruhn and Diestel
  \cite{BD06,Diestel_survey,Bruhn-Stein}
  introduces (by topological means) a notion of {\em infinite}\/ ``cycles'',
  and shows that this notion allows a somewhat cleaner duality theory.
  When $G^\dgg$ is locally finite, the two concepts of duality coincide
  \cite[Lemma~4.7]{BD06}.
  We are therefore entitled to use here the theorems from
  \cite{BD06,Diestel_survey,Bruhn-Stein}
  under the added hypothesis that $G^\dgg$ is locally finite.
}
Abstract duals can be defined for 2-connected graphs,
but in this case the dual may have multiple edges;
we therefore restrict ourselves  for simplicity
to the 3-connected case.

If $G=(V,E)$ is a 3-connected graph,
then an {\em abstract dual}\/ of $G$ is
a connected graph $G^\dgg=(V^\dgg,E^\dgg)$
together with a bijection $E\ni e\mapsto e^\dgg\in E^\dgg$
such that a finite set $C\subq E$ is a cycle in $G$
if and only if $C^\dgg:=\{e^\dgg \colon\: e\in E\}$
is a minimal cutset in $G^\dgg$.
We stress that in this generality the term ``dual''
is something of a misnomer, since $G$ need not be an abstract dual
of $G^\dgg$;
indeed, $G^\dgg$ need not have any abstract dual at all,
even when $G$ is planar and locally finite \cite[p.~266]{Tho80}.
However, if both $G$ and $G^\dgg$ are locally finite,
then the situation becomes particularly nice:

\bt{\bf(Locally finite abstract duals)}\label{T:dual}
Let $G=(V,E)$ be a locally finite 3-connected graph that has a locally finite
abstract dual $G^\dgg=(V^\dgg,E^\dgg)$. Then:
\begin{itemize}
\item[(i)] $G$, with the inverse map $E^\dgg\ni e^\dgg\mapsto e\in E$, is an
abstract dual of $G^\dgg$.
\item[(ii)] $G^\dgg$ is 3-connected.
\item[(iii)] $G^\dgg$ is, up to isomorphism, the only abstract dual of $G$.
\item[(iv)] If $G$ has one end, then so does $G^\dgg$.
\item[(v)] If $G$ is quasi-transitive, then so is $G^\dgg$.
\end{itemize}
\et
\proof
Part~(ii) follows from \cite[Theorem~4.5]{Tho82}.
Then parts~(i) and (iii) follow from \cite[Theorem~9.5]{Tho80}.
Part~(iv) follows from \cite[Theorem~1.1]{Bruhn-Stein},
which states that there is a homeomorphism between the spaces of ends
of $G$ and $G^\dgg$
(considered as subspaces of the Freudenthal compactification,
to be defined in the next subsection);
so in particular $G$ has one end if and only if $G^\dgg$ does.
(Alternatively, this can be deduced from
Proposition~\ref{P:duemb}(iii) below, using Lemma~\ref{L:accum} and
Theorem~\ref{T:geabs}.)

We did not find a reference for part~(v), but this is not hard to prove using
some more results from \cite{Tho80}. We will prove the following, stronger
statement. Let $g\colon\: V\to V$ be a graph automorphism of $G$ and let
$g(\{v,w\}):=\{g(v),g(w)\}$ also denote the induced map $g \colon\: E\to E$ on
edges.  Then there exists an automorphism $g^\dgg$ of $G^\dgg$ such that the
induced map on edges satisfies $g^\dgg(e^\dgg)=g(e)^\dgg$. This shows that two
edges in $G^\dgg$ are of the same type if the corresponding edges in $G$ are
of the same type. Since by part~(i), duality is a symmetric relation, this
``if'' is an ``if and only if''. In particular, $G^\dgg$ is
edge-quasi-transitive if and only if $G$ is (with the same number of types of
edges).

To prove the existence of $g^\dgg$, we need some definitions. Let
$C=(V(C),E(C))$ be a cycle in the graph $G=(V,E)$.
We say that $C$ is an {\em induced cycle}\/
if $C$ is the subgraph of $G$ induced by $V(C)$;
equivalently, this says that $C$ has no {\em diagonals}\/,
i.e., there are no edges in $E\beh E(C)$ that have both endvertices in $V(C)$.
We say that $C$ is a {\em separating cycle}\/ if there
are vertices $v_1,v_2$ in $V\beh V(C)$ that are linked in $G$ but not in
the subgraph of $G$ induced by $V\beh V(C)$.

Now let $G=(V,E)$ be a locally finite 3-connected graph and let
$G^\dgg=(V^\dgg,E^\dgg)$ be a locally finite abstract dual of $G$.
Then \cite[Theorem~9.5]{Tho80} says that there is a one-to-one correspondence
between vertices of $G^\dgg$ and induced non-separating cycles of $G$.
Indeed, for each $v^\dgg\in V^\dgg$, the set $C^\dgg$ of edges in $G^\dgg$
that are incident to $v^\dgg$ has the property that
$C:=\{e \colon\: e^\dgg \in C^\dgg\}$
is an induced non-separating cycle of $G$,
and conversely, every induced non-separating cycle of $G$ arises in this way.

Now let $g$ be a graph automorphism of $G$. Since $g$ maps induced
non-separating cycles into induced non-separating cycles, there is a
bijection $g^\dgg\colon\: V^\dgg\to V^\dgg$ that maps a vertex $v^\dgg$ into a
vertex $w^\dgg$ of $G^\dgg$ if and only if $g$ maps the associated induced
non-separating cycles of $G$ into each other.  Since two vertices of $G^\dgg$
are adjacent if and only of the associated induced non-separating cycles of
$G$ share an edge, we see that $g^\dgg$ is a graph automorphism of $G^\dgg$
such that the induced map $g \colon\: E^\dgg\to E^\dgg$ on edges satisfies
$g^\dgg(e^\dgg)=g(e)^\dgg$.
\qed

Let $G$ and $G^\dgg$ be as in Theorem~\ref{T:dual},
let $v^\dgg\in V^\dgg$ be a vertex in $G^\dgg$,
and let $E^\dgg_{v^\dgg} :=
\{e^\dgg\in E^\dgg \colon\: e^\dgg\mbox{ is incident to }v^\dgg\}$.
Then $E^\dgg_{v^\dgg}$ is a minimal cutset in $G^\dgg$,
which is finite by virtue of the local finiteness of $G^\dgg$;
hence $E_{v^\dgg}:=\{e\in E \colon\: e^\dgg\in E^\dgg_{v^\dgg}\}$
is a cycle in $G$.
We loosely call $v^\dgg$ a {\em face}\/ of $G$
and we call $E_{v^\dgg}$ the {\em boundary}\/ of this face.
Indeed, we will see in the next subsection that for a suitable
embedding of $G$ in the plane $\R^2$,
$v^\dgg$ corresponds to a connected component of $\R^2 \beh G$
and $E_{v^\dgg}$ corresponds to its boundary.
In view of this, we abstractly define a {\em triangulation}\/
(resp.\ {\em quadrangulation}\/) to be a
locally finite 3-con\-nect\-ed graph $G$ that has an abstract
dual $G^\dgg$ in which each vertex has degree~3 (resp.\ 4).

Now assume that $G$ has one end. Then each finite minimal cutset $C$ of $G$
corresponds to a partition $\{V_1,V_2\}$ of $V$ into two connected components,
of which exactly one is infinite. Let $V_1,V_2$ denote the finite and infinite
component, respectively. Since by Theorem~\ref{T:dual}(i), $G$ is an abstract
dual of $G^\dgg$, the set $C^\dgg:=\{e^\dgg \colon\: e\in C\}$ is a cycle in
$G^\dgg$, and each cycle in $G^\dgg$ arises in this way. We call ${\rm
Int}(C^\dgg):=V_1$ and ${\rm Ext}(C^\dgg):=V_2$ the {\em interior}\/ and
{\em exterior}\/ of $C^\dgg$, respectively. We say that $C^\dgg$ {\em
surrounds}\/ a vertex $v\in V$ if $v\in{\rm Int}(C^\dgg)$.

An essential ingredient in our proofs in this paper
is an upper bound (Lemma~\ref{L:circbd})
for certain quasi-transitive triangulations $G$ on the number of cycles in
$G^\dgg$ of a given length surrounding a given vertex $v$ in $G$. To derive
this bound, we need some simple graph-theoretic facts.

\bl{\bf(Distances in a graph and its dual)}\label{L:dualdist}
Let $G=(V,E)$ be a 3-connected graph. Assume that each vertex in $G$ has
degree at most $d_{\rm max}$ and that $G$ has a locally finite abstract dual
$G^\dgg=(V^\dgg,E^\dgg)$. Then for all $e,f\in E$ we have
\be
  d(e^\dgg,f^\dgg) \;\leq\; (\ffrac{1}{2}d_{\rm max}-1)d(e,f)+1 \;,
\ee
where $d(e^\dgg,f^\dgg)$ denotes the distance between $e^\dgg$ and $f^\dgg$ in
the dual graph $G^\dgg$.
\el

\noindent
Note that $d_{\rm max}\geq 3$ since $G$ is 3-connected.

\proofof{Lemma~\ref{L:dualdist}}
If $e=f$ (hence $e^\dgg = f^\dgg$), the statement is trivial,
so consider the case $e\neq f$. 
Let $P$ be a path of minimal length that has $e$ and $f$ as final edges.
For each vertex $v$ of $G$,
let $E^\dgg_v:=\{e^\dgg \colon\: e\mbox{ is incident to }v\}$
denote the boundary of the corresponding face of $G^\dgg$.
Note that $E^\dgg_v$ is a cycle whose length is the degree of $v$.
If some $E^\dgg_v$ and $E^\dgg_w$ share an edge $e^\dgg$,
then $e$ connects $v$ and $w$.
For internal vertices $v,w$ of $P$, by the minimality of $P$,
this is possible only if $v$ and $w$ are adjacent in $P$.
Let $v_1,\ldots,v_n$ be the internal vertices of $P$, where
$n=d(e,f)$, and let $d_1,\ldots,d_n$ denote their degrees  in $G$.
Then the symmetric difference
$D:=E^\dgg_{v_1}\symdif\cdots\symdif E^\dgg_{v_n}$
consists of exactly $\sum_{k=1}^nd_k-2(n-1)$ edges which form a cycle in
$G^\dgg$ containing $e^\dgg$ and $f^\dgg$.
It follows that $G^\dgg$ contains two paths $P^\dgg_1,P^\dgg_2$
which have $e^\dgg$ and $f^\dgg$ as their final edges
and are otherwise edge-disjoint, and whose respective lengths
$k_1,k_2$ satisfy $k_1+k_2-2=\sum_{k=1}^nd_k-2(n-1)$.
We conclude from this that $\min(k_1,k_2) \leq (k_1+k_2)/2
\leq \ffrac{1}{2}(nd_{\rm max}-2n+4)$ and hence
$d(e^\dgg,f^\dgg)\leq(\ffrac{1}{2}d_{\rm max}-1)n+1$.
\qed

Now let $G=(V,E)$ be a locally finite 3-connected quasi-transitive graph
that has a locally finite abstract dual $G^\dgg=(V^\dgg,E^\dgg)$.
By Theorem~\ref{T:dual}(v), $G^\dgg$ is also quasi-transitive,
hence of bounded degree; let $d^\dgg_{\rm max}$ denote the maximal degree
of a vertex in $G^\dgg$.
By Lemma~\ref{L:geodesic}(b), $G$ contains at least one geodesic double ray;
let $V_0$ be the set of all vertices $v \in V$ that lie on
some geodesic double ray.  Then, by quasi-transitivity, the maximal distance
of any vertex in $G$ to the set $V_0$,
\be
   K  \;:=\;  \sup_{w \in V}  \inf_{v \in V_0} d(v,w)
\;,
\ee
is finite.
Note, finally, that by Lemma~\ref{L:geodesic}(a),
at each $v\in V$ there starts at least one geodesic ray.

\bp{\bf(Distance to a surrounding cycle)}\label{P:surround}
Let $G,G^\dgg$,  $d^\dgg_{\rm max}$ and $K$ be as above.
Assume that $G$ has one end.
Let $v\in V$, let $R$ be a geodesic ray in $G$ starting at $v$,
and let $C^\dgg$ be a cycle in $G^\dgg$ of length $L$ surrounding $v$.
Then $C^\dgg$ must cross one of the first $N$ edges of $R$, where
\be\label{Nbd}
N \;:=\; 1+K+\ffrac{1}{2}(\ffrac{1}{2}d^\dgg_{\rm max}-1)\,L.
\ee
\ep

\proof
As discussed above, the cycle $C^\dgg$ corresponds to
a minimal cutset $C$ of $G$,
which divides $G$ into two connected components,
one of which is finite and the other of which is infinite;
moreover, $v$ is contained by hypothesis in the finite component.
Since $R$ is an infinite ray starting at $v$,
it must use somewhere an edge of $C$;
let $f$ be the first such edge,
and let $f^\dgg \in C^\dgg$ be the corresponding dual edge.

Let $w$ be the point in $V_0$ that is closest to $v$,
let $P$ be a path of minimal length  $\ov K\le K$ linking $v$ and $w$,
and let $D$ be a geodesic double ray containing $w$.
Write $D=R_1\cup R_2$ where $R_1,R_2$ are geodesic rays starting at $w$,
and observe that $R'_1:=P\cup R_1$ and $R'_2:=P\cup R_2$ are
rays starting at $v$.
The cycle $C^\dgg$ must cross some edge in $R'_1$ and some edge in $R'_2$.
Let $e_1,e_2$ be the first edges (counting from $v$) in $R'_1,R'_2$
crossed by $C^\dgg$.
We distinguish two cases: I.\ $e_1\neq e_2$ and II.\ $e_1=e_2$.

In case I, $e_1$ and $e_2$ lie on $D$ and are the
$(\ov K+N_1)$-th and $(\ov K+N_2)$-th edge of the rays $R'_1$ and $R'_2$, say.
Then $C^\dgg$ is the union of two paths $P^\dgg_1,P^\dgg_2$,
each of which has $e_1^\dgg$ and $e_2^\dgg$ as their final edges,
and which are disjoint except for their overlap at $e_1^\dgg$ and $e_2^\dgg$.
Let $L_1,L_2$ denote the lengths of these paths in $G^\dgg$,
where $L_1+L_2-2=L$.
Without loss of generality we may assume that $f^\dgg$ lies on $P^\dgg_1$
and is the $M_1$-th edge of $P^\dgg_1$ starting from $e_1$
and the $M_2$-th edge starting from $e_2$, where $M_1+M_2-1=L_1$.
By Lemma~\ref{L:dualdist}
(applied with the roles of $G$ and $G^\dgg$ reversed), we have
\be
  \min(L_1,L_2) \;\geq\; d(e^\dgg_1,e^\dgg_2)+1 
                \;\geq\; \frac{1}{c}\big(d(e_1,e_2)-1\big)+1
                \;=\;    \frac{1}{c}\big({N}_1+{N}_2-2\big)+1  \;,
\ee
where we have abbreviated $c:=\ffrac{1}{2}d^\dgg_{\rm max}-1$.
Therefore
\be\label{MM}
 M_1+M_2 \;=\; L_1+1 \;=\; L+3-L_2 \;\leq\;
   L+2-\frac{1}{c}\big({N}_1+{N}_2-2\big) \;.
\ee
Using Lemma~\ref{L:dualdist}  again,
there exists a path in $G$ of length at most $c(M_1-1)+1$
that has $e_1$ and $f$ as its final edges,
and another path of length at most $c(M_2-1)+1$
that has $e_2$ and $f$ as its final edges.
Combining these paths with the pieces of $R'_1$ and $R'_2$ leading up to $e_1$
and $e_2$, respectively, we find two paths in $G$ starting at $v$ and with $f$
as their final edge, with lengths of at most
\be
{\ov K}+{N}_1+c(M_1-1)\quad\mbox{and}\quad {\ov K}+{N}_2+c(M_2-1) \;,
\ee
respectively. By (\ref{MM}), it follows that the average length of these two
paths is at most
\be\ba{l}
\dis\ffrac{1}{2}\big(2{\ov K}+{N}_1+{N}_2+c(M_1+M_2-2)\big)  \\[5pt]
\dis\quad
\leq\; \ffrac{1}{2}
         \Big[ 2K+{N}_1+{N}_2+c\big(L-\frac{1}{c} ({N}_1+{N}_2-2)\big)
         \Big]
\;=\; 1+K+\ffrac{1}{2}cL  \;.
\ec
Taking the shorter of these two paths, we have found a path of length at most
$1+K+\ffrac{1}{2}cL$ starting at $v$ and having $f$ as its final edge.
But since $R$ is a geodesic ray,
the distance from $v$ to $f$ along $R$ must be at most this.

In case II, $e :=e_1=e_2$ must lie on $P$;
it is the first edge of $P$ (counting from $v$)
that is crossed by some edge in $C^\dgg$.
Then $C^\dgg$ contains two paths
with lengths $L_1,L_2$ satisfying $L_1+L_2-2=L$
that have $e^\dgg$ and $f^\dgg$ as their final edges.
It follows that $d(e^\dgg,f^\dgg)\leq\frac{1}{2}L$.
Therefore, by Lemma~\ref{L:dualdist},
$d(e,f)\leq\frac{1}{2}cL+1$, which means that we can find a path in $G$ of
length at most $\frac{1}{2}cL+2$ that has $e$ and $f$ as its final edges.
Combining this path with the piece of $P$ leading from $v$ to $e$,
we again find a path of length at most $1+K+\ffrac{1}{2}cL$
starting at $v$ and having $f$ as its final edge.
\qed

\subsection{Planar embeddings}\label{subsec.planar}

In this section we collect some results that show that sufficiently ``nice''
embeddings of a graph in the plane give rise to a geometric dual that is also
an abstract dual, and conversely that graphs having a locally finite abstract
dual have ``nice'' embeddings in the plane such that the geometric dual
coincides with the abstract dual.

Embeddings of finite (planar) graphs in the plane are treated in almost any
elementary book on graph theory, but it is more difficult to find a good
reference for infinite planar graphs.
Some articles that we have found useful are
\cite{BIW95,BD06,Diestel-Kuhn_04a,Diestel-Kuhn_04b,Kron_06,RT02,Tho80,Tho82}.

Each graph $G$ gives rise to a topological space
--- which, by a slight abuse of notation, we shall continue to call $G$ ---
that is defined by first assigning a disjoint copy
of $[0,1]$ to each edge of $G$
and a point to each vertex of $G$
and then identifying the endpoints of intervals
with the endvertices of the corresponding edges.\footnote{
  More precisely, given a graph $G=(V,E)$,
  we start from the set $(E \times [0,1]) \cup (V \times \{2\})$,
  and then for each edge $e=xy$ we identify
  $(e,0)$ with $(x,2)$ and $(e,1)$ with $(y,2)$.
}
We equip $G$ with the quotient topology arising from this identification:
thus, a neighborhood base of an inner point on an edge
is formed by the open intervals on the edge containing that
point, while a neighborhood base of a vertex~$x$
is formed by the unions of half-open intervals $[x,z)$
containing $x$, one interval being taken from every edge $[x,y]$ incident to
$x$.\footnote{
    With this topology, $G$ is a 1-dimensional CW-complex
    \cite[pp.~5--6 and pp.~519 ff.]{Hatcher_02}.
}
Such a {\em topological realization}\/ of the graph $G$ is compact if and only
if $G$ is finite, and locally compact if and only if $G$ is locally finite.
It is metrizable if and only if $G$ is locally finite.\footnote{
   Indeed, if $x$ is a vertex of infinite degree,
   then $x$ does not have a countable neighborhood basis
   in the topology we have given $G$.
   (It is possible to equip $G$ with a different topology,
    which is always metrizable, in such a way that a neighborhood base
    of a vertex~$x$ is formed by the unions of half-open intervals $[x,z)$
    using {\em the same}\/ distance $\epsilon = d(x,z)$
    for each incident edge \cite{Geo11a}.
    However, we shall not use this topology.)
}

An {\em embedding}\/ of a graph $G$ in the plane is a continuous injective map
${\phi \colon\, G\to\R^2}$.\footnote{
    Traditionally an embedding is defined
    as a drawing in which the vertices are represented by distinct points
    and the edges are represented by closed continuous arcs joining their
    endvertices, mutually disjoint except possibly at their endpoints.
    Given an embedding in this sense,
    pasting together the continuous mappings
    from $[0,1]$ to $\R^2$ corresponding to the individual edges
    always yields a continuous map ${\phi \colon\, G\to\R^2}$
    (with the topology we have given $G$);
    and the converse is trivial.
    Therefore, our definition of embedding is equivalent to the
    traditional one.
}
A graph $G$ that can be embedded in the plane is called
{\em planar}\/. A {\em plane graph}\/ is a pair $(G,\phi)$ where $G$ is a
graph and $\phi$ is an embedding of $G$ in the plane. We (topologically)
identify the sphere $\Sb$ with the one-point compactification
$\R^2\cup\{\infty\}$ of the plane $\R^2$. Embeddings of graphs in the sphere
are defined analogously to embeddings in the plane; a graph can be
embedded in the sphere if and only if it can be embedded in the plane.

If $G$ is a finite graph,
then $\phi(G)$, being the continuous image of a compact set,
is a closed subset of $\R^2$. Moreover, $\phi$ is necessarily a homeomorphism
to its image, i.e., the inverse map $\phi^{-1} \colon\: \phi(G)\to G$ is also
continuous. Both statements fail in general when $G$ is infinite.
Mainly for these reasons, not much can be said in general about embeddings
of infinite graphs;  one needs extra conditions to proceed.

So let $G$ be a graph with no isolated vertices, and let $\phi\colon\:G\to\Sb$
be an embedding of $G$ in the sphere.  Following an idea of \cite{Kron_06}, we
say that $\phi$ is {\em self-accumulation-free}\/ if no point $z \in \phi(G)$
is an accumulation point of edges that do not contain $z$.\footnote{ More
  precisely, if $x = \phi^{-1}(z)$, we let $E(x) \subseteq G$ be the union of
  all edges containing $x$; we then require that $z$ is not contained in the
  closure of $\phi(G \setminus E(x))$.
  Kr\"on \cite{Kron_06} uses the term ``accumulation-free'',
  but we prefer the term ``self-accumulation-free'' in order to emphasize
  that only accumulation points {\em on the graph itself}\/ are forbidden.
  In this way we clearly distinguish this concept from the standard concepts
  ``VAP-free'' and ``EAP-free'' to be introduced later,
  which forbid accumulation points
  {\em everywhere in the finite plane $\R^2=\Sb\beh\{\infty\}$}\/.
}
We say that $\phi$ is {\em pointed}\/ if the image under $\phi$
of each ray in $G$ converges to a point in $\Sb$.\footnote{
  Observe that a ray $R \subseteq G$ is homeomorphic to $[0,\infty)$
  with its usual topology;  what we are requiring here is that
  $\lim\limits_{x \to +\infty \,(x \in R)} \phi(x)$ exists in $\Sb$.
  Since $\Sb$ is compact, $\phi(x)$ must necessarily have
  {\em at least one}\/ limit point as $x \to +\infty$;
  what we are requiring here is that it have {\em exactly one}\/ limit point.
}
(Clearly, equivalent rays must converge to the same point.
Inequivalent rays may or may not converge to the same point.)

Recall that a compactification $\ov F$ of a topological space $F$
is a compact topological space $\ov F$ such that $F\subq\ov F$ is dense.
We will always require that $\ov F$ be Hausdorff.
As in \cite{RT02}, we say that a compactification $\ov G$
of a connected, locally finite (not necessarily planar) graph $G$
is {\em pointed}\/ if each ray in $G$ converges to some point
in $\ov G\beh G$.\footnote{
  Note that in {\em any}\/ compactification $\ov G$ of $G$,
  the limit points of a ray in $G$ cannot lie in $G$,
  because the topology of $\ov G$ extends that of $G$
  (and rays have no limit points in $G$).
  On the other hand, since $\ov G$ is compact,
  each ray in $G$ necessarily has {\em at least one}\/ limit point in $\ov G$.
  So what we are requiring here is that each ray should have
  {\em exactly one}\/ limit point in $\ov G$.
}
(Clearly, equivalent rays must converge to the same point.
Inequivalent rays may or may not converge to the same point.)
If $G$ is a graph and $\ov G$ is any compactification of $G$,
then an {\em embedding}\/ of $\ov G$ in the sphere is a
continuous injective map $\ov\phi\colon\:\ov G\to\Sb$;
since $\ov G$ is compact, it is necessarily a homeomorphism to its image.

\bl{\bf(Embeddings of compactifications)}\label{L:homeo}
Let $(G,\phi)$ be a locally finite
plane graph with no isolated vertices.
Then the following conditions are equivalent:
\begin{itemize}
\item[{(i)}] $\phi$ is self-accumulation-free.
\item[{(ii)}] $\phi$ is a homeomorphism to its image.
\item[{(iii)}] $\phi$ can be extended to an embedding
 $\ov\phi\colon\:\ov G\to\Sb$ in the sphere
 of some compacti\-fication $\ov G$ of $G$.
 [The extension is of course unique, since $G$ is dense in $\ov G$.]
\end{itemize}
Moreover, under these conditions, the embedding $\phi$
determines the compactification $\ov G$ uniquely
(up to trivial renamings of points in $\ov G\beh G$);
and $\ov G$ is a pointed compactification if and only if
$\phi$ is a pointed embedding.
\el

\proof
If $\phi$ is not a homeomorphism to its image, then we can find
$x_n,x\in G$ such that $\phi(x_n)\to\phi(x)$ but $x_n\not\to x$.
Since $\phi$ is continuous and injective,
the sequence $(x_n)$ cannot have any accumulation points in $G$
other than $x$;
so by passing to a subsequence way may assume that it has
no accumulation points at all.
Since any finite collection of edges is compact,
the sequence $(x_n)$ must visit infinitely many edges.
Since $G$ is locally finite, only finitely many of these edges can contain $x$.
It follows that $\phi$ is not self-accumulation-free.

Conversely, if $\phi$ is not self-accumulation-free,
then there exists $z = \phi(x) \in \phi(G)$
and a sequence $(x_n)$ belonging to edges not containing $x$,
such that $\phi(x_n) \to \phi(x)$.
Clearly $x_n \not\to x$,
so $\phi$ is not a homeomorphism to its image.

If (iii) holds, then $\ov\phi$ is a homeomorphism to its image,
hence so is its restriction $\phi$ to $G$.
Conversely, if $\phi$ is a homeomorphism to
its image, then we may topologically identify $G$ with its image $\phi(G)$.
Then the closure $\ov{\phi(G)}$ of $\phi(G)$ in the sphere $\Sb$ is a
compactification of $\phi(G)$ such that the identity map from $\phi(G)$ to
itself can be continuously extended to $\ov{\phi(G)}$; moreover,
$\ov{\phi(G)}$ is (up to renaming) the only compactification of $\phi(G)$
with this property.

Clearly, if $\ov{\phi(G)}$ is a pointed compactification,
then $\phi$ is a pointed embedding.
Conversely, if $\phi$ is a pointed embedding,
then each ray in $G$ converges to some point $z \in \Sb$;
but since $\phi$ is self-accumulation-free,
we have $z \notin \phi(G)$,
hence $\ov{\phi(G)}$ is a pointed compactification.
\qed

{\bf Remark.} If $G$ is not locally finite, then $\phi$ can {\em never}\/
be a homeomorphism of $G$ (with the topology we have given it)
to its image.  For if $x$ is a vertex of infinite degree,
then it is easy to choose points $x_n$ lying on infinitely many
distinct edges such that $\phi(x_n) \to \phi(x)$;
but $x_n \not\to x$.

\bigskip

Let $\phi\colon\:G\to\Sb$ be any embedding of a graph $G$ in the sphere.
Following \cite{Tho80,Tho82}, let us define a {\em vertex accumulation
  point}\/ (resp.\ {\em edge accumulation point}\/) of an embedding $\phi$ to
be a point in $\Sb$ such that each of its open neighborhoods contains
infinitely many vertices (resp.\ intersects infinitely many edges). We
abbreviate ``vertex accumulation point'' and ``edge accumulation point'' by
VAP and EAP, respectively.
Let us also say that a point $x\in\Sb$ is an
{\em endpoint}\/ of $(G,\phi)$ if there exists a ray in $G$ such that its
image under $\phi$ converges to $x$. The next lemma says that for
self-accumulation-free pointed embeddings of 2-connected graphs,
all these concepts coincide:

\bl{\bf (Endpoints of plane graphs)}\label{L:accum}
Let $G$ be a locally finite 2-connected graph,
and let $\phi\colon\: G\to\Sb$ be a self-accumulation-free pointed embedding
of $G$. Then the following four sets are equal:
\begin{quote}
\begin{enumerate}
  \item The set of vertex accumulation points of $(G,\phi)$.
  \item The set of edge accumulation points of $(G,\phi)$.
  \item The set of endpoints of $(G,\phi)$.
  \item $\ov{\phi(G)}\beh\phi(G)$.
    \end{enumerate}
\end{quote}
\el

\proof 
We will prove the inclusions (iii)$\subq$(i)$\subq$(iv)$\subq$(iii)
and the same with (i) replaced by (ii).
Clearly, every endpoint is also a vertex and edge accumulation point.
Since $\phi$ is self-accumulation-free and locally finite,
each vertex or edge accumulation point lies in $\ov{\phi(G)}\beh\phi(G)$.
For each point $x\in\ov{\phi(G)}\beh\phi(G)$,
we can find a sequence of points $x_n\in\phi(G)$,
all lying on different edges, such that $x_n\to x$.
Since $G$ is 2-connected, it can be shown that there exists a ray $R$
in $G$ whose image under $\phi$
passes through infinitely many of these points.
\footnote{
   Given an infinite set $B\subseteq E$, we can use local finiteness
   to extract an infinite subset $B' \subseteq B$
   that is pairwise vertex-disjoint;
   then by \cite[Proposition~8]{RT02},
   $G$ has a ray that uses infinitely many of the edges in $B'$.
   Given an infinite subset $T \subseteq V$,
   we can obviously choose an infinite set $B$ of edges
   containing all the vertices in $T$, and then proceed as before.
}
Since $\phi$ is pointed, it follows that $x$ is the limit of the $\phi$-image
of $R$.
\qed

  Let $\ov G$ be a pointed compactification of a connected locally finite
 (not necessarily planar) graph $G$.
  Then, by definition, each ray in $G$ converges to some limit in $\ov G\beh G$;
  and conversely, if $G$ is 2-connected, then the argument proving
  (iv)$\subq$(iii) in Lemma~\ref{L:accum} shows more generally
  \cite[p.~4591]{RT02} that each point in $\ov G\beh G$
  is the limit point of some ray in $G$.

Each connected locally finite graph $G$ has a unique (up to renaming)
pointed compactification $\ov G$ such that each point in $\ov G\beh G$
is the limit point of some ray in $G$ and moreover two nonequivalent rays
always converge to different limit points:
this is the {\em Freudenthal compactification}\/ ${\mathcal F}(G)$, see
\cite[Section~8.5]{Diestel_10} or \cite[Section~7]{RT02}.\footnote{
  The Freudenthal compactification is a general construction of
  point-set topology:  it is defined for locally compact Hausdorff spaces,
  or more generally for completely regular rim-compact spaces
  (see e.g.\ \cite{Dickman_88,Kron-Teufl}).
  For {\em locally finite}\/ graphs ---
  which are the only ones we are concerned with ---
  this topological definition coincides with the graph-theoretic definition
  given here or in \cite[Section~8.5]{Diestel_10}.
}
Clearly, ${\mathcal F}(G)\beh G$ is in bijection with the space of ends of $G$.
If $G$ is 2-connected, then, in a sense, ${\mathcal F}(G)$ is the ``largest''
pointed compactification of $G$ (compare the remarks on \cite[p.~4594]{RT02}).
At the other end of the scale, every connected, locally finite,
infinite graph $G$ has a smallest pointed compactification,
namely the one-point compactification $G^\bullet=G\cup\{\infty\}$
in which every infinite sequence of distinct vertices or edges
(hence in particular every ray) converges to the single point $\infty$.

By Lemma~\ref{L:homeo}, an embedding $\phi$ of a connected locally finite
graph $G$ is self-accumulation-free and pointed if and only
if $\phi$ extends to an embedding $\ov\phi$ of a pointed compactification $\ov
G$ of $G$. Richter and Thomassen \cite[Theorems~1 and 13]{RT02} have proven
the following key result concerning the existence and uniqueness of such
embeddings:

\bt{\bf(Embeddings of 3-connected planar graphs)}\label{T:3embed}
Let $G$ be a 3-connected locally finite planar graph.  Then:
\begin{itemize}
\item[(a)] There exists an embedding of
the Freudenthal compactification ${\mathcal F}(G)$ in the sphere $\Sb$.
\item[(b)] If $\ov G$ is any pointed compactification of $G$ and
$\ov\phi_1,\ov\phi_2$ are embeddings of $\ov G$ in the sphere $\Sb$,
then there exists a homeomorphism $h\colon\:\Sb\to\Sb$ such that
$\ov\phi_2=h\circ\ov\phi_1$.
\end{itemize} 
\et

\noindent
The unique (up to homeomorphism) embedding of the
Freudenthal compactification ${\mathcal F}(G)$ in the sphere $\Sb$
will be called the {\em Freudenthal embedding}\/.

Henceforth we assume that $G$ is a 2-connected locally finite graph
and that $\phi\colon\: G\to\Sb$ is a self-accumulation-free pointed embedding
of $G$ in the sphere.
By Lemma~\ref{L:homeo}, $\phi$ extends uniquely to an embedding
$\ov\phi\colon\:\ov G\to\Sb$ of an essentially unique
pointed compactification $\ov G$ of $G$,
and we have $\ov\phi(\ov G) = \ov{\phi(G)}$.
The connected components of the remaining open set $\Sb\beh\ov\phi(\ov G)$
are called the {\em faces}\/ of the plane graph $(G,\phi)$
[and also of $(\ov G,\ov\phi)$].
For any face $f$, we let $\pa f:=\ov f\beh f$
denote its (topological) boundary.
If $X$ is any topological space, we say that a subset $C \subseteq X$
is a {\em circle}\/ if it is homeomorphic to the unit circle $\Sb^1$.
We need the following fundamental facts about the boundaries of faces:

\bt{\bf(Boundaries of faces)}\label{T:facebound}
Let $G$ be a 2-connected locally finite (planar) graph, and let $\phi$ be a
self-accumulation-free pointed embedding of $G$ in the sphere $\Sb$.
Then each face $f$ of $(G,\phi)$ is bounded by a circle,
and exactly one of the following three possibilities holds:
\begin{quote}
\begin{itemize}
\item[(i)] $\pa f$ is the $\phi$-image of a \finite (finite\footnote{
   Of course, cycles in the standard graph-theoretic meaning of the word are
   by definition finite.
   We will always use the word in this standard sense;
   we shall not make explicit use of the ``infinite cycles''
   introduced (by topological means) by Diestel and collaborators
   \cite{Diestel-Kuhn_04a,Diestel-Kuhn_04b,BD06,Diestel_survey,Bruhn-Stein}.
   }$\!$)
cycle in $G$
[hence $\pa f \cap [\ov{\phi(G)}\beh\phi(G)] = \emptyset$].
\item[(ii)] $\pa f \cap \phi(G)$ is
  the disjoint union of a nonempty countable collection of double rays,
  and $\pa f \cap [\ov{\phi(G)}\beh\phi(G)] \neq\emptyset$.
\item[(iii)]  $\pa f \cap \phi(G) = \emptyset$
[hence $\pa f \cap [\ov{\phi(G)}\beh\phi(G)] = \pa f \simeq \Sb^1$].
\end{itemize}
\end{quote}
Moreover, when $\ov G$ is the Freudenthal compactification ${\mathcal F}(G)$,
the set $\pa f\cap\phi(G)$ is dense in $\pa f$;
in particular, case (iii) cannot occur.
\et

\proof
By Lemma~\ref{L:homeo}, $\phi$ extends uniquely to an embedding $\ov\phi$
of an (essentially unique) pointed compactification $\ov G$ of $G$.
Richter and Thomassen \cite[Proposition~3 and Theorem~7]{RT02}
have proven the fundamental result that every face of $(\ov G,\ov\phi)$
is bounded by a circle.
By Lemma~\ref{L:homeo}, such a circle is the image of a circle $C$ in $\ov G$
under the homeomorphism $\ov\phi$.
We shall therefore prove the more general result
that circles in $\ov G$ have properties analogous to those
stated in the theorem
(here $G$ need not be planar).
We first claim that
\begin{quote}
\begin{itemize}
\item[(a)] Whenever $C$ contains an inner point of an edge $e \in G$,
     it contains the entire edge $e$.
\item[(b)] Whenever $C$ contains a vertex $x \in G$,
     it contains precisely two edges of $G$ that are incident to $x$.
\end{itemize}
\end{quote}
For the Freudenthal compactification, (a) and (b) are proven in
\cite[Lemma~2.3]{Diestel-Kuhn_04b}. But since statements (a) and (b) concern
only a neighborhood of a point of $G$, they hold for any Hausdorff
compactification of $G$ (since $G$ is locally compact).
For the Freudenthal compactification, it is moreover proven in
\cite[Lemma~4.3]{Diestel-Kuhn_04a} that $C\cap G$ is dense in $C$.

By (a,b), the union of all edges belonging to $C$ is a 2-regular graph
$C\cap G$, so each connected component of $C\cap G$ is either a \finite cycle
or a double ray.
Since $C\cap G$ is moreover homeomorphic to a subset of $\Sb^1$,
we conclude that it is either empty [case (iii)],
a \finite cycle [case (i)],
or a nonempty disjoint union of double rays [case (ii)].
In the latter case, there can be only countably many double rays
(since $G$ is countable),
and $C$ must also contain some points of $\ov G\beh G$
(in fact, at least as many as there are double rays)
in order to have the topology of a circle.
\qed

\bigskip

\noi
{\bf Examples.}
1.\ The lattices shown in Figure~\ref{fig:quad} have one end, which is mapped
to $\infty$, and the boundary of every face is a cycle in $G$.
As we will see in Proposition~\ref{P:findu} below,
quasi-transitivity and the fact that there is only one end
imply that every face is bounded by a cycle.

2.\ Consider $\N\times\Z$ with nearest-neighbor edges, with its usual embedding
in the plane. This graph has one end, which is mapped to $\infty$.
It has a face that is bounded by a double ray together with
$\{\infty\} = \ov{\phi(G)}\beh\phi(G)$.

\begin{figure}[t]
\begin{center}
\includegraphics[width=10cm,clip]{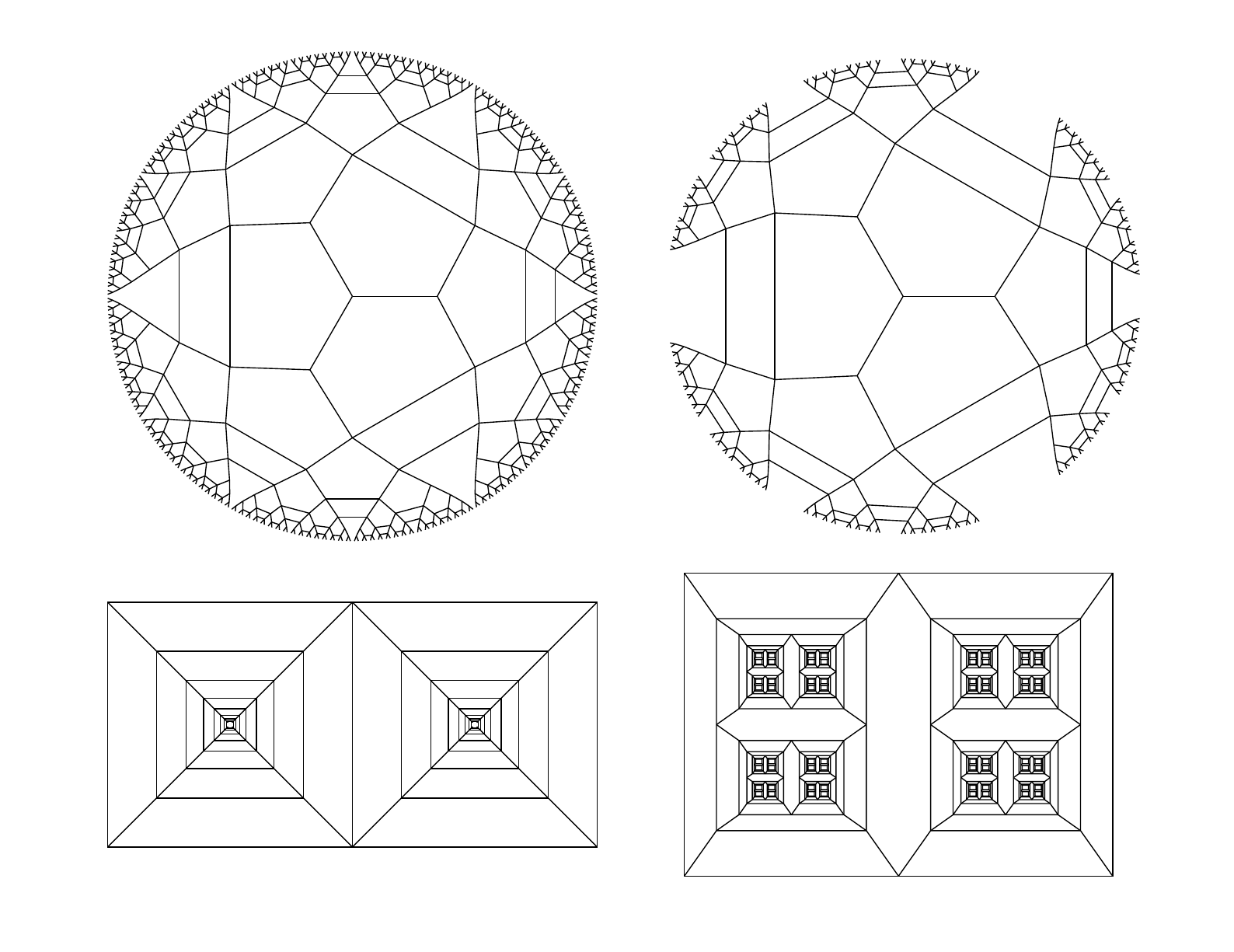}
\caption{\label{F:tree}
   Four examples of self-accumulation-free pointed embeddings
   of 3-connected graphs, demonstrating some of
   the different possibilities for the set of endpoints
   and the boundaries of faces.
}
\end{center}
\end{figure}

3.\ The graph in the top left of Figure~\ref{F:tree} has uncountably many ends,
which are mapped by the embedding $\phi$ onto the unit circle.
This embedding has one face that is bounded by endpoints alone
(namely, the exterior face),
as well as faces that are bounded by \finite cycles
and faces that are bounded by a double ray together with one endpoint.
This graph is constructed in the following way. Consider a 3-regular
tree with origin $\emptyset$.
Each vertex of $G$ is labeled by a finite (possibly empty) sequence
$\ell_0\ell_1\ell_2\cdots\ell_n$ where
$\ell_0\in\{0,1,2\}$ and $\ell_i\in\{0,1\}$ for $i\geq 1$;
and each ray emanating from $\emptyset$ is labeled
by an infinite sequence $\ell_0\ell_1\ell_2\cdots$
of the same type.
The tree can now be embedded in the unit disk
such that the ray $\ell_0\ell_1\ell_2\cdots$ converges to the point
\be\label{binray}
2\pi\cdot\frac{1}{3}\Big(\ell_0+\sum_{i=1}^\infty 2^{-i}\ell_i\Big)
\ee
on the unit circle.
We then make the tree into a 3-connected graph by adding
edges. For each vertex $\ell\neq\emptyset$, we draw edges between the points
$\ell01$ and $\ell10$ as well as between $\ell011$ and $\ell100$. Moreover,
around the origin, we draw the edges
\be
\{01,\, 10\},\, \{011,\, 100\},\, \{11,\, 20\},\, \{111,\, 200\},\,
 \{21,\, 00\},\, \{211,\, 000\}  \;.
\ee
This embedding is not a Freudenthal embedding, since two nonequivalent rays
(e.g.\ $0111\cdots$ and $1000\cdots$) may converge to the same point.
However, we can easily modify this example
to obtain a Freudenthal embedding of the same graph,
by replacing the binary expansion $2^{-i}\ell_i$ in (\ref{binray})
by ternary expansion $2 \cdot 3^{-i}\ell_i$,
as drawn in the top right of Figure~\ref{F:tree}.
In this latter example,
the set of ends is mapped onto a Cantor subset of the unit circle.
There is now a face $f$ of $(G,\phi)$ that contains
the exterior of the unit disc but also extends inside the unit disc;
its boundary consists of a countably infinite collection of
disjoint double rays together with all of $\ov{\phi(G)}\beh\phi(G)$.
Note that here $\pa f\cap\phi(G)$ is dense in $\pa f$.

4.\ The graphs in the bottom row of Figure~\ref{F:tree}
are examples of Freudenthal embeddings of graphs with more than one end,
in which each face is bounded by a \finite cycle.
The left graph has two ends
(and can easily be modified to produce any finite or countable number of ends),
while the right graph has uncountably many ends.

\bigskip

Let us now say that faces $f_1$ and $f_2$
{\em border}\/ each other in an edge $e$
if $e$ lies in the boundary
of both faces, with $f_1$ on one side of $e$ and $f_2$ on the other.
Since $G$ is 2-connected, it is not hard to see that no face borders itself;
rather, each edge~$e$ lies in the common boundary
of precisely two faces, which border each other in~$e$.
If $G$ is moreover 3-connected, then two faces border each other in
at most one edge.\footnote{
  In fact, it is easy to see that 2-edge-connectedness and
  3-edge-connectedness are sufficient, respectively,
  for these statements.
}
So let us assume henceforth that $G$ is 3-connected.
For each edge~$e$ of $G$, we let $e^*:=\{f_1,f_2\}$
denote the pair of faces of $G$ that border each other in~$e$.
We then define the {\em geometric dual}\/ $G^*=(V^*,E^*)$
of the plane graph $(G,\phi)$ to be the graph
whose vertex set $V^*$ is the set of all faces of $(G,\phi)$
that have at least one edge in their boundary
[i.e., fall in case (i) or (ii) of Theorem~\ref{T:facebound}]
and whose edge set is $E^*:=\{e^* \colon\: e\in E\}$.
Note that the edge sets $E$ and $E^*$ are in bijection
under $e \mapsto e^*$;
for any subset $B \subseteq E$
we denote by $B^* = \{e^* \colon\: e \in B\}$
the corresponding subset of $E^*$.

We can now prove that geometric duals, in this generality,
are always abstract duals:

\bt{\bf(Geometric duals are abstract duals)}\label{T:geabs}
Let $G$ be a locally finite 3-connected graph,
and let $\phi\colon\: G\to\Sb$ be a self-accumulation-free pointed embedding
of $G$ in the sphere.
Let $G^*$ be the geometric dual associated with this embedding.
Then:
\begin{itemize}
  \item[(i)] $G^*$ is connected.
  \item[(ii)] $G^*$ is an abstract dual of~$G$.
  \item[(iii)] $G^*$ is locally finite if and only if every face of $G$
     that has at least one edge in its boundary is bounded by a \finite cycle.
\end{itemize}
\et

\proof
For each vertex $v$ of $G$, let $E_v$ denote the set of edges incident to $v$.
Since $E_v$ is finite (say, of cardinality $n$)
and $\phi$ is self-accumulation-free,
basic topological considerations (based on the Jordan curve theorem)
imply that $\phi(v)$ is surrounded by exactly $n$ faces
that border each other pairwise (in cyclic order) in the edges of $E_v$,
and that $E_v^*$ is a cycle in $G^*=(V^*,E^*)$.

(i) Consider any pair $v^*, w^* \in V^*$.
By definition, $v^*$ and $w^*$ are faces of $(G,\phi)$
that contain at least one edge in their boundary;
so let $v$ and $w$, respectively, be any endvertex of any such edge.
Since $G$ is connected, there exists a path
$v = v_1, v_2, \ldots, v_k = w$ in $G$.
Since each cycle $E_{v_i}^*$ is connected,
and $E_{v_i}^* \cap E_{v_{i+1}}^* = \{\{v_i,v_{i+1}\}^*\} \neq \emptyset$,
the set $\bigcup_{i=1}^k E_{v_i}^*$ is connected;
moreover, $v^*$ belongs to $E_{v_1}^*$
and $w^*$ belongs to $E_{v_k}^*$.
So there exists a path in $\bigcup_{i=1}^k E_{v_i}^*$
from $v^*$ to $w^*$.

(ii) Consider any cycle $C$ in $G$.  Then $\phi(C)$ is a circle in $\phi(G)$
that (by the Jordan curve theorem) partitions $V^*$ into two sets (call them
$V^*_1$ and $V^*_2$) corresponding to faces lying in the two components of
$\Sb \beh \phi(C)$.
Since $C$ consists exactly of those edges $e\in E$
that lie on the boundary of some pair of faces
$v_1^*\in V_1^*$ and $v_2^*\in V_2^*$,
it follows that $V^*_1$ and $V^*_2$ are both nonempty,
so that the set $C^*$ is a cutset of $G^*$.

For each $v\in C$, the set $E_v\cap C$ consists of precisely two edges
(call them $e$ and $f$)
such that $e$ borders $v^*_1\in V^*_1$ and $v^*_2\in V^*_2$ and
$f$ borders $w^*_1\in V^*_1$ and $w^*_2\in V^*_2$.
Then $E_v^* \beh C^*$ is the disjoint union
of two paths $P^*_1,P^*_2$ in $G^*$ that link $v^*_1$ and $w^*_1$
in $V^*_1$, and $v^*_2$ and $w^*_2$ in $V^*_2$, respectively.
Doing this for all $v \in C$,
it follows that the set of all vertices in $V^*_1$ (resp.\ $V^*_2$)
that are incident to  an edge of the cutset $C^*$
is connected in $V^*_1$ (resp.\ $V^*_2$).
Therefore $C^*$ is a minimal cutset of $G^*$.

Conversely, if $B$ is a finite subset of $E$
such that $B^*$ is a cutset of $G^*$,
then for each $v\in V$, the cutset $B^*$ must intersect $E_v^*$
(which is a cycle in $G^*$) an even number of times
(that is, $B^* \cap E_v^*$ has even cardinality).
It follows that each vertex of $G$ is incident to
an even number of edges of $B$,
and hence $B$ contains a cycle $C$.
But we have just proven that $C^*$ is a cutset in $G^*$.
Therefore, if $B^*$ is a minimal cutset of $G^*$,
we necessarily have $B=C$.

(iii) By Theorem~\ref{T:facebound},
for every face $f$ of $(G,\phi)$
that has at least one edge in its boundary (i.e., belongs to $V^*$),
its boundary $\pa f$ must either be a \finite cycle or contain a double ray.
Clearly, $f$ has finite degree in $G^*$
in case $\pa f$ is a \finite cycle,
and has countably infinite degree in case $\pa f$ contains a double ray.
\qed

In particular, whenever $G$ is a locally finite 3-connected graph
and $\phi\colon\: G\to\Sb$ is a self-accumulation-free pointed embedding
such that every face of $G$ is bounded by a \finite cycle,
the geometric dual $G^*$ is a locally finite abstract dual of $G$,
so that all the nice properties stated in Theorem~\ref{T:dual} ensue.

Conversely, let us now show that
if a locally finite 3-connected graph $G$
has a locally finite abstract dual,
then $G$ has a unique self-accumulation-free pointed embedding,
and the geometric dual associated with this embedding
coincides with the abstract dual:

\bt{\bf(Embeddings of graphs with locally finite abstract duals)}
 \label{T:duembed}
Let $G=(V,E)$ be a locally finite 3-connected graph that has a locally finite
abstract dual $G^\dgg=(V^\dgg,E^\dgg)$. Then:
\begin{itemize}
\item[(i)] $G$ is planar.
\item[(ii)] In the Freudenthal embedding of $G$, each face is bounded by
 a \finite cycle,
 and the geometric dual $G^*$ of this embedding coincides with the
 (unique) abstract dual $G^\dgg$.
\item[(iii)] The Freudenthal embedding is (up to homeomorphism) the only
 self-ac\-cu\-mu\-la\-tion-free pointed embedding of $G$
 in the sphere.
\end{itemize}
\et

\proof
(i) is proven in \cite[Theorem~9.3]{Tho80}.
As a consequence of (i), Theorem~\ref{T:3embed} guarantees
the existence and uniqueness (up to homeomorphism)
of the Freudenthal embedding.

As a preliminary to (ii) and (iii),
consider any self-accumulation-free pointed embedding of $G$
in the sphere,
and let $G^*$ be the geometric dual associated with this embedding.
By Theorem~\ref{T:geabs}, $G^*$ is an abstract dual of $G$.
But by Theorem~\ref{T:dual}(iii), $G$ has a unique abstract dual $G^\dgg$.
Hence $G^\dgg = G^*$.

(ii)
For the Freudenthal embedding,
Theorem~\ref{T:facebound} guarantees that
every face has at least one edge in its boundary.
And since $G^\dgg = G^*$ is locally finite,
Theorem~\ref{T:geabs}(iii) implies that
every face is bounded by a \finite cycle.

(iii) Let $R_1,R_2$ be inequivalent rays in $G$. Then there
  exists a finite set of edges $B$ in $G$ such that the tails of $R_1$ and
  $R_2$ lie in different connected components of $G\beh B$. Making $B$ smaller
  if necessary, we may assume without loss of generality that $B$ is a minimal
  set with this property. It is not hard to see that $B$ must then be a
  minimal cutset.
Since $G$ is an abstract dual of $G^\dgg$ [by Theorem~\ref{T:dual}(i)],
$B^\dgg$ is a cycle in $G^\dgg$.
Since $G^\dgg = G^*$, $B^*$ is a cycle in $G^*$.
By (ii), each of the vertices of the cycle $B^*$ is a face of $(G,\phi)$
whose boundary is a \finite cycle in $G$.
The sum modulo 2 of all these \finite cycles forms a pair of
disjoint \finite cycles in $G$, each of
which separates the tail of $\phi(R_1)$ from the tail of $\phi(R_2)$.
As a result, these tails necessarily converge to different endpoints.
Since this holds for any pair of inequivalent rays, we conclude
that $\ov G$ must be the Freudenthal compactification,
and $\phi$ the Freudenthal embedding.
\qed

Until now we have defined dual graphs
--- whether in the abstract or geometric sense ---
simply as abstract graphs without a given embedding in the plane.
However, given an embedding of a planar graph,
there is a natural way to embed its dual.
Let $G$ be a locally finite 3-connected graph,
let $\phi \colon\: G \to \Sb$ be a self-accumulation-free pointed embedding
of $G$ in which each face is bounded by a \finite cycle,
and let $G^*$ be the geometric dual of $(G,\phi)$.
If $e$ is an edge of $G$,
we denote by $\mathring{e}$ the interior of the edge $e$
in the topological realization of $G$
(namely, the edge without its endvertices).
We then say that an embedding $\phi^*$ of $G^*$ is a {\em dual embedding}\/
to $(G,\phi)$ if
\begin{quote}
\begin{itemize}
\item[(i)] For each $v^*\in V^*$, we have $\phi^*(v^*)\in v^*$.
\item[(ii)] For each $e\in E$, the arc $\phi^*(e^*)$ intersects
 $\phi(G)$ in a single point, which lies on $\phi(\mathring{e})$.
\end{itemize}
\end{quote}
Less formally, (i) says that each vertex of $G^*$ is represented by a point
lying in the correponding face of $(G,\phi)$,
and (ii) says that two such points that lie in faces that border each other
in an edge $e$ are linked by a dual edge $e^*$ that crosses $e$
in a single interior point and is otherwise disjoint from $\phi(G)$.
We then have:

\bp{\bf(Dual embeddings)}\label{P:duemb}
Let $G$ be a locally finite 3-connected graph,
and let $\phi\colon\: G\to\Sb$ be a self-accumulation-free pointed embedding
of $G$ such that each face of $(G,\phi)$ is bounded by a \finite cycle.
Let $G^*$ be the geometric dual associated with this embedding.
Then there exists a dual embedding $\phi^*$ of $G^*$.
Moreover, for each such $\phi^*$:
\begin{itemize}
\item[(i)] $\phi^*$ is self-accumulation-free and pointed, and each face of 
$(G^*,\phi^*)$ is bounded by a cycle.
\item[(ii)] $G$ is a geometric dual of $(G^*,\phi^*)$ and $\phi$ is a dual
embedding of $G$.
\item[(iii)] The sets of endpoints $\ov{\phi(G)}\beh\phi(G)$ and
$\ov{\phi^*(G^*)}\beh\phi^*(G^*)$ coincide.
\end{itemize}
\ep

\noindent
If $(G,\phi)$ and $(G^\ast,\phi^\ast)$ are as in Proposition~\ref{P:duemb},
then we say that they form a {\em geometric dual pair}\/.

\proofof{Proposition~\ref{P:duemb}}
To prove the existence of a dual embedding $\phi^*$ of $G^*$, we begin by
choosing a point $\phi^*(v^*)\in v^\ast$ for each face $v^\ast$ of $(G,\phi)$, 
and a point $x$ on $\phi(\mathring{e})$ for each edge $e\in E(G)$. 
Let  $x_1,\ldots,x_n$ be the points chosen on the edges of the cycle that bounds $v^*$. 
It follows from basic topological considerations\footnote{
   Indeed, this can be proven by repeated application of the
   Jordan curve theorem and the following basic topological fact:
   Let $C\subs\Sb$ be a circle, which by the Jordan curve theorem
   divides $\Sb$ into two connected open sets $U,V$;
   let $x,z$ be two distinct points on $C$, and let $y\in U$.
   Then there exists a continuous arc from $x$ to $z$ that lies,
   except for its endpoints, entirely in $U$ and passes through $y$.
}
that we can connect $\phi^*(v^*)$ to each of the points $x_1,\ldots,x_n$
by continuous arcs that are disjoint except at their common endpoint
$\phi^*(v^*)$ and  that lie entirely in $v^*$ except for their endpoints
 $x_1,\ldots,x_n$. 
Now if $e\in E(G)$ lies on the boundary of faces $v^\ast_1$ and $v^\ast_2$
and $x$ is the chosen point on $\phi(\mathring{e})$, 
then the concatenation of the arcs from $\phi^*(v^*_1)$ to $x$ and
from $x$ to $\phi^*(v^*_2)$ is an arc from $\phi^*(v^*_1)$ to
$\phi^*(v^*_2)$  that we take as our definition of $\phi^*(e^*)$.

To see that $\phi^*$ is self-accumulation-free, it suffices to show that for
each point $x\in G^*$ we can remove a finite set of edges $E^*_0$ from $G^*$
such that $\phi^*(x)\not\in\ov{\phi^*(G^*\beh E^*_0)}$.
If $x=v^*$ is a vertex of $G^*$, then it suffices to remove
the set $E^*_0$ of all edges incident to $v^*$.
Then $\phi^*(G^*\beh E^*_0)$ is contained in the complement of the face
$v^*$ and hence its closure cannot contain $\phi^*(x)$.
If $x$ lies on an edge $e^*$ connecting two vertices $v^*_1$ and $v^*_2$
of $G^*$, then we remove the set $E^*_0$ of all edges incident to
$v^*_1$ or $v^*_2$. Then $\phi^*(G^*\beh E^*_0)$ is contained in
the complement of the open set formed by the union of
the faces $v^*_1$ and $v^*_2$ and $\phi(\mathring{e})$, while $\phi^*(x)$ lies
inside this open set.

To see that $\phi^*$ is pointed, let $R^*$ be a ray in $G^*$ consisting of
consecutive vertices $v^*_0,v^*_1,\ldots$ connected by edges
$e^*_1,e^*_2,\ldots\;$. Then $v^*_0,v^*_1,\ldots$ are faces of $(G,\phi)$ which
by assumption are bounded by cycles $C_0,C_1,\ldots$ such that $C_{k-1}\cap
C_k=e_k$. Adding these cycles modulo 2 yields a double ray consisting of two
rays $R_1,R_2$ in $\phi(G)$. Moreover, we can  construct a third ray $R_3$ that lies
in the union of the cycles $C_0,C_1,\ldots\,$, passes through each of the edges
$e_1,e_2,\ldots\,$, and has an infinite intersection with both $R_1$ and $R_2$. In
particular, the rays $R_1,R_2$ and $R_3$ are all equivalent and thus, as $\phi$ is pointed, 
they all converge to the same point $x\in\Sb$. It
follows that for each open ball around $x$, there exists an $n$ such that the
sum modulo 2 of the cycles $C_n,C_{n+1},\ldots$ lies in this ball. But this is
a double ray which together with the point $x$ forms a circle in $\Sb$
that contains the tail of the ray $R^*$ in its interior.   
Any open ball around $x$ thus  eventually contains the tail of $R^*$,
which proves that $R^*$ converges to $x$.

This proves not only that $(G^*,\phi^*)$ is pointed, but also that every point
in $\Sb$ that is the limit of some ray in $(G^*,\phi^*)$ is also the limit of
some ray in $(G,\phi)$. It follows that the set of endpoints
of $(G^*,\phi^*)$ is contained in the set of endpoints of $(G,\phi)$:
that is, $\ov{\phi^*(G^*)}\beh\phi^*(G^*)\subq\ov{\phi(G)}\beh\phi(G)$.

To complete the proof, it now suffices to show that each face of
$(G^*,\phi^*)$ is bounded by a cycle [completing the proof of (i)]
and contains a single vertex of $(G,\phi)$ [which implies (ii)].
Then (ii) together with what we already got implies
the reverse inclusion
$\ov{\phi(G)}\beh\phi(G)\subq\ov{\phi^*(G^*)}\beh\phi^*(G^*)$,
so we conclude that (iii) holds.

Since every face of $(G,\phi)$ is bounded by a cycle, the set
$\ov{\phi(G)}\beh\phi(G)$ does not separate $\Sb$, so the same is true for
$\ov{\phi^*(G^*)}\beh\phi^*(G^*)$ which is contained in it. It follows that
each face $f$ of $(G^*,\phi^*)$ has at least one edge in its boundary and
hence by Theorem~\ref{T:facebound} is either a cycle $C^*$ or contains a
double ray $R^*$. For each vertex $v^*$ in $C^*$ or $R^*$ we have that
$\phi^*(v^*)$  lies inside a cycle $C$ of $G$ that crosses two consecutive
edges of $C^*$ or $R^*$. This cycle $C$ can have only one vertex in $f$ since
otherwise there would be an edge of $(G,\phi)$ that lies entirely in $f$,
which contradicts our construction of $\phi^*$. It follows that all edges in
$(G,\phi)$ that cross an edge in $C^*$ or $R^*$ are incident to one and the same
vertex $v$ of $G$ with $\phi(v)\in f$. Since $G$ is locally finite, we can
rule out the double ray so we conclude that $f$ is bounded by a cycle $C^*$
and contains a single vertex of $(G,\phi)$. \qed

Now recall the definition of vertex and edge accumulation points
(just before Lemma~\ref{L:accum}).
Following \cite{Tho80,Tho82}, let us say that an
embedding $\phi\colon\:G\to\R^2\cup\{\infty\}\cong\Sb$ is
{\em VAP-free}\/ (resp.\ {\em EAP-free}\/)
if $\phi(G)\subs\R^2$ and $\phi$ has no VAPs (resp.\ EAPs)
in the finite plane $\R^2$.
Note that if $G$ has at most finitely many isolated vertices
(in particular, if $G$ is connected), then every VAP is also an EAP.
For such graphs, EAP-free embeddings are automatically
self-accumulation-free and pointed.\footnote{
   Since $\phi(G)\subs\R^2$ and $\phi$ has no EAPs in $\R^2$,
   it must be self-accumulation-free.
   Since every ray has at least one accumulation point in $\R^2\cup\{\infty\}$
   and no accumulation points in $\R^2$, all rays converge to $\infty$.
   It seems to us that the concepts
   ``self-accumulation-free'' and ``pointed'' can now
   largely replace the more restricted notion of being EAP-free.
}

An EAP-free embedding $\phi$ of an infinite, connected,
locally finite graph $G$ can be extended to an embedding $\ov\phi$ of the
one-point compactification $G^\bullet$ of $G$ by setting
$\ov\phi(\infty)=\infty$.
Conversely, if the one-point compactification $G^\bullet$ of $G$
can be embedded in the sphere $\Sb\cong\R^2\cup\{\infty\}$,
then without loss of generality we may assume $\ov\phi(\infty)=\infty$,
yielding an EAP-free embedding of $G$. We then have:

\bp{\bf(Graphs with one end)}\label{P:oneend}
Let $G$ be a locally finite 3-connected planar graph.
\begin{itemize}
\item[(i)] If $G$ has at most one end, then it has an EAP-free embedding,
   which is unique up to homeomorphism and coincides with the
   Freudenthal embedding.
\end{itemize}
If in addition $G$ has a locally finite abstract dual $G^\dgg$, then:
\begin{itemize}
\item[(ii)] $G$ has an EAP-free embedding {\em if and only if}
it has at most one end;
and in this case $G$ and $G^\dgg$ can be represented as
a geometric dual pair $(G,\phi)$, $(G^*,\phi^*)$
such that both $(G,\phi)$ and $(G^*,\phi^*)$ are EAP-free.
\end{itemize}
\ep

\proof
(i) If $G$ is finite, then it obviously has an EAP-free embedding.
If $G$ is infinite with one end, then it has an EAP-free embedding
since its one-point compactification
coincides with its Freudenthal compactification,
which by Theorem~\ref{T:3embed}(a) can be embedded in the sphere.
And by Theorem~\ref{T:3embed}(b) this embedding is unique up to homeomorphism.

(ii) If $G$ has a locally finite abstract dual $G^\dgg$,
then Theorem~\ref{T:duembed}(iii) and Lemma~\ref{L:homeo} tell us
that the Freudenthal compactification is the {\em only}\/
pointed compactification of $G$ that is embeddable in the sphere;
therefore, such a graph has an EAP-free embedding
{\em if and only if}\/ it has at most one end.
(This statement can also be found in \cite[Theorem~5.9]{Tho82}.)
By Proposition~\ref{P:duemb}, $G$ and $G^\dgg$ can be represented as
a geometric dual pair $(G,\phi)$, $(G^*,\phi^*)$;
and Proposition~\ref{P:duemb}(iii) then implies in particular
that if $(G,\phi)$ is EAP-free, then so is $(G^\ast,\phi^\ast)$.
\qed

\subsection{Special embeddings}

In this section we discuss embeddings with special ``nice'' properties:
notably, straight-line embeddings and periodic embeddings.

We begin by citing a result of Thomassen \cite{Tho80}
on straight-line embeddings with convex faces:

\bp{\bf(Convex embeddings)}\label{P:convex}
Let $G$ be a locally finite, 3-connected graph with one end.
Then there exists an EAP-free embedding of $G$
in which every edge is a straight line segment and every face is convex.
If in addition $G$ has a locally finite abstract dual $G^\dgg$,
then there exists an EAP-free embedding of $G$ 
in which every face is a convex polygon.
\ep

\proof
By Proposition~\ref{P:oneend}(i), $G$ has an EAP-free embedding in the plane.
Then \cite[Theorems~7.4 and 8.6]{Tho80} imply that
$G$ has an EAP-free embedding
in which every edge is a straight line segment and every face is convex.
If in addition $G$ has a locally finite abstract dual $G^\dgg$,
then Theorem~\ref{T:duembed} guarantees that every face
is bounded by a \finite cycle.
\qed

Using Proposition~\ref{P:convex},
we can deduce that $G$ and $G^\dgg$ can be {\em jointly}\/ embedded
as a geometric dual pair such that edges are represented by
straight-line segments in both graphs:

\bp{\bf(Straight-line embedding of dual pair)}\label{P:straight}
Let $G$ be a locally finite, 3-connected graph
with one end.
Assume that $G$ has a locally finite abstract dual $G^\dgg$.
Then there exist EAP-free embeddings $\phi$ and $\phi^\dgg$
of $G$ and $G^\dgg$
in the (Euclidean) plane such that $(G,\phi)$ and $(G^\dgg,\phi^\dgg)$ form a
geometric dual pair and each edge of $(G,\phi)$ and $(G^\dgg,\phi^\dgg)$ is a
straight line segment.
\ep

\proof
By Theorem~\ref{T:duembed} and Proposition~\ref{P:duemb}, we may embed $G$ and
$G^\dgg$ as a geometric dual pair. Now we can also draw a graph $H$ in the
plane such that the vertex set of $H$ is the union of the vertex sets of $G$
and $G^\dgg$ and two vertices $v\in G$ and $w\in G^\dgg$ are joined by an edge
in $H$ when $v$ and $w$ are the endpoints of an edge in $G$ and its dual edge
in $G^\dgg$;
thus, $H$ is a quadrangulation, with $G$ and $G^\dgg$ being
its sublattices whose edges connect opposing vertices of the
quadrilaterals of $H$.
Then obviously $H$ is planar,
has an EAP-free embedding in the plane,
and has a locally finite geometric (and hence abstract) dual.
Since two faces of $H$ never have the
property that two non-adjacent vertices of $H$ both lie on the boundary of
both faces,
we see that $H$ is 3-connected.
By Proposition~\ref{P:oneend}(ii),
$H$ has one end, so applying Proposition~\ref{P:convex} to $H$ we see that $H$
has an EAP-free embedding such that each face is a convex
polygon (with four corners). Connecting opposite corners of these faces by
straight line segments, we obtain the required straight-line embeddings of
$(G,\phi)$ and $(G^\dgg,\phi^\dgg)$ as a geometric dual pair.
\qed

{\bf Remark.}
We do not know whether $G$ and $G^\dgg$ can be jointly embedded
as a geometric dual pair such that edges are represented by
straight line segments in both graphs
{\em and}\/ faces are convex in both graphs.

\bigskip

We next turn our attention to embeddings of quasi-transitive graphs.
Our main graphs of interest ---
namely, the graph $G$ and its sublattices $G_0$ and $G_1$
arising in Theorem~\ref{T:main} ---
are all locally finite, 3-connected, quasi-transitive, planar graphs
with one end that have a locally finite abstract dual $G^\dgg$.
Remarkably, this latter condition turns out to be superfluous:

\bp{\bf(Quasi-transitive graphs with one end)}\label{P:findu}
Let $G$ be a locally finite, 3-connected, quasi-transitive, planar graph with
one end.
Then:
\begin{itemize}
\item[(i)] $G$ has a locally finite abstract dual $G^\dgg$.
\item[(ii)] The Freudenthal compactification is the only pointed
compactification of $G$.
\item[(iii)] The Freudenthal embedding of $G$ is EAP-free
(when the endpoint is taken to map to $\infty$),
and each face is bounded by a \finite cycle.
\end{itemize}
\ep 

\proof
(ii) is trivial because $G$ has one end (and is 2-connected).
To prove (iii) and (i), consider the Freudenthal embedding of $G$.
By Proposition~\ref{P:oneend}(i), it is EAP-free.
By Theorem~\ref{T:facebound}, no face is bounded by endpoints alone;
and since $G$ has one end, a face boundary containing infinitely many edges
must consist of a single double ray together with the one endpoint.
But quasi-transitivity then implies
(see \cite[Theorem~2.3]{BIW95} or \cite[Theorem~8(1)]{Kron_06})
that no such infinite face can exist,
i.e., every face is bounded by a \finite cycle.
From Theorem~\ref{T:geabs} we then conclude
that $G$ has a locally finite abstract dual $G^\dgg$.
\qed

Quasi-transitivity is essential here, as the example of $\N\times\Z$
(see Example~2 after Theorem~\ref{T:facebound}) shows.
The assumption that $G$ has one end is also essential:
let $H$ be the graph with vertex set $\Z$
and edges $\{n,n+1\}$ and $\{2n,2n+2\}$ for all $n \in \Z$,
let $G$ be the ladder graph $\{0,1\} \times H$,
and let $\phi$ be the obvious embedding.
Then $G$ is quasi-transitive and 3-connected and has two ends;
$\phi$ is EAP-free and maps both ends to the point $\infty$
(i.e., this is a non-Freudenthal embedding);
and $(G,\phi)$ has a pair of faces
that are each bounded by a double ray together with the point $\infty$.

\bigskip

When planar graphs are quasi-transitive, it is natural to ask if they can be
embedded in a periodic way in the plane. This is not true if one restricts
oneself to the Euclidean plane, but remarkably, it turns out to be correct if
one also considers the hyperbolic plane. The following result has
been proven in \cite[Theorem~4.2]{Babai_97} (see also
\cite[Theorem~1]{Servatius_98}):

\bt{\bf(Periodic embeddings)}\label{T:period}
Every locally finite, 3-connected, quasi-transitive planar graph $G$ with one
end can be embedded in the Euclidean plane $\R^2$ or hyperbolic plane $\Hb^2$
such that every automorphism of $G$ corresponds to an isometry of $\R^2$ or
$\Hb^2$, respectively.
\et

We remark that in Theorem~\ref{T:period}, we do not know if
the embedding can be chosen in such a way that, {\em moreover}\/,
edges are represented by straight line segments in $\R^2$ or $\Hb^2$.
Note also that Propositions~\ref{P:convex} and \ref{P:straight}
talk about embeddings such that edges are straight line segments
in the Euclidean geometry. We are not aware of results about straight-line
embeddings in the hyperbolic geometry.

\subsection{Some examples}
Finally, let us describe a method for creating examples of graphs
satisfying the assumptions of Theorem~\ref{T:main}
--- i.e., locally finite, 3-connected, quasi-transitive triangulations
with one end --- and their duals.
All our examples come naturally with a periodic straight-line embedding
in the Euclidean or hyperbolic plane.

Let $p,q\geq 3$ be integers and let $ABC$ be a triangle whose angles (in
anticlockwise order) at the corners $A,B,C$ are $\pi/p$, $\pi/q$, and $\pi/2$,
respectively. Such a triangle can be constructed in either the sphere, the
Euclidean plane, or the hyperbolic plane, depending on whether $1/p+1/q+1/2$
is larger than, equal to, or less than 1, respectively. By reflecting the
triangle $ABC$ in one of its edges and continuing this process, we can
cover the whole space alternately by copies of $ABC$ and its mirror
image \cite[Section~2]{Cox64}. This yields a planar graph with vertices of types
$A$, $B$ and $C$ that are of degree $2p$, $2q$ and $4$, respectively. In
particular, each vertex of type $C$ is adjacent to two vertices of types $A$ and
$B$ each, in alternating order. We may view the $A$ and $B$ sublattices as
planar graphs in their own right by erasing the vertices of type $C$ and
viewing the four edges emanating from $C$ as two straight edges crossing each
other in $C$, where one connects two $A$'s and the other connects two $B$'s.
This yields two regular tesselations that are geometric duals of each other.
In the tesselation formed by the $A$ vertices, each vertex has degree $p$ and
each face is a regular polygon with $q$ edges. This regular tesselation is
denoted by the {\em Schl\"afli symbol}\/ $\{q,p\}$ \cite{Cox64}. Likewise, the
dual $B$ lattice has the Schl\"afli symbol $\{p,q\}$.

In particular, the tesselations with Schl\"afli symbol $\{3,p\}$ (with $p\geq
3$) are regular triangulations of the sphere, the Euclidean plane, or the
hyperbolic plane, depending on whether $p$ is less than, equal to, or larger
than 6, respectively. It is easy to see that $\{3,p\}$, as a graph, is
3-connected and vertex-transitive. It is finite for $p\leq 5$ and infinite for
$p\geq 6$. In particular, Theorem~\ref{T:main} applies when $G_0=\{3,p\}$ with
$p\geq 6$. The case $p=6$, which is the only Euclidean tesselation in this
class, yields $G_0 = $ triangular lattice, $G_1 = $ hexagonal lattice, and $G
= $ diced lattice.  The cases $p > 6$ yield hyperbolic tesselations.  The
graphs $\{3,6\}$ and $\{3,7\}$ and their duals are drawn in
Figure~\ref{fig:quad2}(b,d).

The (dual) tesselations with Schl\"afli symbol $\{p,3\}$ are planar Cayley
graphs in which every vertex has degree three. A full classification of graphs
with these properties can be found in \cite{Geo11}.
In particular, \cite[Table~1, items~12--19]{Geo11}
lists those that are 3-connected and have at most one end.
Note that all these graphs are vertex-transitive.

More general examples of quasi-transitive triangulations satisfying the
assumptions of Theorem~\ref{T:main} can be contructed by starting with any
regular tesselation and dividing the basic polygon into triangles in some
suitable way, so that the resulting graph is 3-connected. It would go to far
to attempt here a full classification of the class of tesselations covered by
Theorem~\ref{T:main}.

\section*{Acknowledgements}
We wish to thank Youjin Deng, Jesper Jacobsen, and Jes\'us Salas for many
helpful conversations over the course of this work.
We also thank Agelos Georgakopoulos, Sebastian M\"uller, Bruce Richter,
Jan Seidler,  and Carsten Thomassen for helpful discussions and correspondence
concerning infinite graphs and planar topology;
Neal Madras and Gordon Slade for correspondence concerning self-avoiding walks;
and Youjin Deng, Kun Chen and Yuan Huang for sharing with us
their preliminary Monte Carlo data.

The research of R.K.~and J.M.S.~was supported in part by
the grants GA\v CR 201-09-1931 and P201/12/2613.
The research of A.D.S.~was supported in part by
U.S.~National Science Foundation grant PHY--0424082.

\vglue1cm

\end{document}